\documentclass[preprint,prd,superscriptaddress,nofootinbib]{revtex4-2}

\usepackage{natbib}

\usepackage[utf8]{inputenc}
\usepackage{amsfonts,bm}
\usepackage{amsmath,amssymb,graphicx,textcomp,xcolor}
\usepackage{hyperref}
\usepackage[caption=false]{subfig}
\usepackage{slashed}
\usepackage{comment}
\usepackage[normalem]{ulem}

\def\msbar{$\overline{\rm MS}$ }

\begin{document}

\newcommand*{\BNL}{Physics Department, Brookhaven National Laboratory, Upton, NY 11973, USA}\affiliation{\BNL}
\newcommand*{\Edinburgh}{School of Physics and Astronomy, The University of Edinburgh, Edinburgh EH9 3FD, UK}\affiliation{\Edinburgh}
\newcommand*{\CU}{Physics Department, Columbia University, New York City, New York 10027, USA}\affiliation{\CU}
\newcommand*{\UConn}{Physics Department, University of Connecticut, Storrs, CT 06269-3046, USA}\affiliation{\UConn}

\newcommand{\todo}[1]{{\color{red}{\bf [TODO: \, #1]}}}
\newcommand{\ques}[1]{{\color{blue} {\bf [Q: \, #1]}}}

\newcommand{\kl}{K_{\textrm{L}}}
\newcommand{\hw}{H_{\textrm{W}}}
\newcommand{\chw}{\mathcal{H}_{\textrm{W}}}
\newcommand{\lla}{\left\langle}
\newcommand{\rra}{\right\rangle}
\newcommand{\tr}{\mathrm{Tr}}
\newcommand{\sinc}{\sin_{\textrm{c}}}
\newcommand{\sumint}[1]{\,\,\,
\mathclap{\displaystyle\int}\mathclap{\textstyle\sum}\,\,\,{}_{#1}}
\newcommand{\gfive}{\Gamma^5}
\newcommand{\gL}{\Gamma^L}
\newcommand{\tsep}{t_{\textrm{sep}}}
\newcommand{\dmax}{\delta_{\rm{max}}}
\newcommand{\wh}{\widehat}
\newcommand{\wt}{\widetilde}
\newcommand{\Res}{\textrm{Res}}
\newcommand{\gf}{G_{\rm F}}

\newcommand{\amend}[2]{{\color{red} { #1}}{\color{blue}{#2}}}

\title{An exploratory calculation of $\kl\to\mu^+\mu^-$ decay from Lattice QCD at physical pion mass}
\author{Peter Boyle}\affiliation{\BNL}\affiliation{\Edinburgh}
\author{En-Hung Chao}\affiliation{\CU}
\author{Norman Christ}\affiliation{\CU}
\author{Ceran Hu}\affiliation{\CU}
\author{Luchang Jin}\affiliation{\UConn}
\author{Yidi Zhao}\affiliation{\CU}

\date{\today}

\begin{abstract}
We compute the complex, long-distance two-photon-exchange amplitude which contributes to the rare $\kl\rightarrow\mu^+\mu^-$ decay from lattice QCD.  We use a $24^3\times 64$ physical-pion-mass gauge field ensemble at an inverse lattice spacing of $1.023$ GeV and a QED${}_\infty$-based formalism.
Our implementation strategies for all five non-SU$(3)$-flavor-suppressed diagram topologies are given in detail.
We achieve a 25\% statistical precision on the dispersive part of this long-distance amplitude.
This calculation is carried out with 2+1 quark flavors and therefore requires the addition of counter terms to compensate for the absence of the Glashow–Iliopoulos–Maiani mechanism.  These counter terms are not included in the current calculation and will be the subject of a second paper.
The precision of our results is limited by the reconstruction of the physical contribution of the $\eta$ intermediate state, for which various strategies are tested and compared.
\end{abstract}

\maketitle

\tableofcontents

\newpage

\section{Introduction}
\label{sec:intro}
The decay of a long-lived kaon into a pair of charged muons is a rare strangeness-changing-neutral current process which is forbidden at first order in the standard model.  Despite its rarity, clean signals have been obtained from experiment.
In particular, the state-of-the-art measurement from the E871 experiment at the Brookhaven National Laboratory~\cite{E871:2000wvm} has achieved a percent-level precision with the branching ratio~\cite{ParticleDataGroup:2022pth}
\begin{equation}\label{eq:br}
\textrm{Br}(\kl\rightarrow\mu^+\mu^-) = 6.84(11)\times 10^{-9}\,.
\end{equation}
Because this decay is forbidden at tree-level, it should provide a precision test of the
standard model that is sensitive to physics from a high-energy scale.
At one-loop in the electroweak interaction, this decay results from the exchange of two $W$-bosons and from two $W$-bosons and one $Z$-boson, entering at O($G_{\rm F}^2$), where $G_{\rm F}$ is the Fermi decay constant.  This is a short distance (SD) contribution that can be accurately calculated in electroweak and QCD perturbation theory.  
Currently, the best estimate from perturbation theory for this SD part includes the effects from the charm quark calculated to next-to-next-to-leading order, which alone leads to a branching ratio of $\textrm{Br}(\kl\to\mu^+\mu^-)_{\rm SD} = 0.79(12)\times 10^{-9}$~\cite{Gorbahn:2006bm}.

However, a direct comparison with experiment is obscured by a significant contribution to this decay of order $\gf\alpha^2$, where $\alpha$ is the fine-structure constant.  This third-order electroweak process involves the exchange of a single $W$ boson and two photons, a process which we will abbreviate as LD2$\gamma$.  An accurate determination of this two-photon exchange process is required if $\kl\rightarrow\mu^+\mu^-$ decay is to be used to test the high-energy structure of the standard model at one loop. This two-photon exchange amplitude is a long-distance (LD) process involving energies at the 1 GeV scale and below.  This amplitude is non-perturbative and its first-principles determination requires a Lattice Quantum Chromodynamics (LQCD) calculation.  

Unlike the O$(\gf^2$) SD  part, which contributes only to the dispersive part of the decay amplitude and will be referred as the SD$\gf^2$ amplitude below, the two-photon exchange contribution is the sum of a dispersive and an absorptive parts. 
Throughout this work, we adopt the phase convention where the latter is purely imaginary and its algebraic sign will be specified later in the context of our lattice calculation.
With this choice, the dispersive part of the LD$2\gamma$ decay amplitude is purely real.

The absorptive part of the LD2$\gamma$ decay amplitude -- and therefore its imaginary part -- can be determined directly by using the optical theorem and the known $\kl\rightarrow\gamma\gamma$ decay rate~\cite{Martin:1970ai}.  This imaginary part alone results in a branching ratio of $\textrm{Br}(\kl\rightarrow\mu^+\mu^-)_{\rm abs} = 6.59(5)\times 10^{-9}$, representing 96\% of the experimental $\kl\to\mu^+\mu^-$ decay rate.  Since to a good approximation the LD2$\gamma$ process gives the entire imaginary part of the decay amplitude, the magnitudes of both the real and imaginary parts of the $\kl\to\mu^+\mu^-$ decay amplitude $\mathcal{A}_{\kl\mu\mu}$ are known from experiment:
\begin{eqnarray}
\left|\mathrm{Re}\mathcal{A}^\mathrm{expt}_{\kl\mu\mu}\right| &=& 1.53(14) \times 10^{4} \textrm{ MeV}^{3}\,, \\
|\mathrm{Im}\mathcal{A}^\mathrm{expt}_{\kl\mu\mu}|              &=& 7.10(3) \times 10^{4} \textrm{ MeV}^{3},
\end{eqnarray}
using conventions defined below [Eq.~\eqref{eq:rate2amp}].

In contrast, the real part of the LD2$\gamma$ amplitude cannot be determined from other measurable physical processes.
Over many decades there have been attempts to determine the real part from phenomenological models~\cite{DAmbrosio:1997eof, Knecht:1999gb, Isidori:2003ts}.
Very recently, an analysis of the $\kl\rightarrow\gamma^*\gamma^*$ transition form factor based on dispersive techniques has been performed~\cite{Hoferichter:2023wiy}, where the authors find a mild tension of at least 1.7$\sigma$ between the standard model and experiment from the real part of the $\kl\to\mu^+\mu^-$ decay amplitude. 
However, the relative sign between the LD2$\gamma$ and SD$\gf^2$ amplitudes cannot be resolved by any of these approaches.
In fact, from the available phenomenological estimates, the real part of the combined LD2$\gamma$ and SD$\gf^2$ amplitudes has a different sign depending on whether the interference between these two amplitudes is constructive or destructive.
Thus, a calculation from first principles which determines the sign of the LD2$\gamma$ amplitude relative to the SD$\gf^2$ term will be especially valuable in a comparison between the standard model and experiment.

Fortunately, with major advances in computer technology and substantial algorithmic developments, this complex LD2$\gamma$ process is now accessible to a first-principles, LQCD calculation.
That such a complex process might be accessible to a LQCD calculation is suggested by the success of LQCD calculations of the hadronic light-by-light (HLbL) scattering contribution to the anomalous magnetic moment ($g-2$) of the muon.  In this process the most important Feynman diagram contains a muon propagator joined to a closed quark loop by three exchanged photons~\cite{Blum:2015gfa, Blum:2023vlm, Asmussen:2022oql, Chao:2021tvp, Chao:2022xzg}, a process similar in complexity to the LD2$\gamma$ process being studied here.  

As for the entire calculation of $g-2$ for the muon, the HLbL calculation is naturally performed in Euclidean space and is thus well suited to LQCD.  In contrast, the LD2$\gamma$ process involves a physical Minkowski-space decay and is described by a complex amplitude.  It is not a process that is usually studied using LQCD methods.  However, since the imaginary part comes from the two on-shell photons, particles within the perturbative part of the process, a LQCD calculation of the full complex amplitude is possible~\cite{Christ:2020dae, Christ:2020bzb} and LQCD calculations of the similar but less complex $\pi^0\to e^+ e^-$ decay have been successfully performed~\cite{Christ:2022rho}.  

An important purpose of this earlier work was to develop methods that could be applied to carry out the calculation presented here.  Since QED can be accurately treated in an infinite-volume space-time continuum using perturbation theory, we combine that formulation of QED with the finite-volume lattice treatment of QCD, a method referred to as QED${}_{\infty}$.  This detailed approach to the calculation of the LD2$\gamma$ amplitude was proposed and the accompanying systematic errors analyzed in detail in Ref.~\cite{Chao:2024vvl}. 

The usual Wick rotation that allows Minkowski-space quantities to be computed using a Euclidean metric faces two obstacles in case of the LD2$\gamma$ amplitude.  The first is the presence of propagator poles in the perturbative part of the amplitude that obstruct the usual Wick rotation of the contour over which an explicit integration of an internal energy variable is performed.  As in the case of $\pi^0\to e^+ e^-$ decay, the contribution of these poles can also be obtained from Euclidean-space QCD amplitudes, allowing a LQCD calculation of the entire complex amplitude.

The second difficulty arises from the large-time behavior of the hadronic Green's functions when these functions are analytically continued in time during the Wick rotation of the integral over time that accompanies the Wick rotation of the internal loop energy integral.  This difficulty can be treated by evaluating the decay amplitude resulting from the Schr\"odinger evolution over a finite time $T_{\mathrm{max}}$ and then taking the limit $T_{\mathrm{max}}\to\infty$.  Intermediate states in the decay process with energy less than the kaon mass, such as a single intermediate off-shell pion ($\kl\to\pi^0\to\mu^+\mu^-$), result in a Wick-rotated time contour made up of individual segments which separately diverge exponentially at large $T_{\mathrm{max}}$, although their sum remains finite.

Specifically, the Wick-rotated time integration contour can be divided into two parts.  The first is the usual integral in Euclidean time, $0 \le t_0 \le T_{\mathrm{max}}$ which, because of these states, grows exponentially as $T_{\mathrm{max}}\to\infty$.  The second is the circular arc portion of the complex $t_0$ integral with a radius $T_{\mathrm{max}}$ joining the original $-it_0=T_{\mathrm{max}}$ end point on the real, Minkowski-time axis to the $t_0=T_{\mathrm{max}}$ end point of the rotated contour on the imaginary, Euclidean-time axis.  While normally neglected as $T_{\mathrm{max}}\to\infty$, when states $|n\rangle$ with $E_n < M_K$ are present, this semicircular contour also grows exponentially as $T_{\mathrm{max}}\to\infty$, canceling the exponential growth in the Euclidean time integration.  

In a finite-volume calculation, such QCD energy eigenstates $|n\rangle$ are discrete.  Their contributions are simple exponential functions of $t_0$.  As a result, the contributions from the integrals over these semicircular contours are simple exponentials in $T_{\mathrm{max}}$.  These terms can be directly evaluated from the product of explicit matrix that can be separately calculated using LQCD and then used to remove the exponentially growing terms from a LQCD calculation in which such exponential growth appears.

In Ref.~\cite{Chao:2024vvl}, the authors identify the intermediate states which are less energetic than the kaon and require the special treatment described above. 
For physical kinematics, the most important such intermediate states are a single pion ($\pi^0$) at rest and a finite-volume two-pion state with non-zero spatial momentum combined with a photon ($\pi\pi\gamma$) whose total energy in the kaon rest system is less than $M_K$.  The $\pi^0$ state is straightforward to deal with as in previous calculations of the $K_{\rm L}$--$K_{\rm S}$ mass difference, $\Delta M_K$~\cite{Bai:2014cva, Huo:2025bhq}.

The $\pi\pi\gamma$ intermediate state offers more serious challenges.  First, with our QED$_\infty$ treatment of electromagnetism, there can be one or more finite-volume $\pi\pi$ states with non-zero spatial momentum in the kaon rest system with a recoiling infinite-volume photon carrying the opposite momentum.  (In fact, because of momentum non-conservation in QED$_\infty$, the $\pi\pi$ and photon momenta need not be equal and opposite, although such terms are suppressed by powers of $1/L$  and likely unimportant~\cite{Chao:2024vvl}.)  For free-particle kinematics $L\ge 11.8$~fm is required if the energy of a $\pi\pi\gamma$ state is to fall below $M_K$. If momentum non-conservation is considered then such effects will enter only for $L> 6$ fm. Thus, for our $L=4.7$ fm volume no exponentially growing terms resulting from finite-volume $\pi\pi\gamma$ states will occur.  For the future studies of larger volumes, one may need to deal with multiple, $\pi\pi\gamma$ states that each give exponentially growing behavior.

Once the exponential growing effects of states with energy less than $M_K$ have been properly managed, the result is the desired Minkowski-space LD2$\gamma$ decay amplitude computed in finite volume.  If the finite-volume hadronic amplitude being computed using LQCD contains no multi-particle intermediate states with energy below $M_K$, then the resulting position-space amplitude is exponentially localized and our QED$_\infty$ result will have finite-volume errors that decrease exponentially in the linear size of that volume.  However, this condition is not obeyed by the $\pi\pi\gamma$ state discussed above.

The two pions and photon in the $\pi\pi\gamma$ state with energy below $M_K$ can be described as making up a ``propagating" state with only power-law suppressed finite-volume effects.  Unlike the propagating two-pion intermediate states which enter the calculation of $\Delta M_K$ where a correction for the finite-volume effects of the $\pi\pi$ intermediate states is known~\cite{Christ:2015pwa}, a correction for the potentially large finite-volume effects arising from the three-particle $\pi\pi\gamma$ states for our coordinate-space-based framework is at present unknown.  However, a recent paper of Feng and Tuo~\cite{Tuo:2024bhm} provides a valuable analysis of this problem. Fortunately, for the limited energy at or below $M_K$ such three-particle states are expected to be phase-space suppressed and contribute little the LD2$\gamma$ amplitude being studied.  

In Ref.~\cite{Chao:2024vvl} a detailed phenomenological estimate of the size of the low-energy $\pi\pi\gamma$ state contribution is performed, concluding that these effects are at or below the 10\% level.  Thus, in this paper we will neglect the effects of these states and include a 10\% systematic error resulting from this approximation.  There are Green's functions in the calculation presented here in which the effects of a $\pi\pi\gamma$ might be seen.  Our failure to observe such effects is consistent the expectation that they can indeed be neglected.

In this work, we apply the proposed computational strategy to a $24^3\times 64$, inverse lattice spacing $1/a=1.023$ GeV physical-pion-mass domain wall fermion ensemble (`24ID') with $N_{\rm f} = 2+1$ dynamical quark flavors from the RBC/UKQCD collaboration. 
In the present work, we include three quark flavors in the valence sector. In the absence of the Glashow–Iliopoulos–Maiani (GIM) mechanism, the computed amplitude will contain unphysical, logarithmically divergent terms~\cite{Isidori:2005tv}, which must be corrected by the addition of explicit counter terms.  This issue will be addressed briefly in Section~\ref{sect:amp-lat}, but a complete treatment of those counter terms will be the subject of a future publication.

In spite of potentially large discretization effects due to the coarse lattice spacing, such an exploratory study allows us to develop and demonstrate our numerical implementation strategies and recognize possible challenges to overcome.

One such difficulty that is studied at length is the contribution of an intermediate $\eta$-meson. 
Although we do not have an exponentially growing contribution in the limit of large Euclidean time for the case of the $\eta$, owing to the typical temporal extent of the lattice and the expected unfavorable signal-to-noise ratio from a lattice calculation with quark-line disconnected diagrams, any practical truncation in the time integration will leave a large unphysical exponentially-falling $\eta$ contribution ($\propto \exp[{-(M_\eta-M_K)T_{\max}}]/(M_\eta-M_K)$), since the mass of the $\eta$ is very close to that of the kaon.
This unphysical $\eta$ contribution needs to be carefully removed with its physical part preserved.

As will be argued in Section~\ref{sect:form}, with exact isospin symmetry, the electromagnetic (EM) current can be written as a linear combination of vector currents transforming as operators with isospin $I=0$ and $I=1$.
Consequently, the product of two EM currents can be decomposed into combinations with $I=0$ and 2 and $I=1$.  Since the $I=1$ combination does not couple to the $\eta$, this term can be precisely determined and used to search for the effects of a possible $\pi\pi\gamma$ intermediate state.

This paper is structured as follows. 
In Section~\ref{sect:form}, the notation and conventions used will be specified and the theoretical framework reviewed. 
The simulation details and numerical results will be discussed in Section~\ref{sect:res}.
In particular, strategies are developed and compared to reconstruct the physical contribution of the noisier $I=0,2$ operator product, a contribution which dominates our error budget.
In Section~\ref{sect:sign}, we address the relative sign between the computed LD$2\gamma$ amplitude and the earlier result for the SD$\gf^2$ amplitude.
Finally, we conclude in Section~\ref{sect:concl} with a discussion of future plans and the limited conclusions that can be drawn from our calculation performed at a single, relatively coarse lattice spacing.

The implementation details of the Wick-contractions and the (approximate) low Dirac eigenmodes which allow us to improve the quality of the simulated data will be deferred to Appendices~\ref{sect:comp} and~\ref{sect:a2a}.
The subtleties in obtaining O($a^2$)-accurate results while treating the unphysical intermediate states are explained in Appendix~\ref{sect:unphys}.
A supplementary investigation of the effects of three-pion intermediate states is reported in Appendix~\ref{sec:3pi}.  Except where stated otherwise, all numerical quantities in this paper are expressed in lattice units.

\section{Theoretical Framework}\label{sect:form}
\begin{figure}[h]
\centering
\includegraphics[width=0.5\textwidth]{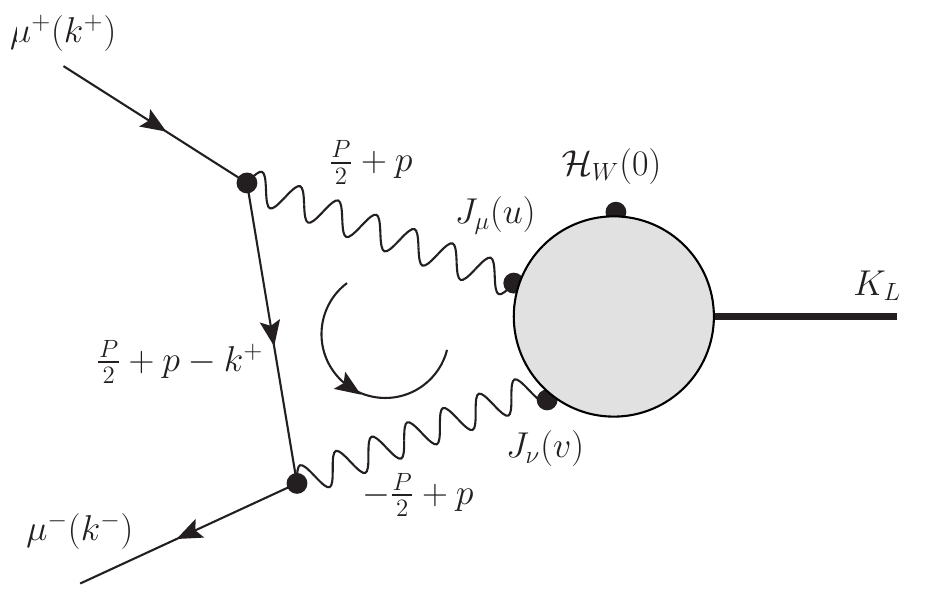}
\caption{Feynman graph representing the LD2$\gamma$ amplitude computed.}
\label{fig:Feynman}
\end{figure}

In our calculation, we neglect CP violation and therefore treat all CKM matrix elements as real. 
Likewise, we also use a $\kl$ state with the definite $CP=-1$, which then requires the final muon pair to be in a parity odd and charge conjugation even ${}^1S_0$ state, where both the muon and antimuon have the same helicity.  Figure~\ref{fig:Feynman}, reproduced from Ref.~\cite{Chao:2024vvl}, shows the combination of perturbative muon and photon propagators and QCD four-point function that must be combined to obtain the LD2$\gamma$ decay amplitude.  Here and in the remainder of the paper we will work in the rest system of the kaon.

Although Euclidean-space lattice QCD is being used to compute this two-photon decay process, the result of this calculation is presented as a conventional Minkowski-space amplitude obtained from an expansion of the expression:
\begin{eqnarray}
\mathcal{A}_{\mathrm{LD2}\gamma}\,(2\pi)^3\delta^3(\textbf{k}^+ + \textbf{k}^- - \textbf{P}_K) && \label{eq:A-def}\\
&& \hskip -0.6 in = \left\langle\mu^+(\textbf{k}^+)\,\mu^-(\textbf{k}^-)\left| \textrm{T}\left\{\exp\left[-i \int_{-\infty}^\infty dt\, H_\mathrm{EM}(t)\right]H_W\right\}\right|\kl(\textbf{P}_K)\right\rangle 
\nonumber
\end{eqnarray}
to fourth order in the EM Hamiltonian $H_\mathrm{EM}$.  The QCD energy and momentum eigenstates in Eq.~\eqref{eq:A-def} obey the normalization condition $\langle \textbf{p}\,'|\textbf{p}\rangle = 2E(\textbf{p})(2\pi)^3\delta^3(\textbf{p}\,'-\textbf{p})$ and $H_W$ is the effective $\Delta S=\pm1$ weak Hamiltonian.  Both muons have the same helicity, $h = \pm \frac{1}{2}$.

\subsection{Euclidean-space evaluation of the two-photon exchange amplitude}

In this section we apply the strategy developed in Ref.~\cite{Chao:2024vvl} and present the Euclidean-space Green's functions needed to compute the LD2$\gamma$ decay amplitude.  We will neglect the effects of the low-energy $\pi\pi\gamma$ intermediate states but remove the unphysical, exponentially-growing Euclidean-space contribution of the $\pi^0$.\footnote{
In Appendix~\ref{sec:3pi}, a supplementary study on the effects of the three-pion intermediate states is presented.
These intermediate states are often neglected in both phenomenological applications and in lattice calculations because of phase-space suppression.
Our findings in the Appendix~\ref{sec:3pi} provide numerical evidence to support the latter statement.}

This calculation can be expressed concretely as the difference of two amplitudes.  The first, $\mathcal{A}^{\mathrm{lat}}$, results from the lattice calculation of the Euclidean-space four-point function with an upper limit, $\delta_{\mathrm{max}}$, imposed on the time difference between the smaller time $v_0$ and the time at which the weak operator $\mathcal{H}_W(0)$ is evaluated.  Here we break the symmetry between the two currents by assuming the time order $u_0 \ge v_0$. 

The second amplitude $\mathcal{A}^{\mathrm{sub}}$ can be computed from three-point functions and is also a function of $\delta_{\mathrm{max}}$.  This second amplitude provides the subtraction that removes the unphysical exponentially-growing part of $\mathcal{A}^{\mathrm{lat}}$.

The complete physical LD2$\gamma$ amplitude can then be written~\footnote{The structure of Eqs.~\eqref{eq:Result_4pt}, \eqref{eq:Result_sub} and \eqref{eq:Result_total} appears similar to that of Eqs. (16), (18) and (19) in Ref.~\cite{Chao:2024vvl}.  However, the details are different since in the theoretical Eq.~(16) in Ref.~\cite{Chao:2024vvl} all contributions from states with energy below $M_K$ have been removed while in the more practical lattice Eq.~\eqref{eq:Result_4pt} no states have been removed and only an upper limit on the time, $\delta_{\max}$, has been imposed.} as:
\begin{equation}
\mathcal{A}_{\mathrm{LD2}\gamma} =\mathcal{A}^{\rm lat} - \mathcal{A}^{\rm sub}\,,  \label{eq:Result_total}
\end{equation}
where
\begin{eqnarray}
    \mathcal{A}^{\mathrm{lat}}(\delta_{\max},\delta_{\min},R_{\max}, \tsep) &=& \int_{-\delta_{\min}}^{\delta_{\max}} dv_0\int_V d^3\textbf{v}
  \int_{v_0}^{R_{\max}+v_0} du_0\int_V d^3\textbf{u}\; e^{M_K(u_0+v_0)/2}
  \label{eq:Result_4pt}\\
  && \hskip 0.4 in \times L_{\mu\nu}(u-v) 
 \mathcal{N}e^{M_K t_\mathrm{sep}} \langle \textrm{T}\left\{ J_\mu(u)J_\nu(v) \chw(0) \kl(-\tsep)\right\} \rangle\,,
  \nonumber
\end{eqnarray}
\begin{eqnarray}
    \mathcal{A}^{\mathrm{sub}}(\delta_{\max},\delta_{\min},R_{\max},\tsep) &=& \sum_n\int_V d^3\textbf{v}
  \int_{0}^{R_{\max}} dw_0\int_V d^3\textbf{u}\; e^{M_K w_0/2}\left[\frac{e^{-(E_n-M_K)\delta_{\rm max}}}{M_K-E_n}\right]
  \label{eq:Result_sub}\\
  && \hskip -1.5 in  
  \times L_{\mu\nu}(\textbf{u}-\textbf{v},w_0)\mathcal{N}e^{M_K t_\mathrm{sep}} \langle \textrm{T}\left\{ J_\mu(\textbf{u},w_0)J_\nu(\textbf{v},0)\right\}|n\rangle
  \langle n|\textrm{T}\left\{\chw(0) \kl(-\tsep)\right\} \rangle .
  \nonumber
\end{eqnarray}
Here the operators $J_\mu(u)$ and $J_\nu(v)$ are the EM currents, $\mathcal{H}_W(0)$ is the weak Hamiltonian, $\kl(-t_{\rm sep})$ a kaon
interpolating operator and $\mathcal{N}$ a factor which normalizes that operator.  The calculation is performed in the spatial volume $V$ with a time extent $T$.

We refer to the factor $L_{\mu\nu}(u-v)$ as the leptonic kernel.  It is obtained from the muon and photon propagators shown in Figure~\ref{fig:Feynman} after the integral over the loop momentum $p$ is evaluated using an appropriately deformed contour for the loop energy $p_0$ so that the latter integral, performed over the difference of the time coordinates $u_0-v_0$, converges.  This is discussed at greater length in Refs.~\cite{Chao:2024vvl, Christ:2022rho} and the leptonic kernel $L_{\mu\nu}(u-v)$ is given explicitly in Eq.~(20) in Ref.~\cite{Chao:2024vvl}.

Equation~\eqref{eq:Result_4pt} provides the details of the space-time integrations performed when the lattice four-point function is evaluated.  Both the positions $\textbf{u}$ and $\textbf{v}$ are integrated over $V$ and the kaon interpolating operator is separated in time from the weak Hamiltonian by $t_{\rm sep}$.  We must take care that the most negative value of $v_0$, $-\delta_{\min}$, is sufficiently large to capture the complete contribution to the decay amplitude but also small enough that the difference $t_{\rm sep}-\delta_{\min}$ is sufficiently large to guarantee that only a kaon state can be produced by $K_L(-t_{\rm sep})$. The quantity $R_{\rm max}$ is the largest time separation allowed between the two currents and must be chosen large enough to capture the full contribution to the decay amplitude.  A failure to observe a plateau as $R_{\rm max}$ is increased may be a sign of an unexpected contribution from a low-energy $\pi\pi\gamma$ state.

Equation~\eqref{eq:Result_sub} provides the unphysical part of Eq.~\eqref{eq:Result_4pt} which grows exponentially with increasing $\delta_{\max}$.  That equation can be obtained from Eq.~\eqref{eq:Result_4pt} in three steps.  First, a sum over intermediate states $|n\rangle\langle n|$ with $E_n < M_K$ is inserted between $J_\nu(v)$ and $\mathcal{H}_W(0)$.  Second, the integral over $v_0$ is performed over the interval $[0,\delta_{\max}]$ while the difference $w_0 = u_0-v_0$ is held fixed.  Third, the factor resulting from that $v_0$ integration, $(e^{-(E_n-M_K)\delta_{\rm max}}-1)/(M_K-E_n)$ is altered by dropping the physical, ``$-1$'' term in the numerator.  In our treatment of the states with $E_n < M_K$ only the intermediate $\pi^0$ state is introduced. 

The result, given in Eq.~\eqref{eq:Result_sub}, is the unphysical term which grows exponentially as $\delta_{\max}$ is increased.  This term must be subtracted from the lattice result given in Eq.~\eqref{eq:Result_4pt} to obtain the finite physical result.  Their combination, given in Eq.~\eqref{eq:Result_total}, should show a plateau as $\delta_{\rm max}$ is increased.  

A similar approach can be used to more accurately treat the $\eta$ intermediate state.  If the $\eta$ state is included in the right-hand side of Eq.~\eqref{eq:Result_sub}, the unphysical term that results will decrease slowly with increasing $\delta_{\max}$.  For practical values of $\delta_{\max}$ this unphysical term would be too large to neglect so that a similar subtraction is required.

The detailed treatment of Eqs.~\eqref{eq:Result_4pt} and \eqref{eq:Result_sub} for the case of a lattice calculation when the integral over time is replaced by a sum over discrete time values while avoiding O($a$) discretization errors is given in Appendix~\ref{sect:unphys}.

Henceforth, we will in addition use a more convenient definition of the $R_{\max}$-regulated leptonic kernel on the lattice which respects the Bose symmetry of the decay amplitude
\begin{equation}\label{eq:kmunu}
K_{\mu\nu}\left(r,R_{\max}\right)\equiv
\begin{cases}
2L_{\mu\nu}(r)e^{M_K r_0/2}\,,\quad\textrm{if $0 < r_0 \leq R_{\max}$,} \\
L_{\mu\nu}(\textbf{r},0)\,,\quad\textrm{if $r_0 = 0$,} \\ 
0 \quad\textrm{otherwise.}
\end{cases}
\end{equation}

\begin{figure}
\includegraphics[scale=1.0]{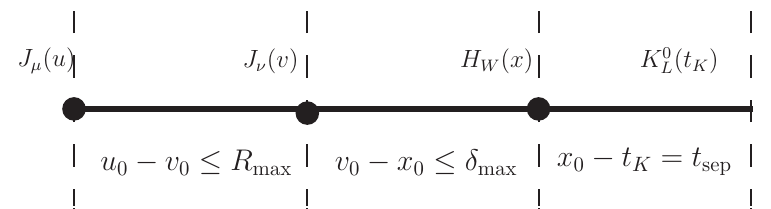}
\caption{Relevant separations involved in our formalism. Here we denote the position of the weak Hamiltonian with `$x$' instead of using the position `0' as in Eqs.~\eqref{eq:Result_4pt} and \eqref{eq:Result_sub}. This more general location for $\chw(x)$ is also used for clarity in some other places in the paper, e.g. Appendix~\ref{sect:comp}.}
\label{fig:dist}
\end{figure}

\subsection{Decay amplitude from the lattice}\label{sect:amp-lat}

In the continuum, the strangeness-changing $\Delta S = 1$ process can be represented by a weak effective Hamiltonian obtained by using QCD perturbation theory and the renormalization group to run down from the 100 GeV scale to a scale $\mu$ below the charm-quark mass~\cite{Buchalla:1995vs}
\begin{equation}
\chw^{\rm eff} = \frac{G_{\rm F}}{\sqrt{2}}V^*_{us} V_{ud}\sum_{i=1}^{10} C_i(\mu) Q_i^\text{\msbar} + \textrm{h.c.}
\label{eq:Buras}
\end{equation}
where $G_{\rm F}$ is the Fermi decay constant, the $V_{ij}$ are elements of the CKM matrix, the $C_i(\mu)$ are Wilson coefficients at the scale $\mu$. The operators $Q^\text{\msbar}_i$ are the ten linearly-dependent four-quark $\Delta S=1$ operators, which make up the traditional physical basis and are defined using QCD perturbation theory in the \msbar renormalization scheme.  Both the Wilson coefficients and the operators $Q^\text{\msbar}_i$ depend on the \msbar renormalization scale $\mu$.

At energy scales of 1.5 GeV and above where the quark masses and chiral condensate might be neglected, three-flavor QCD becomes symmetric under $SU(3)_L\times SU(3)_R$ chiral flavor group and it is convenient to work with what is known as the \textit{chiral} basis, a set of seven linearly independent four-quark operators each of which transforms in a specific representation of $SU(3)_L\times SU(3)_R$. Specific equations defining the seven operators in the chiral basis and for the ten operators in the physical basis can be found in Ref.~\cite{Lehner:2011fz} where the physical basis used here is referred to as \textit{basis II}. In order to increase the accuracy of perturbation theory, as in previous RBC/UKQCD work, we will consider the three-flavor operators $Q_i$ and their Wilson coefficients $C_i$ at energies above the charm scale.  When doing so we keep within the three-flavor theory and do not include a charm quark. 

In order to express the weak Hamiltonian $\chw^{\rm eff}$, written in the physical basis with Wilson coefficients obtained from QCD and from electroweak perturbation theory, in terms of operators defined directly using LQCD we apply the same renormalization procedure as appears in the RBC/UKQCD Collaboration paper describing a $K\to\pi\pi$ calculation performed using the same 24ID ensemble~\cite{RBC:2023ynh}.  In a series of steps the physical-basis \msbar operators given in Eq.~\eqref{eq:Buras} is transformed into an operator written in terms of the seven chiral-basis operators defined in the lattice formulation of QCD for the specific 24ID ensemble used in our calculation.  These seven lattice operators, ${Q_i^{\prime}}^\textrm{lat}$ are multiplied by chiral-basis, lattice-renormalized Wilson coefficients ${C_i^\prime}^\textrm{lat}$.

Schematically, the Wilson coefficients multiplying the seven chiral-basis weak operators defined on the lattice can be obtained from those computed in the continuum using the physical basis in the \msbar scheme at a given high renormalization scale $\mu_h$ by equating the lattice and continuum operators:
\begin{eqnarray}
\sum_{i=1}^{10} C_i(\mu_h) Q_i^\text{\msbar} &=& \label{eq:renorm}
\sum_{i=1}^{10} C_i(\mu_h)\Biggl\{
  \sum_{j=1}^7\Bigl[T+\frac{\alpha_s(\mu_h)}{4\pi}\Delta T\Bigr]_{ij} \Biggl\{
   \sum_{k=1}^7 Z^{\text{\msbar}\to\textrm{RI}}_{jk}(\mu_h) \\
&& \hskip 1.2 in \Biggl\{\sum_{\ell=1}^7 S_{k\ell}^{\mu_h\to\mu_l} \Biggl\{ 
\sum_{m=1}^{7}Z_{\ell m}^{\textrm{RI}\rightarrow\textrm{lat}}(\mu_l,a){Q_m^{\prime\,\textrm{lat} }}\Biggr\} \Biggr\} \Biggr\} \Biggr\} \nonumber \\
&=& \sum_{i=1}^7C_i^{\prime\,\textrm{lat}}Q^{\prime\,\textrm{lat}}_i\,.
\end{eqnarray}
Equating the coefficients of the seven operators $Q^{\prime\,\textrm{lat}}_i$ in these two equations then determines the Wilson coefficients $C^{\prime\,\textrm{lat}}_i$.

Reading from left to right on the right side of Eq.~\eqref{eq:renorm} we begin with the ten physical basis \msbar operators with their Wilson coefficients defined at the high \msbar energy scale $\mu_h = 4.006$ GeV, chosen sufficiently large that perturbation theory should be reliable. The first matrix $T+\frac{\alpha(\mu_h)}{4\pi}\Delta T$, found in Ref.~\cite{Lehner:2011fz} converts those \msbar operators to the chiral basis.  The second matrix $Z(\mu_h)^{\text{\msbar}\to\textrm{RI}}$, computed in perturbation theory in Ref.~\cite{Lehner:2011fz}, converts those \msbar chiral basis operators into the regularization-independent (RI) SMOM($\slashed{q},\slashed{q}$) scheme~\cite{RBC:2023ynh} at the same high energy scale $\mu_h$. 

Next the step-scaling matrix $S^{\mu_h\to\mu_l}$, is defined by
\begin{equation}
S_{k\ell}^{\mu_h\to \mu_l} = \sum_{n=1}^7 Z(\mu_h,a')_{k,n}^{\textrm{RI}\to\textrm{lat}} \left[\left(Z(\mu_l,a')^{\textrm{RI}\to\textrm{lat}}\right)^{-1}\right]_{n,m}.
\label{eq:step-scaling}
\end{equation}
This matrix expresses the seven chiral-basis operators defined at the perturbative scale $\mu_h$ in terms of those same operators defined at a lower energy scale $\mu_l$, accessible from our coarse 24ID ensemble.  The two factors in Eq.~\eqref{eq:step-scaling} are computed non-perturbatively using a lattice with an inverse lattice spacing $1/a' = 3.1$ GeV in Ref.~\cite{RBC:2023ynh}.  The intermediate lattice energy scale $1/a'$ is chosen to be larger than the scale of the 24ID ensemble to reduce the discretization errors associated with the higher RI scale $\mu_h$. 

The final matrix $Z_{\ell m}^{\textrm{RI}\rightarrow\textrm{lat}}(\mu_l,a)$ expresses the SMOM($\slashed{q},\slashed{q}$) operators defined at the scale $\mu_l$ in terms of the lattice operators whose matrix elements we compute on the 24ID ensemble. Thus, $1/a=1.023$ GeV, the inverse lattice spacing of the 24ID ensemble.  

Since the precision of our calculation is limited at the 10\% level by our lack of control of the finite-volume errors associated with the $\pi\pi\gamma$ intermediate state, we chose to compute the matrix elements of only the most important ``current-current'' physical-basis operators, $Q^\textrm{lat}_1$ and $Q^\textrm{lat}_2$.  These should be the lattice operators with the largest Wilson coefficients.  The renormalization procedure outlined above provides the Wilson coefficients for the seven chiral-basis lattice operators $Q_i^{\prime\textrm{lat}}$.  Because of the linear dependence of the ten physical basis operators there is ambiguity in determining the coefficients of the ten physical basis operators in terms of the seven $C_i^{\prime\textrm{lat}}$.  We arbitrarily resolve this ambiguity by choosing the ten, physical basis lattice coefficients of $\{Q_i^\textrm{lat}\}_{1 \le i \le 10}$ to be those for which the coefficients of the three operators $Q_4$, $Q_9$ and $Q_{10}$ vanish.  We find
\begin{equation}
C^{\rm lat}_1 = -0.312(14) \,,\quad
C^{\rm lat}_2 = 0.718(14)\,,\quad
C^{\rm lat}_3 = 0.018(14)\,,
\end{equation}
where the uncertainties are purely statistical.
All of the other Wilson coefficients (QCD and electroweak penguins) are at most 0.5\% in magnitude, supporting our choice to compute only the matrix elements of $Q^\textrm{lat}_1$ and $Q^\textrm{lat}_2$.

Although we expect the systematic error arising from the neglect of the operators $Q^\textrm{lat}_{3-10}$ to be small, it is important to confirm this statement with more complete numerical evidence, especially because of the partial cancellation between the contributions of $Q_1$ and $Q_2$.  Such elaboration is left for future work.

To evaluate the matrix elements required in Eq.~\eqref{eq:Result_4pt} and Eq.~\eqref{eq:Result_sub}, we use a Coulomb-gauge-fixed wall-source for the kaon interpolating operator to improve the overlap with the ground state.
We use a local operator for the EM current, which is multiplicatively renormalized,
\begin{equation}
J^{\rm lat}_\nu(v) = Z_V \left(\mathcal{Q}_u \bar{u}\gamma_\nu u(v) 
+ \mathcal{Q}_d \bar{d}\gamma_\nu d(v)
+ \mathcal{Q}_s \bar{s}\gamma_\nu s(v)
\right)\,,
\end{equation}
with physical EM charges
\begin{equation}\label{eq:chargefac}
\mathcal{Q}_u = \frac{2}{3}\,,\quad
\mathcal{Q}_d = -\frac{1}{3}\,,\quad
\mathcal{Q}_s = -\frac{1}{3}\,.
\end{equation}
We neglect $SU(3)$-flavor-suppressed diagrams, \textit{ie}., those with a self-contracted loop (`tadpole') at one of the EM currents.
The contribution of these diagrams is expected to be small as is the case in the $\pi^0\to e^+e^-$ calculation~\cite{Christ:2022rho}.

\begin{figure}[h!]
\includegraphics[scale=0.8]{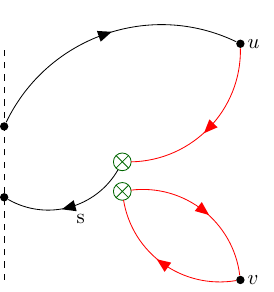}
\includegraphics[scale=0.8]{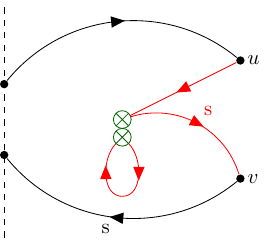}
\includegraphics[scale=0.8]{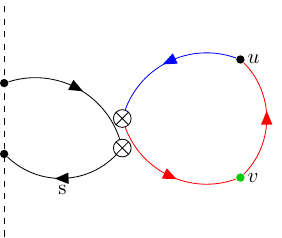}
\\
\includegraphics[scale=0.8]{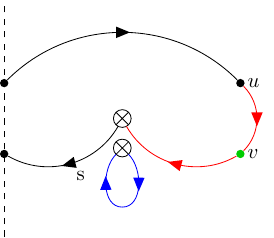}
\hspace{8pt}
\includegraphics[scale=0.85]{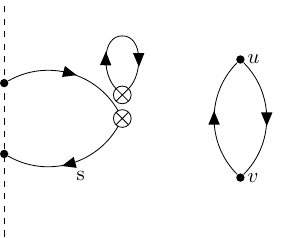}
\caption{The five types of Wick-contraction topology evaluated in this paper. From left to right, top to bottom: Type-1, Type-2, Type-3, Type-4 and Type-5.  The vertical dashed line represents the Coulomb-gauge-fixed kaon interpolating operator, the pair of circled crosses the four-quark weak operator and the vertices labeled $u$ and $v$ the two EM currents.}
\label{fig:diagrams}
\end{figure}

We classify the Feynman diagrams representing the fermion Wick contractions needed to evaluate the four-point correlator appearing in Eq.~\eqref{eq:Result_4pt} into two categories, the \textit{connected} diagrams (Type-1,-2,-3,-4), where all four operators in the correlation function are connected to each other by quark lines, and the \textit{disconnected} diagrams (Type-5) otherwise.
The computed contraction classes are shown in Fig.~\ref{fig:diagrams} with one representative diagram per topology.
Note that each type includes several different flavor combinations and either of the operators $Q_1$ and $Q_2$. 
A complete list of Wick contractions can be found in Appendix~\ref{sect:comp}.

Finally, we should recognize the ultraviolet divergences present in Type-1 and Type-3 contractions when evaluated in our 2+1 flavor theory, in spite of the renormalization procedure that has already been applied to the weak Hamiltonian. 
The treatment of these divergences is complicated by our use of non-conserved, local electromagnetic currents which will not obey the standard Ward identities for internal momenta close to the lattice scale.
Since our ensemble has a lattice spacing too large to allow the introduction of a valence charm quark, the physical effects of the charm quark in this 2+1 flavor calculation must be represented by counter terms which can be determined without reference to perturbation theory as suggested in Section~V~B of Ref.~\cite{Christ:2015aha}.  The determination of these counter terms and their matrix elements which enter this decay will be the subject of a future paper.

\subsection{Isospin decomposition}\label{sect:isospin}

Since the $\pi^0$ and $\eta$ intermediate states behave very differently in our calculation it is convenient to have a mathematical technique to distinguish the amplitudes to which each contributes.  Decomposing the product of the two EM currents according to their total isospin symmetry allows us to do that.  In this way the statistical fluctuation from the unphysical $\pi^0$ does not interfere with the analysis of the isospin component where the $\pi^0$ is absent.
For three flavors with the physical charge factors in Eq.~\eqref{eq:chargefac}, we can decompose the electromagnetic current $J_\mu$ into an isovector part $J_\mu^3$ and an isoscalar part $J_\mu^8$,
\begin{equation}
J_\mu = \frac{1}{2}J_\mu^3 + \frac{1}{6} J_\mu^8\,,
\end{equation}
where
\begin{equation}
J_\mu^3 = \bar{\psi}\lambda^3\psi\,,\quad
J_\mu^8 = \sqrt{3}\bar{\psi}\lambda^8\psi\,.
\end{equation}
Here the $\lambda^i$ are the conventional Gell-Mann matrices and $\psi=(u,d,s)^\intercal$.
We can then decompose the product of the two EM currents into two isospin components, one with $I=1$ and the other with $I=0$ or 2:
\begin{equation}
J_\mu(u)J_\nu(v) = \mathcal{J}_{\mu\nu}^{I=1}(u,v) + \mathcal{J}_{\mu\nu}^{I=0,2}(u,v)\,,
\end{equation}
where
\begin{equation}
\mathcal{J}_{\mu\nu}^{I=1}(u,v) = 
\frac{1}{12}\left( J_\mu^3(u)J_\nu^8(v) + J_\mu^8(u)J_\nu^3(v)\right)\,, 
\end{equation}
\begin{equation}
\mathcal{J}_{\mu\nu}^{I=0,2}(u,v) = 
\frac{1}{4}\left(
J_\mu^3(u) J_\nu^3(v) + \frac{1}{9}J_\mu^8(u) J_\nu^8(v)\right)\,.
\end{equation}
The amplitudes in Eqs.~\eqref{eq:Result_4pt} and \eqref{eq:Result_sub} can be decomposed according to this strategy leading to the isospin-specific amplitudes $\mathcal{A}^{\textrm{lat}/\textrm{sub}}_{I=1}$ and $\mathcal{A}^{\textrm{lat}/\textrm{sub}}_{I=0,2}$.
The $\pi^0$ intermediate state contributes only to $\mathcal{A}^{\textrm{lat}/\textrm{sub}}_{I=1}$ while the $\eta$ contributes only to $\mathcal{A}^{\textrm{lat}/\textrm{sub}}_{I=0,2}$.

\section{Numerical results}\label{sect:res}
Here we discuss our numerical results for the $I=1$ and $I=0,2$ amplitudes on the $24^3\times 64$, physical quark mass ensemble, `24ID', generated by the RBC-UKQCD collaboration in separate sub-sections.
This ensemble is generated with $2+1$ flavor M\"obius domain wall fermions, whose simulation parameters can be found in Section~II of Ref.~\cite{RBC:2023ynh}.
The inverse lattice spacing for this ensemble is $1/a = 1.023$ GeV, and the pseudo-scalar meson masses are $M_\pi = 142$ MeV and $M_K = 515$ MeV.
The renormalization factor for the vector current is $Z_V = 0.72672(35)$,  determined in Ref.~\cite{Tu:2020vpn}.

The present calculation makes extensive use of propagator data generated as part of the RBC/UKQCD HLbL project~\cite{Blum:2019ugy}.
In particular, 64 Coulomb-gauge-fixed wall-source and 512 randomly-distributed point-source propagators were made available for this work for all of the 110  configurations analyzed.
These existing propagators reduce the computation cost of the current project, where all but the type-3 diagrams and a portion of the type-5 diagrams can be calculated without performing further inversions. See Appendix~\ref{sect:comp} for detailed descriptions of the computational strategy for each diagram.

In both the $I=1$ and $I=0,2$ cases, we compute the total contribution to Eqs.~\eqref{eq:Result_4pt}, partially-summed over the earliest time $v_0$ up to $\delta_{\max}$ for $0\leq\delta_{\max}\leq T/2$, and \eqref{eq:Result_sub} for several values of the parameters $\tsep$ and $R_{\max}$.
Depending on the diagram, we collect data for 3 to 6 values of $\tsep$ ranging from 6 to 16 lattice units and perform an error-weighted average over the results for all values of $\tsep$ that are computed in order to improve the quality of the signal.

Having different values of $R_{\max}$ serves two purposes.  First, we need to tune $R_{\max}$ to achieve a balance between omitting large values of $u_0$ that contribute to our amplitude and those that contribute only noise.
Since $R_{\max}$ is the largest distance separating the two EM currents, we expect the propagation of a $\rho$-meson to dominate the $R_{\max}$ behavior.  For physical kinematics, the mass of the $\rho$-meson in lattice units is about $0.8$, hence one would expect a rapid saturation of the $u_0$-integral. 
However, having a value for $R_{\max}$ that is too large mainly introduces noise into the calculation.

Second, as discussed in Ref.~\cite{Chao:2024vvl}, in a QED${}_\infty$ calculation the inequality of QCD and QED volumes leads to the non-conservation of momentum which allows a residual exponentially-growing contribution in $R_{\max}$ which is power-law-suppressed in the volume. 
Having results for different, sufficiently large values of $R_{\max}$ should help us determine if this effect is small enough to be neglected, as claimed in Refs.~\cite{Chao:2024vvl}. 

Our implementation of Type-3 diagrams utilizes the all-to-all technique described in Appendix~\ref{sect:t3} with 1000 Z-M\"obius-approximated low modes with reduced extent in the fifth direction $L_s = 12$ (see Ref.~\cite{Mcglynn:2015uwh} and references therein) and 16 hits for every time-diluted wall-source with $\mathbb{Z}_2\times \mathbb{Z}_2$ noise.
For the tadpole term appearing in Type-4 and Type-5 [Eq.~\eqref{eq:lxx}], we use a similar technique but with 64 spin-color diluted volume-source random vectors per configuration.

For the $I=0,2$ case, the contribution of the intermediate $\eta$ at rest needs special care.
Although formally converging at the $\delta_{\max}\to \infty$ limit as $M_\eta > M_K$, the unphysical contribution Eq.~\eqref{eq:Result_sub} falls off very slowly due to the small separation between $M_\eta$ and $M_K$. 
As the $I=0,2$ part of the amplitude involves disconnected diagrams, the signal deteriorates well before before the integral over the slowly falling $\eta$ contribution has become saturated. 
In Section~\ref{sect:ieq0}, we discuss several strategies to reconstruct the physical $\eta$ contribution.

The numerical results on the decay amplitude $\mathcal{A}$ in this section are related to the physical decay rate $\Gamma_{\kl\to\mu^+\mu^-}$ via
\begin{equation}\label{eq:rate2amp}
\Gamma_{K_{\rm L}\rightarrow\mu^+\mu^-} =\frac{\beta}{8 \pi M_K } \left| e^4V_{ud}V^*_{us}\frac{G_{\rm F}}{\sqrt{2}}
\mathcal{A} \right|^2\,,\quad
\beta = \sqrt{1-\frac{4m_\mu^2}{M_K^2}}\,,
\end{equation}
where $e$ is the fundamental electric charge.
All the numerical values for the constants and parameters appearing above will be taken from Ref.~\cite{ParticleDataGroup:2022pth}.
We compute all the five types of diagram on the same 110 configurations separated by at least ten molecular-dynamics units in order to form the isospin components with the correlation between different types of diagram consistently accounted for.

\subsection{$I=1$ Amplitude}\label{sect:ieq1}

We have obtained a very clear signal for the $I=1$ channel since it contains only connected diagrams.
The unphysical-pion subtraction needs to be done with care.
As explained in detail in Appendix~\ref{sect:unphys} to avoid O($a$) errors while integrating, the trapezoid rule is carefully followed at the boundary of the integration domain when evaluating the sum over $v_0$ in Eq.~\eqref{eq:Result_4pt}.  We then subtract the exponentially growing term given by Eq.~\eqref{eq:Result_sub}.  This continuum-determined subtraction is meaningful when the upper limit $\delta_{\max}$ for the $v_0$ integral is large.  If $\delta_{\max}$ is one or two lattice spacings this continuum expression loses meaning.  In order to plot the integrated expression determining the result in Eq.~\eqref{eq:Result_total} even for small $\delta_{\max}$ we adopt a reasonable but \textit{ad hoc} subtraction for each $v_0$.  This subtraction under the sum sign, given in Eq.~\eqref{eq:Result_total-sch2}, is chosen so that when summed to a large value of $\delta_{\max}$ an O($a^2$)
accurate result is obtained.  Note that this subtraction is needed only when $v_0\geq 0$.

\begin{figure}[h!]
\centering
\begin{minipage}{0.45\textwidth}
\centering
\includegraphics[scale=0.7]{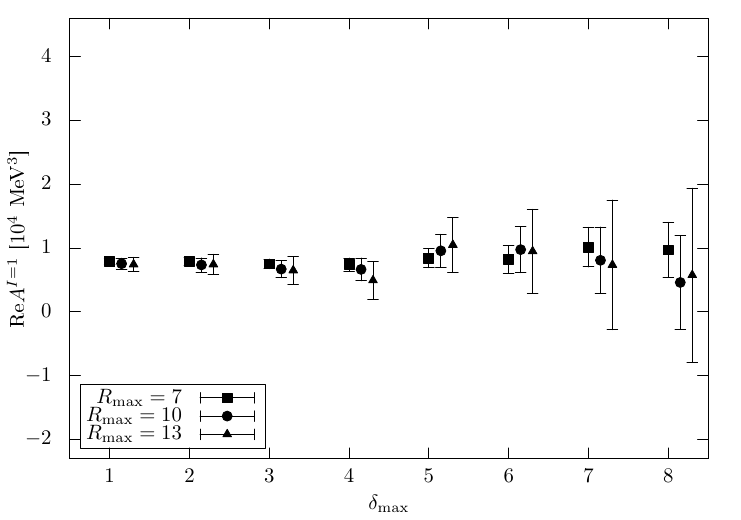}
\end{minipage}
\hspace{32pt}
\begin{minipage}{0.45\textwidth}
\centering
\includegraphics[scale=0.7]{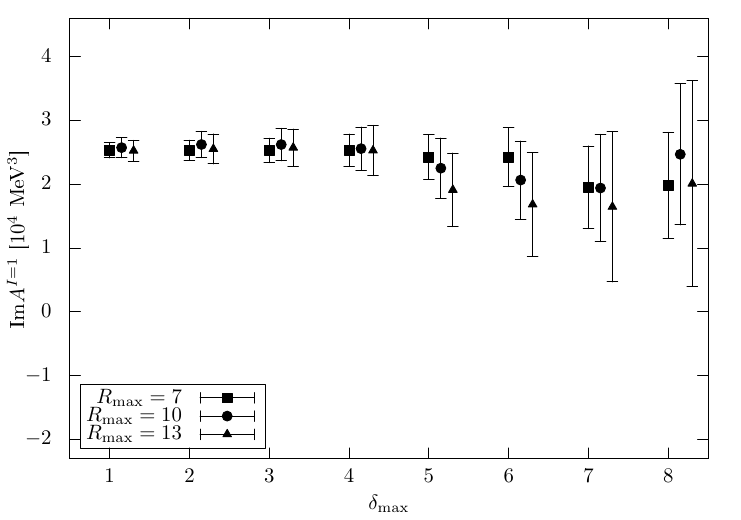}
\end{minipage}
\caption{The real (left) and imaginary (right) parts of the $I=1$ amplitude as a function of $\delta_{\max}$, the upper limit on the integral over the time separation between the closest EM current and the weak operator.}
\label{fig:ieq1}
\end{figure}

The results for the partially-integrated amplitudes are given in Figure~\ref{fig:ieq1} as a function of the upper limit $\delta_{\max}$ for both the real and imaginary parts at three different values of $R_{\max}$ (7, 10 and 13). 
We see that plateaus are formed at small $\delta_{\max}>0$, which indicates the insensitivity to long-distance physics of this observable.
All results with different values of $R_{\max}$ are consistent with each other within the quoted statistical errors, suggesting that our smallest $R_{\max}=7$ should be sufficiently large to capture the non-zero hadronic contributions.  
Within the quoted error, we do not see an obvious change in the central value as $R_{\max}$ is increased, confirming the expectation that for the current spatial volume there are no propagating two-pion states that must be accommodated, a central assumption of our computational strategy.

We also do not see any sign of residual linearly- or exponentially-growing behavior in the plotted results as $\delta_{\max}$ increases.  This suggests that only the pion state gives an unphysical contribution.  In particular there is no suggestion of any contribution from a propagating three-pion state that is visible above the 10\% statistical errors present for both the real and the imaginary parts when $R_{\max}=7$ and $\dmax=4$.  We will use these values to obtain our final results.

\subsection{$I=0,2$ Amplitude}\label{sect:ieq0}

\subsubsection{Type-5 diagrams}

The $I=0,2$ channel involves disconnected diagrams (Type-5), which require substantially more measurements than the other diagrams in order to achieve the same statistical precision. 
We use a low-mode-improved method to calculate the Type-5, Diagram 1 (T5D1) as described in Appendix~\ref{sect:t5}, where the low-mode part is constructed using the 1000 eigenmodes of the Z-M\"obius-approximated Dirac operator with the smallest eigenvalues. 

The high-mode correction  [Eq.~\eqref{eq:pilowhigh}] that must be added to the low-mode contribution is obtained from 2048 measurements per configuration using point-source propagators whose sources are located at the position of one of the EM currents and whose sinks positions are the location of the second EM current.  These propagators are evaluated with a ``sloppy'' stopping condition (a relative error of $10^{-4}$) when performing a conjugate gradient inversion of the Dirac operator.  
With the M\"obius-Accelerated Domain Wall Fermion (MADWF) solver~\cite{Yin:2011np}, the computational cost of these sloppy inversions is three times less than if a higher precision of $10^{-10}$ were chosen.
One quarter of these sloppy measurements are performed using the source positions used for the 512 higher precision (relative error of $10^{-8}$) propagators that were available.  The difference between the results using these higher-precision and sloppy propagators is used to correct for the bias present in the sloppy inversions.
With this strategy, the bias correction is about an order of magnitude smaller than the statistical fluctuations~\cite{Chao:2024cnu}, which makes this an efficient strategy.

By themselves, the Type-5 diagrams exhibit exponentially-growing behavior due to the $\pi^0$-pole which is present in the  disconnected amplitude and is canceled after the proper $I=0,2$ combination with Type-3 and Type-4 diagrams has been included.
With the current statistics, we see that the final, combined $I=0,2$ amplitude has a rather clear signal without an unphysical $\pi^0$ contribution. (See the data points labeled `$\eta$-unsub.' in Figure~\ref{fig:ieq0}.)

\subsubsection{Treatment of the $\eta$ intermediate state}
\label{sec:eta}
We propose two classes of methods (three different methods in total) to remove the slowly decreasing unphysical contribution of the $\eta$-intermediate state.
The first class is referred to as \textit{direct subtraction} and corresponds to a single method (\textbf{Method 1}), which is exactly the same procedure as used to treat the unphysical $\pi^0$ state in the $I=1$ case. 

In the second class the weak Hamiltonian is modified by subtracting a quark-bilinear operator whose on-shell matrix element is zero  but whose coefficient can be chosen to make the off-shell $\langle\eta|\chw^\prime|K_L\rangle$ matrix element vanish:
\begin{equation}
    \chw^\prime = \chw+c_s\left(\overline{s}d+\overline{d}s \right)\,.
\end{equation}
Such a modification of $\chw$ will not change the physical result, but with the $\eta$ contribution removed will permit truncating the $v_0$-integral at an early time before the statistical noise has become unacceptably large.

The on-shell matrix element of the operator $\overline{s}d+\overline{d}s$ vanishes because it is the divergence of the vector current $J^{sd}_\mu$ which generates
the flavor rotation~\cite{Bernard:1985wf}
\begin{equation}
\begin{pmatrix}
d \\ s
\end{pmatrix}
\rightarrow 1+\varepsilon i \tau_2
\begin{pmatrix}
d \\ s 
\end{pmatrix}\,,
\quad\mbox{where}\quad
\tau_2 \equiv 
\begin{pmatrix}
0 & -i \\
i & 0 
\end{pmatrix}\,
\label{eq:flavor_trans}
\end{equation}
is the conventional Pauli matrix.  Using $q$ to represent the two-component column vector of quark fields on the right-hand side of Eq.~\eqref{eq:flavor_trans}, we can write:
\begin{equation}
    J^{sd}_\mu = \overline{q} \gamma^\mu \tau_2 q\,.
    \label{eq:J-sd}
\end{equation}

The usual proof that the on-shell matrix element of the operator $\overline{s}d+\overline{d}s$ vanishes exploits the Ward identity obtained by replacing $\chw$(x) in the hadronic Green's function given in Eq.~\eqref{eq:Result_4pt} with the current $J^{sd}_\mu$ and taking the divergence:
\begin{eqnarray}
\frac{\partial}{\partial x_\lambda}\left\langle 0|  \textrm{T}\left\{J^a_\mu(u)J^b_\nu(v) J^{sd}_\lambda(x) \kl(t_K)\right\}|0\right\rangle &&  \nonumber \\ 
&& \hskip -2.2 in =(m_s-m_d)\left\langle 0|\textrm{T}\left\{ J^a_\mu(u) J^b_\nu(v)  \left(\overline{s}d+\overline{d}s\right)(x) \kl(t_K)\right\}|0 \right\rangle \label{eq:wict}\\
&& \hskip -2.4 in  -i\left\langle 0|\textrm{T}\left\{ \delta^4(x-u) \left[Q^{sd},J^a_\mu(u)\right]J^b_\nu(v)K(t_K) + 
\delta^4(x-v)J^a_\mu(u) \left[Q^{sd},J^b_\nu(v)\right]\kl(t_K)\right\}|0\right\rangle
\nonumber
\end{eqnarray}
where we have adopted the Hilbert space expression for the Ward identity since the two delta function terms can be easily written in terms of commutators with the conserved charge operator $Q^{sd}\equiv \int d^3\textbf{x}\; J_0^{sd}(\textbf{x},x_0)$. 
We have omitted a third delta function term involving the kaon interpolating operator $K(t_K)$ since in our calculation $\tsep = x_0-t_K$ is always strictly greater than zero.  Here we have introduced the labels $a$ and $b$ on the two currents to enable a separation of the EM current into $I=0$ and $I=1$ components.

If the two $J^i$'s are taken to be the full EM currents with equal charges for the down and the strange quarks, then $\left[Q^{sd},J_\mu^{\rm em}\right] = 0$ and we will arrive at the usual conclusion that the combination $\bar{s}d+\bar{d}s$ is a total derivative (\textbf{Method 2}).
However, if one considers products of the individual $I=0$ and 1 components separately in the product of the two EM currents in Eq.~\eqref{eq:wict} (\textbf{Method 3}), one needs to include additional contact terms due to the non-vanishing commutators $\left[Q^{sd},J_\mu^i\right]$ when $i=3$ or 8.
Simple algebra leads to
\begin{equation}
-i\frac{1}{3}\left[Q^{sd},J^8_\mu\right](u) = i \left[Q^{sd},J^3_\mu\right](u)
= \left(\bar{s}\gamma_\mu d+ \bar{d}\gamma_\mu s\right)(u)\equiv \widetilde{K}^*_\mu(u)\,.
\label{eq:contact}
\end{equation}
The right-most term in Eq.~\eqref{eq:contact} can be viewed as an interpolating operator for a $|{K^0}^*\rangle + |{\overline{K}^0}^*\rangle$ state.

The Ward identity appearing in Eq.~\eqref{eq:wict} is a continuum equality and may not be obeyed by operators defined in lattice field theory.  However, it is likely that if the continuum current $J_\mu^{sd}$ is replaced by a conserved Mobius-fermion lattice current~\cite{Boyle:2015mb}, that an analogous lattice Ward identity could be derived, with the $\widetilde{K}^*_\mu$ contact term possibly becoming non-local.  We will neglect such refinements as being of O($a^2$).

The use of this Ward-identity leads to the second class of $\eta$ intermediate state treatments, referred to as \textit{$\bar{s}d$-subtraction} methods, which consist in modifying each weak operator $Q_i$ in our master equation 
Eqs.~\eqref{eq:Result_4pt} and \eqref{eq:Result_sub} according to
\begin{equation}\label{eq:csi1}
Q_i^{\rm sub}(0) = Q_i(0) + c_{s,i} \left(\bar{s}d(0) + \bar{d}s(0)\right),
\end{equation}
where
\begin{equation}
c_{s,i} \equiv -\frac{\lla \eta| Q_i(0)| \kl\rra}{\lla \eta | \bar{s}d(0) + \bar{d}s(0) | \kl\rra}\,.
\label{eq:csi2}
\end{equation}
In the case that the full EM currents are used, no contact terms appear in the Ward identity and we do not change the LD2$\gamma$ amplitude if replace the operator $Q_i$ with $Q_i^\mathrm{sub}$ which will ensure that the $\eta$ intermediate state does not contribute.

For the case where we are separating the product of EM currents into $I=1$ and $I=0,2$ components, Eqs.~\eqref{eq:csi1} and \eqref{eq:csi2} are unchanged but the $c_{s,i} \left(\bar{s}d(0) + \bar{d}s(0)\right)$ term added to $Q_i$ in Eq.~\eqref{eq:csi1} is no longer equal to a total divergence.  We can still use this strategy to remove the $\eta$ intermediate state but must compute and explicitly remove the matrix element of the contact term that appears in Eq.~\eqref{eq:wict}.  The contribution of the contact term can be expressed as a product of the usual leptonic kernel $L_{\mu\nu}$, the coefficient $c_{si}$ and a three-point hadronic Green's function involving the kaon interpolating operator and the product of two currents obtained from the sum of the commutators $\left[Q^{sd}(u_0), J_\mu^a(u)\right] J_\nu^b(v)+J_\mu^a(u)\left[Q^{sd}(v_0), J_\nu^b(v)\right]$, for an appropriate combination of the indices $a$ and $b$.  The weak Hamiltonian $H_W(x)$ does not enter this simpler Green's function. In both cases, Eq.~\eqref{eq:csi2} determines the coefficients $c_{s,i}$ so that there will be no physical nor unphysical contributions from an $\eta$ intermediate state.

In both classes of $\eta$ intermediate state treatments, we need to compute matrix elements related to the $\eta$, which are very noisy because of the presence of quark-line disconnected diagrams.
To address this difficulty, we perform analyzes using smaller $\eta$ propagation times where the noise is less pronounced which requires that we also include the more massive $\eta'$ state. Thus, we will work with two interpolating operators: an SU(3) octet for the $\eta$ and an SU(3) singlet operator for the $\eta'$. Extracting the $\eta$ mass from a two-point function now becomes a $2\times 2$ generalized eigenvalue problem (GEVP)~\cite{Blossier:2009kd}.
For this purpose, we consider two operators 
\begin{equation}
O_l \equiv \frac{i}{\sqrt{2}}\left( \bar{u}\gamma_5 u + \bar{d}\gamma_5 d \right) \,,
\end{equation}
\begin{equation}
O_s \equiv i \bar{s}\gamma_5 s \,.
\end{equation}
The GEVP consists in finding two pairs of eigenvalues and 2-dimensional vectors $(\lambda_n,V_n)$ for $n=1,2$ with $\lambda_1 \leq \lambda_2$ which solve the linear system
\begin{equation}
C(t) V_n(t,t_0) = \lambda_n(t,t_0) C(t_0) V_n(t,t_0)\,,
\end{equation}
where
\begin{equation}
C(t) = \begin{pmatrix}
\lla O_l(t)O_l^\dagger(0)\rra & \lla O_s(t)O_l^\dagger(0)\rra \\
\lla O_l(t)O_s^\dagger(0)\rra & \lla O_s(t)O_s^\dagger(0)\rra \\
\end{pmatrix}\,.
\end{equation}
The solutions $V_{1,2}$ and $\lambda_{1,2}$ provide the linear combinations of operators with optimal overlap with the $\eta$ and $\eta^{\prime}$ states and give their masses respectively. 

\begin{figure}
\centering
\includegraphics[scale=0.8]{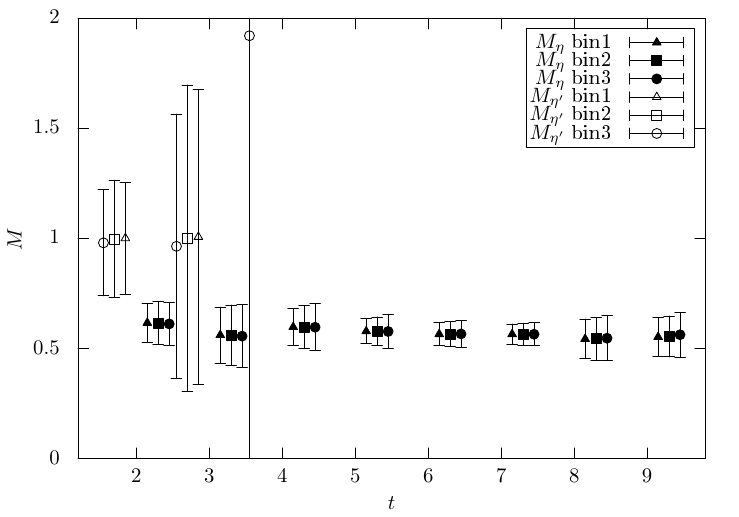}
\caption{The effective masses for $\eta$ and $\eta^\prime$ measured by solving the GEVP with different bin sizes.}
\label{fig:meff}
\end{figure}

We obtain the following values for the effective masses of the $\eta$ and $\eta^\prime$ in lattice units from 519 configurations, which do not include the 110 configurations used for the four-point function calculations,
\begin{equation}
M_\eta = 0.569(19)\,,\quad
M_{\eta^\prime} = 0.996(60)\,.
\end{equation}
These results were obtained by fitting the time dependence of the eigenvalues $\lambda(t,t-1)$ over the range $t\in[4,8]$ for $M_\eta$ and $t\in[2,4]$ for $M_{\eta^\prime}$.
The effective-mass plot for the $\eta$ and $\eta^\prime$ is given in Figure~\ref{fig:meff}.
The 519 configurations used in this calculation were separated by 4 Molecular-Dynamics time units and we have tested for autocorrelations by also binning the results into bins of 1, 2 and 3 configurations.  The consistency among results shown in Figure~\ref{fig:meff} for different bin sizes suggests that the auto-correlation is small.

Note that although the signal is lost for the $\eta^\prime$ beyond the fitted region, including the $\eta^\prime$ in the GEVP analysis helps stabilize the effective mass obtained for the $\eta$ at larger time-separations.
The masses obtained are then used to extract the matrix elements $\lla \eta|Q_{1,2}|\kl\rra$ and $\lla \eta|\left(\bar{s}d+\bar{d}s\right)|\kl\rra$ from the three-point correlation functions $\lla O_{l,s}(t_\eta) Q_{1,2}(0)\kl(-t_K) \rra$ and $\lla O_{l,s}(t_\eta) \left(\bar{s}d(0)+\bar{d}s(0)\right)\kl(-t_K)\rra$.
Recognizing that it is better to take advantage of the much smaller statistical fluctuations when $O_s$ is used alone as the interpolating operator for the $\eta$ ground state, we perform a two-state fit to dependence on time separation $t_\eta$ between the operator and the sink for each three-point function after projecting to the kaon ground state at a reasonably large $t_K$
\begin{equation}\label{eq:2sf}
A(t_\eta) = A_\eta e^{-M_\eta t_\eta} +A_{\eta^\prime} e^{-M_{\eta^\prime} t_\eta}\,,
\end{equation}
to extract the desired matrix elements $A_\eta$ and $A_{\eta^\prime}$.

\begin{figure}[h! ]
\includegraphics[scale=0.5]{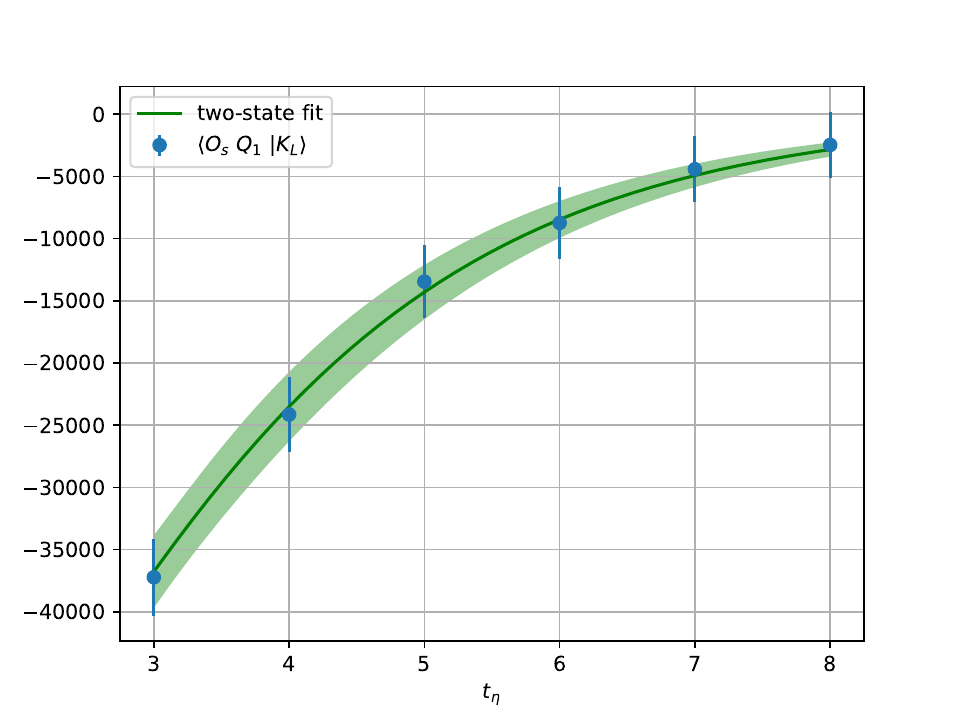}
\includegraphics[scale=0.5]{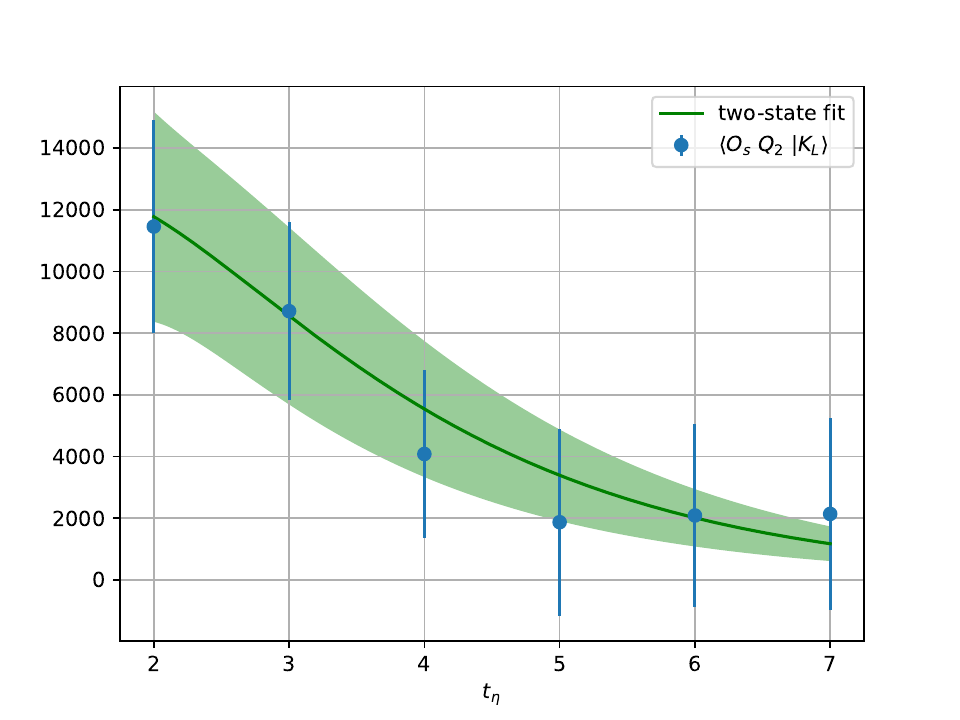}
\includegraphics[scale=0.5]{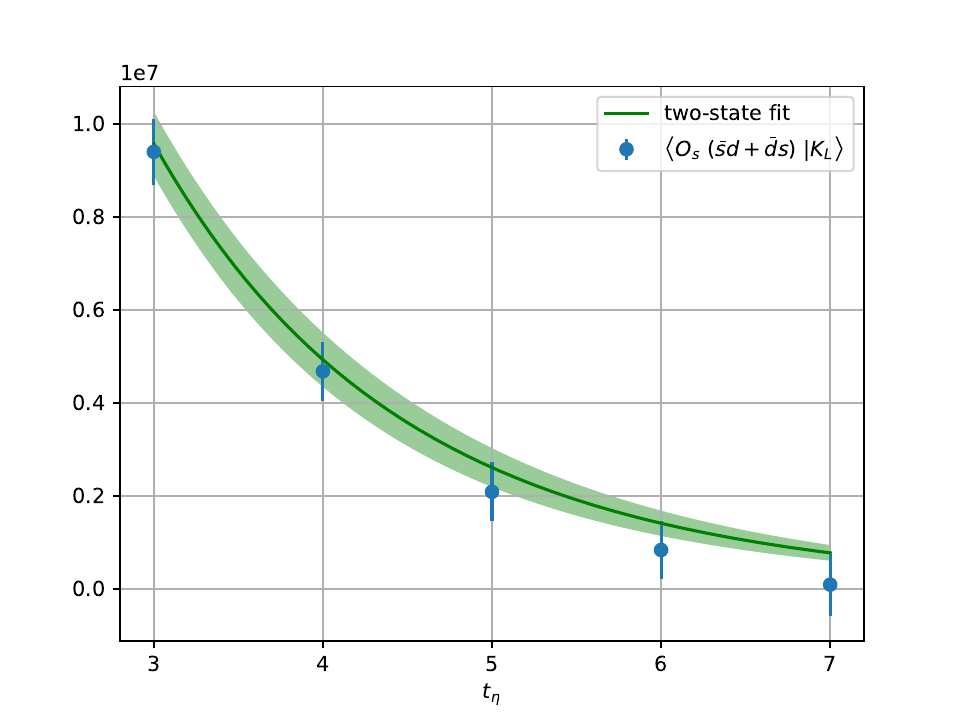}
\caption{Fitting $\lla O_s(t_\eta) O(0) |\kl\rra$ with $O = Q_1$, $Q_2$ and $\bar{s}d+\bar{d}s$ to the ansatz given in Eq.~\eqref{eq:2sf} with $t_{K} = 8$.}
\label{fig:k2e-fit}
\end{figure}

\begin{figure}[h!]
\includegraphics[scale=0.5]{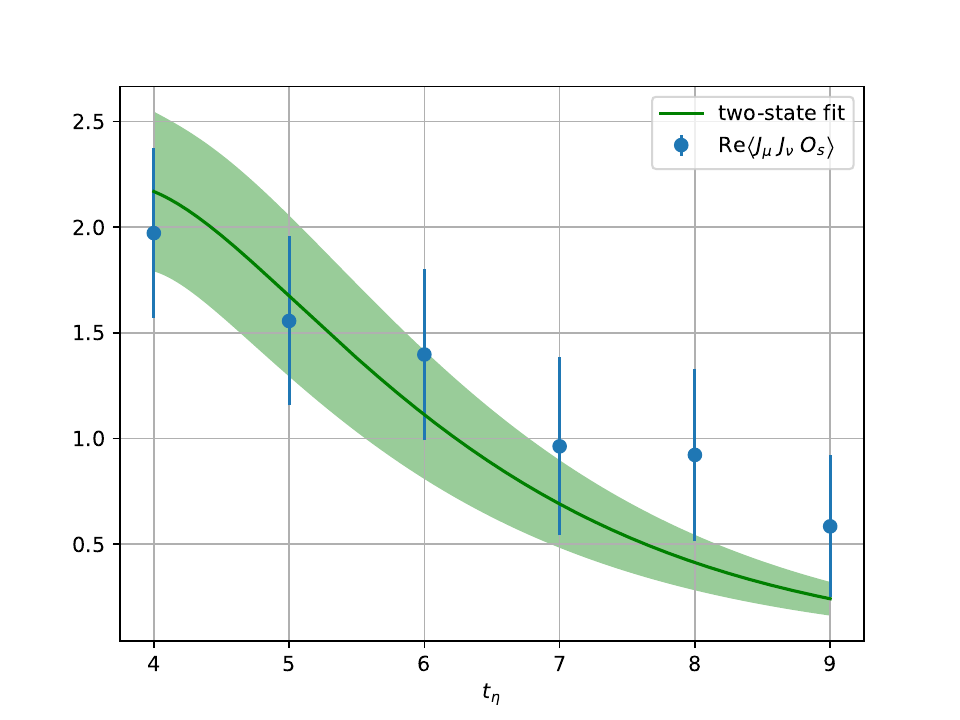}
\includegraphics[scale=0.5]{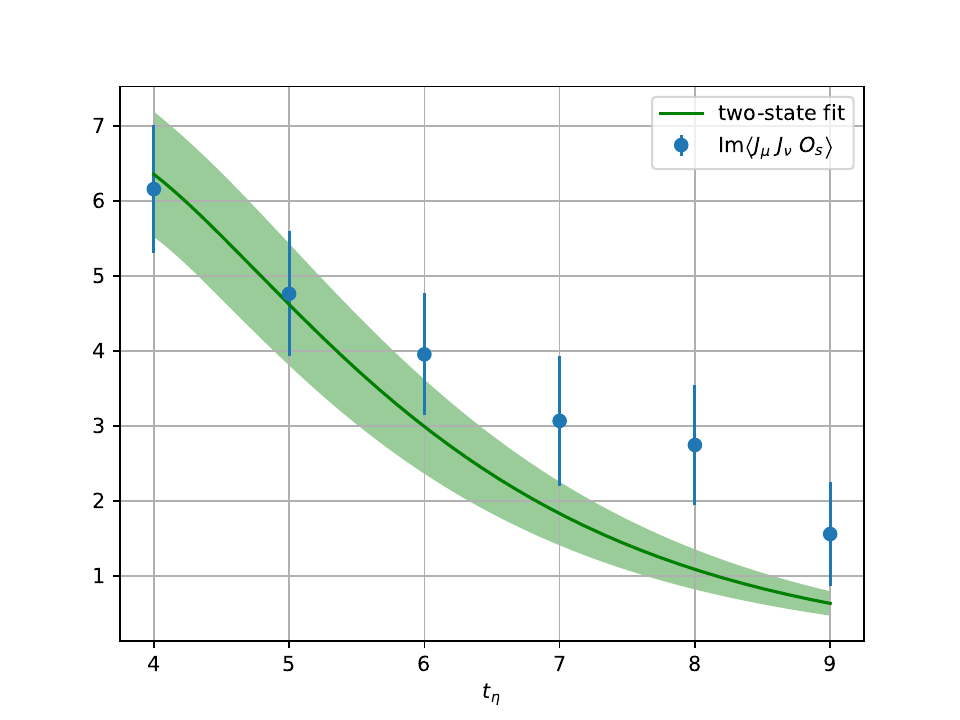}
\caption{Fitting $W(R_{\max}=7)$ (cf. Table~\ref{tab:e3pt}) to the ansatz Eq.~\eqref{eq:2sf}.}
\label{fit:jje-fit}
\end{figure}

This procedure is repeated at several different values of $t_K$.
The final quoted value for each three-point function is determined from a correlated fit to the outcomes from all these different two-state fits to a constant.\footnote{This fitting strategy differs from that used for the four-point functions where the average over $t_\mathrm{sep}$ is performed before fitting. However, this difference is not important.}
The same fitting strategy Eq.~\eqref{eq:2sf} is also applied to determine the real and imaginary parts of 
$\int d^4 r K_{\mu\nu}\left(r,R_{\max}\right)\lla 0|J_{\mu}(r)J_{\nu}(0)|\eta\rra$ from $\int d^4 r K_{\mu\nu}\left(r,R_{\max}\right)\lla 0|J_{\mu}(r)J_{\nu}(0)O_s(-t_\eta)\rra$.  This matrix element is needed for the direct subtraction Method 1, with the difference that one needs to consider only the separation between the $\eta$ interpolating operator and the earliest EM current.
The fitting qualities are illustrated in Figure~\ref{fig:k2e-fit} and Figure~\ref{fit:jje-fit} and the results for the matrix elements summarized in Table~\ref{tab:e3pt}.
All the three-point functions are measured on the same 110 configurations as the four-point functions so that correlations between the statistical fluctuations arising from the $\eta$ intermediate state can be exploited. 
These results lead to the following values for the coefficients $c_{s,i}$ [Eq.~\eqref{eq:csi2}] needed for $\bar{s}d$-subtraction Method 2 and Method 3:
\begin{equation}
c_{s,1} =  0.0097(32)\,,\quad
c_{s,2} = -0.0020(11)\,.
\end{equation}

\begin{table}[h!]
\centering
\begin{tabular}{|c|c|c|c|c|}
\toprule
Quantity & Value & $t_\eta$ & $t_K$ & $\chi^2/\rm{dof}$ \\
\hline\hline
$\lla\eta| Q_1(0) |\kl\rra$ & -0.0118(18)  & $[3,8]$ & $[6,17]$ & 0.86 \\
$\lla\eta| Q_2(0) |\kl\rra$ &  0.00243(89) & $[2,7]$ & $[6,17]$ & 0.59 \\
$\lla\eta| \bar{s}d(0)+\bar{d}s(0)|\kl\rra$ &  1.22(32)    & $[3,7]$ & $[6,19]$ & 0.67 \\
\hline
$\textrm{Re}\{W( 7)\}$ & 0.0103(37) & $[4,9]$ & --- & 0.85 \\
$\textrm{Re}\{W(10)\}$ & 0.0094(38) & $[4,9]$ & --- & 0.45 \\ 
$\textrm{Re}\{W(13)\}$ & 0.0122(48) & $[4,9]$ & --- & 0.67 \\ 
\hline
$\textrm{Im}\{W( 7)\}$ & 0.0269(75) & $[4,9]$ & --- & 1.56 \\
$\textrm{Im}\{W(10)\}$ & 0.0283(81) & $[4,9]$ & --- & 1.70 \\
$\textrm{Im}\{W(13)\}$ & 0.0268(85) & $[4,9]$ & --- & 1.14 \\
\botrule
\end{tabular}
\caption{Results for the $\eta$-related matrix elements needed for the various proposed $\eta$ subtraction schemes. All the values are given in lattice units and the states are normalized to unity. Here $W(R_{\max})\equiv \left(m_\mu\alpha_{\rm QED}^2\right)^{-1}\sum_{r\in \Lambda} K_{\mu\nu}\left(r,R_{\max}\right) \langle 0 | J_{\mu}(r)J_\nu(0)|\eta\rangle$ with everything in lattice units.}
\label{tab:e3pt}
\end{table}

\begin{figure}[h!]
\centering
\begin{minipage}{0.45\textwidth}
\centering
\includegraphics[scale=0.7]{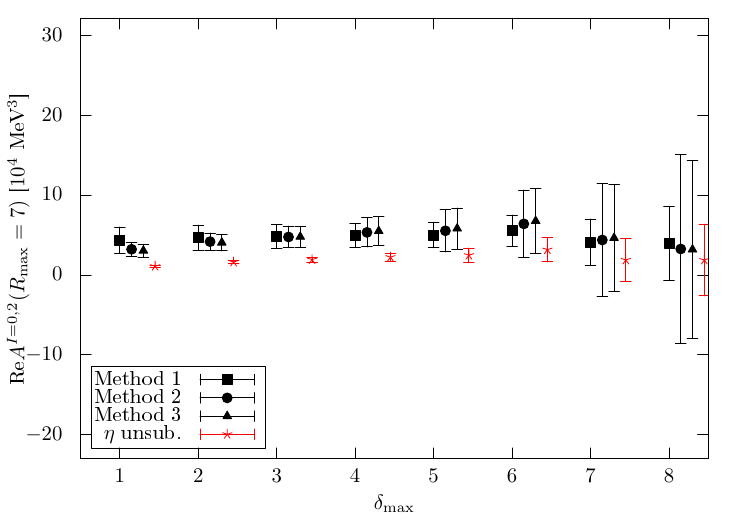}
\end{minipage}
\hspace{32pt}
\begin{minipage}{0.45\textwidth}
\centering
\includegraphics[scale=0.7]{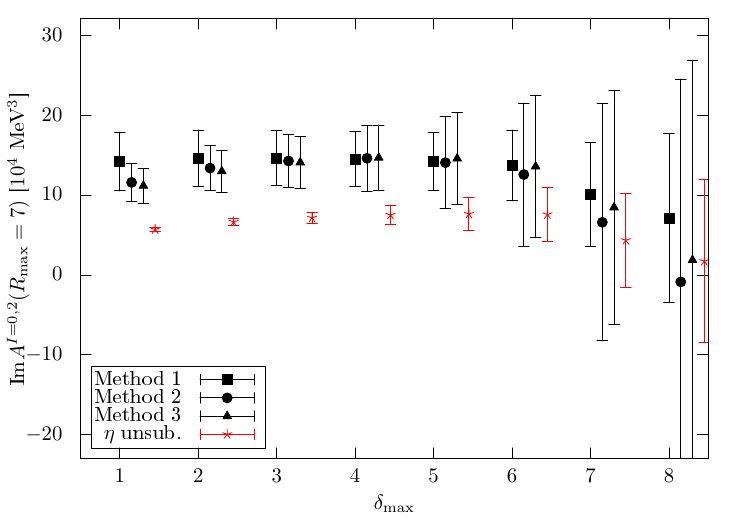}
\end{minipage}
\caption{The $I=0,2$ amplitude as a function of the temporal cut-off $\delta_{\max}$.}
\label{fig:ieq0}
\end{figure}

The results from the three different treatments of the $\eta$ intermediate state and the raw four-point function data before any treatment for the $\eta$ are plotted in Figure~\ref{fig:ieq0} for comparison.
All three methods are very consistent with each other in the plateau region starting around $\dmax =3$, but the signal begins to deteriorate rapidly when $\dmax \geq 6$.
The consistency among these different treatments of the $\eta$ is a strong and non-trivial check for our numerical implementation, as a sizable upward shift in the central values compared to the statistical error after the removal of the $\eta$ is observed. 
This large observed shift also suggests that the physical contribution of the $\eta$ could be important and the pion-pole dominance approximation commonly considered in the literature might not be accurate~\cite{GomezDumm:1998gw, Isidori:2003ts}.
However, such a statement needs to be supported by a calculation which includes a continuum limit.

\subsection{Combined results}
\begin{figure}
\centering
\begin{minipage}{0.45\textwidth}
\includegraphics[scale=0.65]{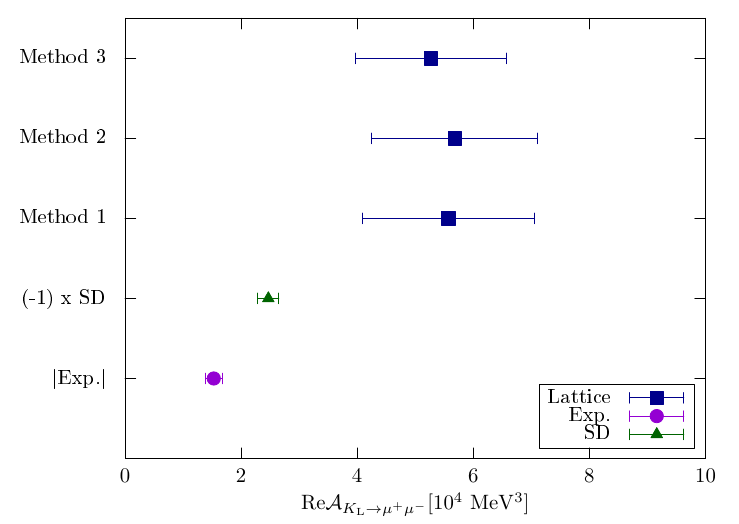}
\end{minipage}
\hspace{16pt}
\begin{minipage}{0.45\textwidth}
\includegraphics[scale=0.65]{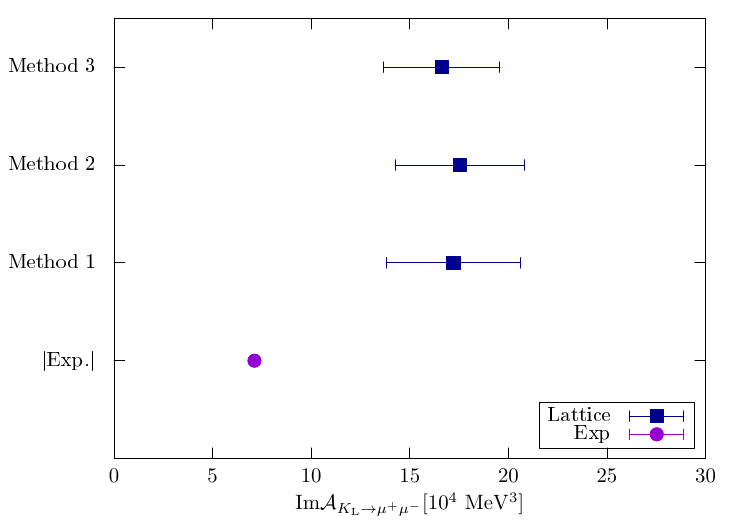}
\end{minipage}
\caption{Comparison between the real and imaginary parts of the LD2$\gamma$ decay amplitude computed on the 24ID ensemble to the SD$\gf^2$ part as well as the magnitudes of the real and imaginary parts of the decay amplitude extracted from experiment. See the text for the sign conventions used here.}
\label{fig:combined}
\end{figure}

We combine our $I=1$ and $I=0,2$ results from the three different treatments for the $\eta$, including correlations, and compare them to the SD$\gf^2$ result and the experimental values in Figure~\ref{fig:combined}.  Since the phase of our lattice result for $\mathcal{A}_{\kl\mu\mu}$ depends on the unphysical phase of the kaon source, we adjust that phase so that the absorptive part of $\mathcal{A}_{\kl\mu\mu}$ is purely imaginary as stated earlier, and that the imaginary part of the decay amplitude is positive.
Because only the magnitudes of the experimental values are known, we arbitrarily choose their signs to agree with our lattice results.  

Using the adopted phase of our kaon source we directly evaluate the matrix element of the effective Hamiltonian given in Eq.~(110) in Ref.~\cite{Buchalla:1993wq} in lattice calculation.  The result, also shown in Figure~\ref{fig:combined} and Table~\ref{tab:res-total}, is given by a consistent calculation of the SD$\gf^2$ amplitude that can be directly combined with the LD$2\gamma$ result computed here.  A more detailed discussion of our evaluation of the standard model result for the SD$\gf^2$ amplitude and its sign is given in Section~\ref{sect:sign}, where we conclude that the interference between the LD2$\gamma$ and the SD$\gf^2$ dispersive parts is destructive, at least for this calculation in which necessary renormalization counter terms have not yet been included.

For each of the three treatments of the $\eta$ shown in Figure~\ref{fig:combined}, we obtain the value by performing a correlated fit to the data for $\dmax\in [4,6]$.
Here all three methods are very consistent with comparable statistical errors, but the values obtained for the imaginary part are about 3$\sigma$ above the experimental result deduced from the $\kl\to\gamma\gamma$ process. 
For our final results for each isospin channel and their total, tabulated in Table~\ref{tab:res-total}, we quote the central values and statistical errors from Method 3 and take the standard deviation among the three different methods as the systematic error from the treatment of the $\eta$-intermediate state.

Given the exploratory nature of this first lattice calculation, a detailed discussion of systematic errors would be premature.  However, there are at least four likely sources of systematic error. The 24ID ensemble used here with an inverse lattice spacing of $1/a=1.023$ GeV leads to uncertain but possibly large discretization errors.  For a similar calculation of the electromagnetic difference between the $\pi^+$ and $\pi^0$ masses~\cite{Feng:2021zek} the result obtained from the 24ID ensemble was 20\% below the experimental value.  The finite-volume errors associated with low-energy $\pi\pi\gamma$ states was estimated to be at or below 10\%~\cite{Chao:2024vvl} and errors as large as 10\% might result from our omission of all but two of the seven weak operators present in the $\Delta S=1$ effective weak Hamiltonian.  Finally, the 30\% statistical error present in our treatment of the $\eta$ intermediate state could well obscure systematic errors of the same size, in spite of the agreement between the three different methods we tried.

\begin{table}[h!]
\begin{tabular}{c|c|c}
\hline\hline
  &  $\mathrm{Re}\mathcal{A}\times 10^{-4}$ [MeV${}^3$] & $\mathrm{Im}\mathcal{A}\times 10^{-4}$ [MeV${}^3$] \\
  \hline
$I=1$   & 0.69(9) & 2.64(21) \\ 
$I=0,2$ & 5.06(1.41)(0.09) & 14.31(3.11)(0.22) \\
Total(LD$2\gamma$)   & 5.27(1.30)(0.17) & 16.61(2.96)(0.38) \\
\hline
SD$\gf^2$ & $- 2.47(18)$ & --- \\
\hline
exp. & $\pm 1.53(14)$ & $\pm 7.10(3)$
\\
\hline\hline
\end{tabular}
\caption{Summary table of our results on the 24ID ensemble compared to the prediction for the short-distance contribution~\cite{Gorbahn:2006bm} and the experimental values. Here we have adopted the phase convention that the absorptive part of the LD2$\gamma$ amplitude comes only from the imaginary part of the decay amplitude and that this imaginary part is positive, which determines our choice for the phase of the kaon interpolating operator (see Section~\ref{sect:sign} for details). 
The values for $I=0,2$ and the total are quoted from Method 3, where the rightmost errors are systematic, estimated from the standard deviation among the results from the three different treatments of the $\eta$. Of course, a physically meaningful comparison between the experimental and theoretical results in this table must await the inclusion of the omitted renormalization counter terms in the lattice result.}
\label{tab:res-total}
\end{table}

\section{Interference between the LD$2\gamma$ and SD$\gf^2$ amplitudes}\label{sect:sign}

We begin this section by discussing our lattice QCD evaluation of the SD$\gf^2$ amplitude based on the earlier standard-model prediction presented in Refs.~\cite{Buchalla:1993wq, Buchalla:1995vs, Gorbahn:2006bm}, including both its magnitude and sign, using the same conventions adopted when computing the LD$2\gamma$ amplitude which is the subject of this paper.  The second part of this section examines a series of issues that require special attention when computing the sign of LD$2\gamma$ amplitude.

\subsection{Lattice QCD evaluation of the SD$\gf^2$ amplitude}

In order to determine the relative sign between the SD part and the LD part, we follow closely the conventions of Buchalla and Buras~\cite{Buchalla:1993wq}.
Their effective Hamiltonian describing the O($\gf^2$) contribution to $K_L\to\mu^+\mu^-$ provides an accessible summary of their calculation which we can easily exploit to determine the sign of this SD$\gf^2$ amplitude:
\begin{equation}\label{eq:asd}
\chw^{\rm SD} = -\frac{G_{\rm F}}{\sqrt{2}}\frac{\alpha_{\rm em}}{2\pi \sin^2\theta_{\rm W}}\Big[\left(V_{cs}^*V_{cd}Y_{\rm NL}+V^*_{ts}V_{td}Y(x_t)\right)(\bar{s}d)_{V-A}(\bar{\mu}\mu)_{V-A}+\textrm{h.c.}\Big]\,.
\end{equation}
The quantities $Y_{\rm NL}$ and $Y(x_t)$ are known functions which have been computed with the next-to-next-to-leading order charm quark effects included in Ref.~\cite{Gorbahn:2006bm}.  Here $\alpha_{\rm em}=\alpha_{\rm em}(M_Z)$ and $\sin^2\theta_{\rm W}$ are evaluated at the electroweak scale~\cite{Buchalla:1997kz} while $x_t$ is the ratio of the squares of the masses of the top quark and $W$ boson, $m_t^2/M_W^2$.

We can now evaluate the matrix element of $\mathcal{H}_W^{\rm SD}$ given in Eq.~\eqref{eq:asd} between the same perturbative $\mu^+\mu^-$ and non-perturbative $|\kl\rangle$ states as were used to compute the $\mathcal{A}_{\kl\mu\mu}$ in Section~\ref{sect:res}, giving the relative sign of these two amplitudes physical meaning.  

While experimentally inaccessible, in the conventional theoretical framework of our calculation we can also determine the relative sign between these two $\kl\to\mu^+\mu^-$ amplitudes and $f_K$ which appears in the QCD matrix element $\langle 0|\left[(\bar{s}d)_{V-A}+\textrm{h.c.}\right]|\kl\rangle$ that enters as a factor in the matrix element of $\chw^\textrm{SD}$.  For this relative sign with respect to $f_K$ to be meaningful, we must describe explicitly the perturbative $\mu^+\mu^-$ state used in our calculation of $\mathcal{A}_{\kl\mu\mu}$.

Our $\mu^+\mu^-$ state can be specified as follows.  For the perturbative QED sector, we choose to work in the Weyl basis for the gamma matrices and use the representation for the spinor fields with momentum parallel to the $z$-direction
\begin{equation}\label{eq:diracspinor}
u^\pm(p) =\begin{pmatrix}
\sqrt{E-p_z}\;\phi_\pm\\
\sqrt{E+p_z}\;\phi_\pm
\end{pmatrix}\,,
\quad
v^\pm(p) =\begin{pmatrix}
\sqrt{E-p_z}\;\chi_\pm\\
-\sqrt{E+p_z}\;\chi_\pm
\end{pmatrix}\,,
\end{equation}
where
\begin{equation}
\phi^+ = \begin{pmatrix}
1 \\ 0
\end{pmatrix}\,,\quad
\phi^- = \begin{pmatrix}
0 \\ 1 
\end{pmatrix}\,,\quad
\chi^+ = \begin{pmatrix}
0 \\ 1
\end{pmatrix}\,,\quad
\chi^- = \begin{pmatrix}
1 \\ 0 
\end{pmatrix}\,.
\end{equation}
With this choice, the SD$\gf^2$ matrix element becomes
\begin{eqnarray}
\mathcal{A}_{\textrm{SD}\gf^2} &=&
 \lla\mu^+(\textbf{k}^+)\mu^-(\textbf{k}^-)|\chw^{\rm SD}(0)|\kl\rra \label{eq:ASD1} \\ 
 &=&2m_\mu \frac{G_{\rm F}}{\sqrt{2}}\frac{\alpha_{\rm em}(M_Z)}{2\pi \sin^2\theta_{\rm W}}{\rm Re}\left(V_{cs}^*V_{cd}Y_{\rm NL}+V^*_{ts}V_{td}Y(x_t)\right) \label{eq:ASD2} \\
 && \hskip 2.0 in \times \lla 0|\left[\bar{s}(0)\gamma^0\gamma^5d(0)+\textrm{h.c.}\right]|K_{\rm L}\rra\,, \nonumber
\end{eqnarray}
where the $\mu^+$ destruction operator appears to the left of that for the $\mu^-$ in the matrix elements defined in Eqs.~\eqref{eq:A-def} and \eqref{eq:ASD1}. Here the rightmost (renormalized) matrix element is to be computed in a QCD-only background.
This matrix element is proportional to the decay constant of the kaon, which is necessarily computed, including its sign in our conventions, when Eq.~\eqref{eq:ASD2} is evaluated.  We find
\begin{eqnarray}
\lla 0|\left[\bar{s}(0)\gamma^0\gamma^5d(0)+\textrm{h.c.}\right]|\kl\rra
     &=&-1.173(8)\times 10^5\mbox{ MeV}^2 \\
     &=& \sqrt{2}M_K f_K.
\end{eqnarray}
Here we have defined the phase of $f_K$~\cite{ParticleDataGroup:2022pth} as that obtained from a matrix element with a kaon state whose phase is chosen to make the imaginary part of the amplitude $\mathcal{A}_{\mathrm{LD}2\gamma}$ positive.

\subsection{Using lattice QCD to compute analytically-continued Minkowski-space Green's functions}

Consider a general Minkowski-space QCD Green's function $G(x^{[N]})$ that depends on $N$ space-time positions $x^{[N]}\equiv\{x_j\}_{1 \le j \le N}$.  If sums over intermediate QCD energy eigenstates are inserted between the $N$ operators appearing in that Green's function then the individual terms in the sums over intermediate states depend exponentially on the time components of those positions.  As a result the entire  Green's function is an analytic function of the times $\{(x_j)^0\}_{1 \le j \le N}$ if small negative imaginary parts are added to those $N$ times to ensure that the sums over intermediate states converge. 

If each of these $N$ time variables is analytically continued from their original real values to negative imaginary values $(x_j)^0 \to \exp\{-i\phi\}x_j^0$ with $\phi:0\to\frac{\pi}{2}$ and $(x_{j})^0 = \exp\{-i\frac{\pi}{2}\}(x_{\textrm{E},j})^4$ then we can define the result as the Euclidean Green's function $G_{\rm E}(x^{[N]}_{\rm E})$ depending on the $N$ Euclidean four-vectors $x^{[N]}_{\rm E} \equiv\{x_{\textrm{E},j}\}_{1\leq j \leq N}$ with $x_{\textrm{E},j} = (\vec x_{\textrm{E},j}, (x_{\textrm{E},j})^4) = (\vec x_{j}, i(x_{j})^0)$. Of course, $G_{\textrm{E}}(x_{\textrm{E}}^{[N]})$ could have been obtained from a Schr\"odinger-picture theory by constructing time-dependent operators $\mathcal{O}_{\textrm{E}}(x_\textrm{E}^{[N]})$ using a Heisenberg representation in which $\exp\{-iH_\mathrm{QCD}x^0\}$ is replaced by $\exp\{-H_\mathrm{QCD}x^4\}$.  If that Schr\"odinger-picture theory is relativistically invariant then the $G(x^{[N]})$ will be Lorentz covariant while $G_{\textrm{E}}(x_{\textrm{E}}^{[N]})$ will be O$(4)$-covariant.  

In the calculation reported here such an analytic continuation has been performed to transform a Minkowski-space Green's function obeying our Minkowski-space conventions into the Euclidean-space Green's functions appearing in Eqs.~\eqref{eq:Result_4pt}, \eqref{eq:Result_sub} and \eqref{eq:ASD2}.  We must be careful that the conventions used in our Euclidean-space lattice calculation are those that would result from the analytic continuation that has been performed.

The Green's functions that must be reproduced in Euclidean space involve four operators, the electromagnetic current $J^\mu(x)$, the effective weak Hamiltonian density $\chw(x)$, the $\Delta S= \pm 1$ axial current $(J_{sd}^5)^\mu(x)\equiv \bar{s}(x)\gamma^\mu\gamma^5 d(x) + \textrm{h.c.}$ and the kaon interpolating $\kl(t)$.  The operator $\kl(t)$ appears linearly in all of the Green's functions of interest and does not carry a physically meaningful phase so we need not be concerned that the operator $\kl(t)$ may not even have a familiar Minkowski-space expression. The Minkowski- and Euclidean-space representations for the operators $J^\mu(x)$, $\mathcal{H}_W(x)$ and $(J_{sd}^5)^\mu(x)$ can be easily compared knowing how the Dirac matrix conventions are related.

Recall that the Minkowski-space Dirac operator for a particle of mass $m$ in our conventions can be re-expressed after the analytic continuation to Euclidean-space as follows:
\begin{eqnarray}
D &=& i\gamma^\mu \frac{\partial}{\partial x^\mu} - m                                  \label{eq:dirac1} \\
  &=&  i\left[ \gamma^0 \frac{\partial}{\partial(-ix^4)}  +\sum_{j=1}^3 \gamma^i \frac{\partial}{\partial x^j}\right] -m \label{eq:dirac2} \\
  &=&  -\left[ \gamma^0 \frac{\partial}{\partial x^4}  -i\sum_{i=1}^3 \gamma^j \frac{\partial}{\partial x^i}\right] -m , \label{eq:dirac2}
\end{eqnarray}
which takes the form used in Euclidean space,
\begin{equation}
    D_{\rm E}=\sum_{i=\mu}^4\gamma_{\rm E}^\mu \frac{\partial}{\partial x_{\rm E}^\mu}+m \label{eq:dirac2}
\end{equation}
if we define:
\begin{equation}
    \gamma_{\rm E}^4 = \gamma^0\,, \quad \gamma_{\rm E}^j = -i\gamma^j .
\end{equation}

Since only the three spatial gamma matrices are changed when writing the analytically continued Green's function in Euclidean notation, both $\gamma^0$ and $\gamma^5$ need not be altered. Adopting this prescription for the gamma matrices used in our lattice calculation, we can reproduce the desired Minkowski-space matrix elements from the following formulae:
\begin{eqnarray}
    J^0 &=& J_{\rm E}^4\,, \quad J^j = iJ_{\rm E}^j\,, \\
    (J^5_{sd})^0 &=& (J^5_{sd,\rm E})^4\,,\quad (J^5_{sd})^j = i(J_{sd, \rm E}^5)^j\,, \\
    \overline{\psi}_a\gamma^\mu(1-\gamma^5)\psi_b \overline{\psi}_c\gamma_\mu(1-\gamma^5)\psi_d
  &=& 
    (\overline{\psi}_{\rm E})_a\gamma_{\rm E}^\mu(1-\gamma_{\rm E}^5)(\psi_{\rm E})_b (\overline{\psi}_{\rm E})_c\gamma_{\rm E}^\mu(1-\gamma_{\rm E}^5)(\psi_{\rm E})_d\,.
\end{eqnarray}
Here the $\psi_i$ represent different spinor fields. The Euclidean and Minkowski Grassmann spinor variables are unchanged except for their space-time labels -- no rotation of spinor indices is needed.   
The validity of these statements can be readily checked by analytically continuing some simple position space Green's functions in the time.

The Euclidean-space gamma matrix conventions derived above are very close to those used in lattice QCD, for example those found in Refs.~\cite{Montvay:1994cy,Gattringer:2010zz}.
The conventions in those two references are mathematically similar to those in the Grid~\cite{Boyle:2016lf,Yamaguchi:2022feu}  lattice code which we use.  
The Grid convention is shared with the Columbia Physics System~\cite{Jung:2014XW}, with Chroma~\cite{Edwards:2004sx}, and is as documented in the USQCD QDP++ documentation on spin conventions~\cite{qdpxx}.
However, in all these cases the $\gamma_{\rm E}^5$ matrix that is introduced is the negative of the $\gamma_{\rm E}^5$ matrix that results from the analytic continuation in the time that we are making.  Thus, when using the result for the matrix element of the axial current $(J^5_{sd})^0$ we must introduce an extra minus sign to obtain the matrix element of interest, in apparent conflict with the projection operators defined in Eq.~(8.11) of Ref.~\cite{Montvay:1994cy}.  The only other sign that must be recognized when interpreting our lattice matrix element as providing Minkowski-space results is the product $J^iJ^j$ which appears in Eqs.~\eqref{eq:Result_4pt} and \eqref{eq:Result_sub} where a minus sign must be introduced,
since our lattice calculation contains the factors $(\gamma_{\rm E}^i)(\gamma_{\rm E}^j) = -(\gamma^i)(\gamma^j)$ and the factors $+(\gamma^i)(\gamma^j)$ are what should occur in our Minkowski-space calculation.

\section{Conclusion}\label{sect:concl}

The formalism developed in Ref.~\cite{Chao:2024vvl} allows for a first-principles determination of the long-distance two-photon contribution to the $\kl\to\mu^+\mu^-$ decay amplitude from lattice QCD. 
Although our method does not accurately treat the contribution from low-energy $\pi\pi\gamma$ states, in Ref.~\cite{Chao:2024vvl} we show that these effects are at or below the ten-percent level.
In addition to the possibility for the systematic improvement of lattice QCD results, another appealing aspect of this \textit{ab initio} treatment of QCD is the ability to determine the relative sign of the two-photon exchange amplitude and the amplitude arising from the one-loop short-distance electro-weak contribution.

In this paper we present our numerical implementation of the formalism of Ref.~\cite{Chao:2024vvl} and the first result using a $24^3\times 64$ M\"obius Domain Wall fermion ensemble at physical pion mass and an inverse lattice spacing of $1/a=1.023$ GeV.
We have included all Feynman diagrams with the exception of SU$(3)$-flavor-suppressed tadpole terms in which an electromagnetic current attaches to a closed quark loop.
However, we should emphasize that while extensive, the calculation presented in the paper is not complete and cannot yet be directly compared with experiment.  Additional counter terms needed to represent the effects of the GIM mechanism have not been included.  This second part of the lattice QCD calculation of the two-photon contribution to $K_L\to\mu^+\mu^-$ will be presented in a second paper.

In addition to determining the needed renormalization counter terms, the natural next target for a more accurate result is the $48^3\times 96$ physical-pion-mass ensemble with $1/a=1.73$ GeV (`48I') from the RBC/UKQCD collaboration~\cite{RBC:2014ntl}, a calculation that is now also underway.
Since this ensemble is significantly larger than the current 24ID ensemble, the reconstruction of the $\eta$ contribution, which is at present the largest source of statistical error should be improved by the average over a larger volume and, because of the smaller lattice spacing, the increased number of data points in the region where the $\eta$ signal can be seen.

\section*{Acknowledgements}
We thank our RBC and UKQCD Collaboration colleagues for discussion and ideas, in particular Chulwoo Jung for technical support.
This research used resources of the National Energy Research Scientific Computing Center (NERSC), a Department of Energy User Facility using NERSC awards HEP-ERCAP0023253, 0026351, 0031326. 
P.B. was supported in part by US DOE Contract DESC0012704(BNL) and the Scientific Discovery through Advanced Computing (SciDAC) program LAB 22-2580.
E.-H.C. and N.H.C. were supported in part by the U.S. Department of Energy (DOE) grant No.~DE-SC0011941. 
L.C.J. acknowledges support by DOE Office of Science Early Career Award No.~DE-SC0021147 and DOE Award No.~DE-SC0010339.

\newpage

\appendix

\section{Computation strategies}\label{sect:comp}
We give the implementation details of the Wick contractions to get $\mathcal{A}^{\rm lat}$ [Eq.~\eqref{eq:Result_4pt}] considered in this work.
For the ones needed for $\mathcal{A}^{\rm sub}$ [Eq.~\eqref{eq:Result_sub}] and the $\eta$-removal described in Section~\ref{sect:ieq0}, we simply follow the strategies reported in Ref.~\cite{Zhao:2022njd}.
We denote the whole set of lattice sites by $\Lambda$ and the hyperplane corresponding to the time-slice $t$ by $\Lambda_0(t)$. 
The coordinate-space variables $u$ and $v$ indicate the locations of the EM currents (with time order $v_0\leq  u_0$), $x$ that of the weak Hamiltonian and $t_K$ that of the kaon wall, in accordance with Section~\ref{sect:amp-lat}.

While computing the amplitude Eq.~\eqref{eq:Result_4pt}, translational-invariance allows us to set the position of one of the three local operators to be the reference point and sum over the two other positions.
The reference point corresponds to the source location of point-source propagators needed for the contraction diagram to be evaluated.
Calculations will be repeated for many different reference point to improve statistics.
For each contraction class, we evaluate the associated \textit{reduced amplitudes} defined as follows on each gauge configuration.
For Type-1 and Type-2 diagrams, we find it convenient to set the reference point to $x$, 
\begin{equation}\label{eq:ampT12}
\bar{\mathcal{A}}^{\mathcal{D}}_i(\delta_{\max}, R_{\max}, \tsep) = 
\sum_{\substack{u,v\in\Lambda,\\ d \leq \delta_{\max},\, u_0-v_0\leq R_{\max}}}\delta_{v_0-x_0,d}\;e^{M_K (v_0-t_K)}\ K_{\mu\nu}(u-v,R_{\max})\; \mathcal{C}^{\mathcal{D}}_{i,\mu\nu} (u,v,x,t_K)\,,
\end{equation}
and to $v$ for Type-3 and Type-4 diagrams and Type-5 Diagram 2 (T5D2), 
\begin{equation}\label{eq:ampT34}
\bar{\mathcal{A}}^{\mathcal{D}}_i(\delta_{\rm max}, R_{\max}, \tsep)=
\sum_{\substack{u,x\in\Lambda,\\ d \leq \delta_{\max},\, u_0-v_0 \leq R_{\max}}}\delta_{v_0-x_0,d}
\;e^{M_K (v_0-t_K)}\ K_{\mu\nu}(u-v,R_{\max})\; \mathcal{C}^{\mathcal{D}}_{i,\mu\nu} (u,v,x,t_K)\,.
\end{equation}
The case of Type-5 Diagram 1 (T5D1) is a bit different as we use a low-mode-improved version of Eq.~\eqref{eq:ampT34}, which will be discussed in detail in Section~\ref{sect:t5}.
In the above, the subscript $i$ is the index for the corresponding weak operator and the superscript $\mathcal{D}$ indicates the associated diagram.
The contraction functions $\mathcal{C}^{\mathcal{D}}$ will be given in each dedicated subsection of this appendix.
The isospin components as defined in Section~\ref{sect:isospin} can be written as linear combinations of the reduced amplitudes Eq.~\eqref{eq:ampT12} and Eq.~\eqref{eq:ampT34}
\begin{equation}\label{eq:amp-iso-decomp}
\mathcal{A}^{\rm lat }_{i}(\delta_{\max},R_{\max},\tsep) = \bar{\mathcal{N}}\sum_{\substack{i=1,2 \\ \mathcal{D}\in\textrm{diagrams}}} b_i^{\rm I}(\mathcal{D}) C_i\lla\bar{\mathcal{A}}_i^{\mathcal{D}}(\delta_{\max},R_{\max},\tsep)\rra_U\,,
\end{equation}
where the charge factors $b_i^{\mathcal{D}}$ are given in Table~\ref{tab:bfac}, the $C_i$'s are the Wilson coefficients, $\bar{\mathcal{N}}$ is the normalization factor to get the kaon ground state in the relativistic normalization and $\langle\cdot\rangle_{U}$ is the gauge-ensemble average.

\begin{table}[h!]\centering
\begin{tabular}{c|c|c|c|c|c|c|c|c}
\toprule
$b_i^{I=1}$ & T1D1 & T1D2 & T2D1 & T3D1 & T4D1 & T4D2 & T5D1 & T5D2 \\
\hline\hline
$Q_1$ & --- & -1/6 & 1/6  & 1/6  & -1/6 & --- & --- & --- \\
$Q_2$ & --- & -1/6 & -1/6 & -1/6 & 1/6  & --- & --- & --- \\
\hline\hline
$b_i^{I=0,2}$ & T1D1 & T1D2 & T2D1 & T3D1 & T4D1 & T4D2 & T5D1 & T5D2 \\
\hline\hline
$Q_1$ & -2/9 & -1/18 & -1/18 & 5/18, & 5/18 & 1/9 & -5/9 &  -1/9 \\
$Q_2$ & -2/9 & -1/18 & 1/18  & -5/18 & -5/18& -1/9& 5/9  &   1/9 \\
\hline
\end{tabular}
\caption{The charge factors defined in Eq.~\eqref{eq:amp-iso-decomp}}
\label{tab:bfac}
\end{table}

The $R_{\max}$-regulated kernel $K_{\mu\nu}(u-v,R_{\max})$ is defined in Eq.~\eqref{eq:kmunu}.
We will keep the $R_{\max}$-dependence implicit in the remainder of this appendix.
The variable $\tsep$ is defined as the distance between the weak operator and the kaon wall,
\begin{equation}
\tsep \equiv x_0-t_K\,.
\end{equation}
In our calculations, we aim at computing each diagram at several values of $(\tsep, R_{\max})$.
For each $R_{\max}$, our computer program constructs a different kernel with points with $u_0-v_0 > R_{\max}$ set to zero.

In the rest of this appendix, we will use the notations
\begin{equation}
\gfive_\mu\equiv \gamma_5\gamma_\mu\,,\quad
\gL_\mu\equiv \gamma_\mu(1-\gamma_5)\,.
\end{equation}
The Lorentz indices $\mu$ and $\nu$ are associated with the currents at $u$ and $v$ respectively.
A light (respectively, strange)-quark point-source propagator with source at $y$ and sink at $x$ is denoted by $L(x,y)$ (respectively, $H(x,y)$).
$L(x,t_K)$ (respectively, $H(x,t_K)$) denotes a light (respectively, strange)-quark Coulomb-gauge-fixed wall-source propagator which carries zero spatial momenta and has its source located at time-slice $t_K$.
Finally, for the naming convention, a diagram `a' is related to its counterpart `b' by swapping the locations of the two currents, \textit{i.e.} $(u,\mu)\leftrightarrow(v,\nu)$.
For simplicity, we will make the coordinate-space arguments in some of the expressions implicit, while keeping them consistent with Eqs.~\eqref{eq:ampT12} and \eqref{eq:ampT34}.

\subsection{Type-1}

\begin{figure}
\centering
\includegraphics[scale=0.8]{./Figures/type1d1a}
\includegraphics[scale=0.8]{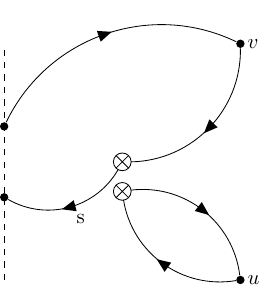}
\includegraphics[scale=0.8]{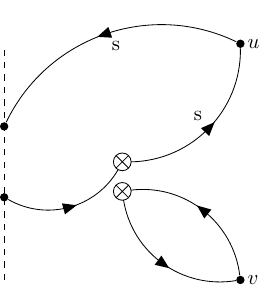}
\includegraphics[scale=0.8]{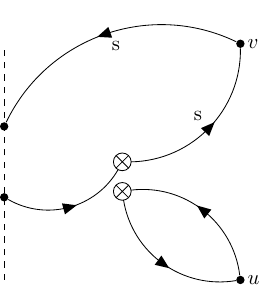}
\caption{Type-1 diagrams. From left to right: T1D1a, T1D1b, T1D2a and T1D2b.}
\label{fig:t1diag}
\end{figure}

For the Type-1 topology, there are four diagrams: `T1D1a', `T1D1b', `T1D2a', `T1D2b', as shown in Figure~\ref{fig:t1diag}.
The contraction functions entering Eq.~\eqref{eq:ampT12} are:
\begin{equation}
\begin{split}
\mathcal{C}_{1,\mu\nu}^{\mathrm{T1D1a}} =  \tr[\gL_\rho L(x,v)\gamma_\nu L(v,x)] 
\,
\tr\left[\gL_\rho
\left\{ L(x,u)\gamma_\mu L(u,t_K)\gamma_5 H(t_K, x) +\mathrm{h.c.}\right\}\right]\,,
\end{split}
\end{equation}
\begin{equation}
\begin{split}
\mathcal{C}_{1,\mu\nu}^{\mathrm{T1D2a}} = 
\tr[\gL_\rho L(x,v)\gamma_\nu L(v,x) ]
\,
\tr\left[\gL_\rho \left\{ L(x,t_K)\gamma_5 H(t_K,u)\gamma_\mu H(u,x)
+\mathrm{h.c.}\right\}
 \right] \,,
\end{split}
\end{equation}
\begin{equation}\label{eq:c2t1d1a}
\begin{split}
\mathcal{C}_{2,\mu\nu}^{\mathrm{T1D1a}} =  \tr_{\rm C}\Big[ \tr_{\rm S}[\gL_\rho L(x,v)\gamma_\nu L(v,x)]
\,
\tr_{\rm S}\left[\gL_\rho
\left\{ L(x,u)\gamma_\mu L(u,t_K)\gamma_5 H(t_K, x) +\mathrm{h.c.}\right\}\right]\Big]\,,
\end{split}
\end{equation}
\begin{equation}\label{eq:c2t1d2a}
\begin{split}
\mathcal{C}_{2,\mu\nu}^{\mathrm{T1D2a}} = 
\tr_{\rm C}\Big[\tr_{\rm S}[\gL_\rho L(x,v)\gamma_\nu L(v,x) ]
\,
\tr_{\rm S}\left[\gL_\rho \left\{ L(x,t_K)\gamma_5 H(t_K,u)\gamma_\mu H(u,x)
+\mathrm{h.c.}\right\}
 \right]\Big] \,,
\end{split}
\end{equation}
plus those for the diagram `b's, obtained with $(u,\mu)\leftrightarrow (v,\nu)$.
In the above, $\tr_{\rm C}$ (respectively, $\tr_{\rm S}$) indicates tracing with regard to the color (respectively, spin) indices, whereas $\tr$ is the trace over both spin and color indices.

The advantage of adopting the parametrization Eq.~\eqref{eq:ampT12} should become clear at this point: by taking the reference point at $x$, we only need a pair of wall-source propagators and a pair of point-source propagators for both the light and strange quarks per reference point.
To get the reduced amplitude Eq.~\eqref{eq:ampT12}, we convolve the four-point functions with the leptonic kernel. 
With this observation, the sum over $u$ for all $v\in\Lambda$ can be done efficiently via Fast Fourier Transform.
We re-write the four-point functions in blocks to compute them for different $(\tsep, R_{\rm max})$ with reduced overhead.
We define the `bubble'-blocks as 
\begin{equation}
G_{\nu\rho}(v,x) \equiv \tr_{\rm S}\left[\gL_\rho A_\nu \right]\,,\quad A_\nu\equiv L^\dagger(v,x)\gfive_\nu L(v,x)\,,
\end{equation}
and the `triangle'-blocks as
\begin{equation}
F^{1}_{\mu\rho}(u,x,\tsep) = (-1)\times\tr_{\rm S}\left[\gL_\rho \left(B_1(u) + B_1^\dagger(u)\right) \right]\,,\quad B_1\equiv L^\dagger(u,x)\gfive_\mu L(u,t_K)H^\dagger(x,t_K)\,,
\end{equation}
\begin{equation}
F^{2}_{\mu\rho}(u,x,\tsep) = \tr_{\rm S}\left[\gL_\rho \left(B_2(u) + B_2^\dagger(u)\right) \right]\,,\quad B_2(u)\equiv L(x,t_K)H^\dagger(u,t_K) \gfive_\mu H(u,x)\,.
\end{equation}
Our program loops over the kernels first.
In each iteration, we need the kernel-dependent quantities, obtained by convolving the above quantities with different regulated kernels
\begin{equation}
\hat{G}_{\nu\rho}(v,x) \equiv (L_{\mu\nu} \ast G_{\mu\rho})(v,x)\,,
\end{equation}
\begin{equation}
\hat{F}_{\nu\rho}(v,x,\tsep) \equiv (L_{\mu\nu} \ast F_{\mu\rho})(v,x,\tsep)\,,
\end{equation}
where the convolution $\ast$ is defined via
\begin{equation}
(K_{\mu\nu}\ast F_{\mu\rho}) (v) \equiv \sum_{u\in\Lambda} K_{\mu\nu}(u-v)F_{\mu\rho}(u) = 
\mathcal{F}^{-1}\left[\mathcal{F}[K_{\mu\nu}](-p) \mathcal{F}[F_{\mu\rho}](p)\right]\,,
\end{equation}
with $\mathcal{F}$ denoting a Fourier transform.
Noting that the bubble-blocks do not depend on $\tsep$, we can reuse the bubble-blocks while looping over $\tsep$'s.
Taking the color-trace properly, we get the reduced amplitudes
\begin{equation}
\bar{\mathcal{A}}_1^{\textrm{T1D1a}} = 
(-1)\times\sum_{\substack{v\in\Lambda\;|\;d\leq\delta_{\rm max}}} \delta_{v_0-x_0, d}\;
e^{M_K (v_0-t_K)}\ \tr_{\rm C}\left[ \hat{F}_{\nu\rho}^1(v,x,\tsep)\right]\tr_{\rm C}\left[ G_{\nu\rho}(v,x) \right]\,,
\end{equation}
\begin{equation}
\bar{\mathcal{A}}_1^{\textrm{T1D1b}} = 
(-1)\times\sum_{\substack{v\in\Lambda\;|\;d\leq\delta_{\rm max}}}\delta_{v_0-x_0, d}\; 
e^{M_K (v_0-t_K)}\ \tr_{\rm C}\left[ F_{\nu\rho}^1(v,x,\tsep)\right]\tr_{\rm C}\left[ \hat{G}_{\nu\rho}(v,x) \right]\,,
\end{equation}
\begin{equation}
\bar{\mathcal{A}}_2^{\textrm{T1D1a}} = 
(-1)\times\sum_{\substack{v\in\Lambda\;|\;´d\leq\dmax}} 
\delta_{v_0-x_0, d}\;
e^{M_K (v_0-t_K)}\ \tr_{\rm C}\left[ \hat{F}_{\nu\rho}^1(v,x,\tsep)G_{\nu\rho}(v,x) \right]\,,
\end{equation}
\begin{equation}
\bar{\mathcal{A}}_2^{\textrm{T1D1b}} = 
(-1)\times\sum_{\substack{v\in\Lambda\;|\;d\leq\dmax}} 
\delta_{v_0-x_0, d}\;
e^{M_K (v_0-t_K)}\ \tr_{\rm C}\left[ F_{\nu\rho}^1(v,x,\tsep) \hat{G}_{\nu\rho}(v,x) \right]\,,
\end{equation}
and the `D2' diagrams are obtained by substituting $F^1$ with $F^2$.

\subsection{Type-2}

\begin{figure}[h!]
\centering
\includegraphics[scale=1.]{./Figures/type2d1a}
\includegraphics[scale=1.]{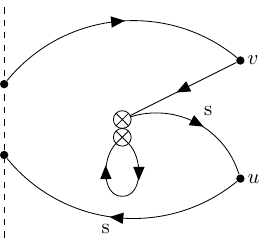}
\caption{Type-2 diagrams. From left to right: T2D1a and T2D1b.}
\label{fig:t2diag}
\end{figure}
The diagrams for this contraction class are `T2D1a' and `T2D1b' as shown in Figure~\ref{fig:t2diag}. 
The four-point functions for this contraction class are
\begin{equation}
\mathcal{C}_{1,\mu\nu}^{\mathrm{T2D1a}} = 
\tr[\gL_\rho L(x,x) ]
\,
\tr\left[\gL_\rho \left\{ L(x,u)\gamma_\mu L(u,t_K)\gamma_5 H(t_K, v)\gamma_\nu H(v,x) - \mathrm{h.c.} \right\}
 \right] \,,
\end{equation}
\begin{equation}
\mathcal{C}_{2,\mu\nu}^{\mathrm{T2D1a}} = 
\tr\Big[\gL_\rho L(x,x)
\gL_\rho \left\{ L(x,u)\gamma_\mu L(u,t_K)\gamma_5 H(t_K, v)\gamma_\nu H(v,x) - \mathrm{h.c.} \right\}
 \Big] \,,
\end{equation}
plus the `b' diagrams obtained with $(u,\mu)\leftrightarrow(v,\nu)$.
The parametrization Eq.~\eqref{eq:ampT12} is chosen for the same reason as in the Type-1 case.

Similarly, we can write the quantities to compute in terms of sub-blocks to reduce computation overhead.
We define
\begin{equation}
F_\mu(u,x,\tsep)\equiv \gamma_5 L^\dagger(u,x) \gfive_\mu L(u,t_K)\,,
\end{equation}
\begin{equation}
G_\nu(v,x,\tsep)\equiv H^\dagger(v,t_K) \gfive_{\nu} H(v,x)\,,
\end{equation}
and their counterparts convolved with the kernel
\begin{equation}
\hat{F}_\nu(v,x,\tsep)\equiv (K_{\mu\nu}\ast F_\mu )(v,x,\tsep)\,,
\end{equation}
\begin{equation}
\hat{G}_\nu(v,x,\tsep)\equiv (K_{\mu\nu}\ast G_\mu )(v,x,\tsep)\,.
\end{equation}
With
\begin{equation}
T^1_\rho(x)\equiv \gL_\rho L(x,x)\,,
\end{equation}
\begin{equation}
T^2_\rho[B,C](z,x,\tsep)\equiv \gL_\rho \left[
B_\nu(z,x,\tsep) C_\nu(z,x,\tsep) - \textrm{h.c.}
\right]\,,
\end{equation}
we then have
\begin{equation}
\bar{\mathcal{A}}_{1}^{\mathrm{T2D1a}} = 
\sum_{\substack{v\in\Lambda\;|\;d\leq\dmax}} \delta_{v_0-x_0,d}\;
e^{M_K (v_0-t_K)}\ \tr\left[T^1_\rho(x)\right]\tr\left[T^2_\rho(\hat{F},G;v,x,\tsep)\right]\,, 
\end{equation}
\begin{equation}
\bar{\mathcal{A}}_{1}^{\mathrm{T2D1b}} = 
\sum_{\substack{v\in\Lambda\;|\;d\leq\dmax}} \delta_{v_0-x_0,d}\;
e^{M_K (v_0-t_K)}\ \tr\left[T^1_\rho(x)\right]\tr\left[T^2_\rho(F,\hat{G};v,x,\tsep)\right]\,, 
\end{equation}
\begin{equation}
\bar{\mathcal{A}}_{2}^{\mathrm{T2D1a}} = 
\sum_{\substack{v\in\Lambda\;|\;d\leq\dmax}} \delta_{v_0-x_0,d}\;
e^{M_K (v_0-t_K)}\ \tr\left[T^1_\rho(x) T^2_\rho(\hat{F},G;v,x,\tsep)\right]\,, 
\end{equation}
\begin{equation}
\bar{\mathcal{A}}_{2}^{\mathrm{T2D1b}} = 
\sum_{\substack{v\in\Lambda\;|\;d\leq\dmax}} \delta_{v_0-x_0,d}\;
e^{M_K (v_0-t_K)}\ \tr\left[T^1_\rho(x) T^2_\rho(F,\hat{G};v,x,\tsep)\right]\,. 
\end{equation}

\subsection{Type-3}\label{sect:t3}
\begin{figure}[h!]
\centering
\includegraphics[scale=1.]{./Figures/type3d1a}
\includegraphics[scale=1.]{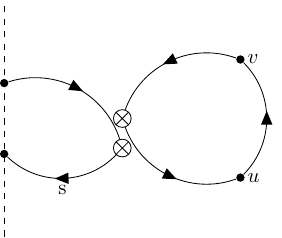}
\caption{Type-3 diagrams. From left to right: T3D1a and T3D1b.}
\label{fig:t3diag}
\end{figure}

The Type-3 diagrams are `T3D1a' and `T3D1b', depicted in Fig.~\ref{fig:t3diag}.
The corresponding four-point functions are
\begin{equation}
\begin{split}
\mathcal{C}_{1,\mu\nu}^{\mathrm{T3D1a}} = 
\tr[\gL_\rho L(x,u)\gamma_\mu L(u,v)\gamma_\nu L(v,x) ]
\,\tr[\gL_\rho \left\{L(x,t_K)\gamma_5 H(t_K,x) -\mathrm{h.c.} \right\}]\,,
\end{split}
\end{equation}
\begin{equation}
\begin{split}
\mathcal{C}_{2,\mu\nu}^{\mathrm{T3D1a}} = 
\tr\Big[\gL_\rho L(x,u)\gamma_\mu L(u,v)\gamma_\nu L(v,x) 
\gL_\rho\left\{L(x,t_K)\gamma_5 H(t_K,x) -\mathrm{h.c.} \right\}
\Big]\,,
\end{split}
\end{equation}
plus the `b' diagrams with $(u,\mu)\leftrightarrow(v,\nu)$.
We adopt the parametrization Eq.~\eqref{eq:ampT34} for the Type-3 diagrams with reference point at $v$ so that the `bubble' formed by the kaon wall and the weak operator is summed over the whole spacetime volume, presumably leading to smaller noise for this part.
The reduced amplitudes are then given by
\begin{equation}\label{eq:a1t3d1a}
\bar{\mathcal{A}}_1^{\rm{T3D1a}} =\sum_{\substack{u,x\in\Lambda\;|\;d\leq\dmax}} \delta_{v_0-x_0,d}\;
e^{M_K(\tsep+d)}\;\tr[\gL_\rho F(x,v)]\tr[\gL_\rho G(x,t_K)]\,,
\end{equation}
\begin{equation}
\bar{\mathcal{A}}_1^{\rm{T3D1b}} =\sum_{\substack{u,x\in\Lambda\;|\;d\leq\dmax}} \delta_{v_0-x_0,d}\;
e^{M_K(\tsep+d)}\;\tr[\gL_\rho F^\dagger(x,v)]\tr[\gL_\rho G(x,t_K)]\,,
\end{equation}
\begin{equation}
\bar{\mathcal{A}}_2^{\rm{T3D1a}} =\sum_{\substack{u,x\in\Lambda\;|\;d\leq\dmax}} \delta_{v_0-x_0,d}\;
e^{M_K(\tsep+d)}\;\tr[\gL_\rho F(x,v)\gL_\rho G(x,t_K)]\,,
\end{equation}
\begin{equation}\label{eq:a2t3d1b}
\bar{\mathcal{A}}_2^{\rm{T3D1b}} =\sum_{\substack{u,x\in\Lambda\;|\;d\leq\dmax}} \delta_{v_0-x_0,d}\;
e^{M_K(\tsep+d)}\;\tr[\gL_\rho F^\dagger(x,v)\gL_\rho G(x,t_K)]\,,
\end{equation}
where
\begin{equation}
F(x,v) \equiv S(x,v) \gamma_5 L^\dagger(x,v)\,,
\end{equation}
\begin{equation}
G(x,t_K) \equiv L(x,\tilde{t}_K)H^\dagger(x,\tilde{t}_K)\gamma_5- \textrm{h.c.}\,.
\end{equation}
\begin{equation}\label{eq:Sseq}
S(x,v)\equiv L(x,u)\gamma_\mu L(u,v)\gamma_\nu K_{\mu\nu}(u-v)\,,
\end{equation}
Compared to the previous cases, extra propagators are needed in addition to the available point-source and wall-source ones due to the summation over two of the coordinate-space variables. 
Instead of using the sequential-propagator technique, we adopt the all-to-all propagator technique~\cite{Foley:2005ac} to reduce the overhead on the inversion of the Dirac operator while computing with different $R_{\max}$-regulated leptonic kernels.
In our implementation, we use a low-mode improved stochastic estimator, where the Z-M\"obius-approximated low-mode part of the Dirac operator is implemented exactly from the eigenvectors of the preconditioned Dirac operator and high-mode part reconstructed stochastically.
The method works as follows.
Let $\xi$ be a random variable satisfying the usual orthonormality condition on the entire spacetime
\begin{equation}
\mathbb{E}[\xi(x)\xi^\dagger(y)] = \delta_{x,y}\,,
\end{equation}
where $\mathbb{E}$ is the expectation value over the probability distribution of $\xi$. 
With $\{\xi_j\}_{1\leq j \leq N_{\rm hits}}$ a set of realizations of $\xi$ and two sets of spin-color vectors $\{v_i\}_{1\leq i\leq N_{\rm ev}}$ and $\{w_i\}_{1\leq i \leq N_{\rm ev}}$, we can construct an unbiased estimator for the light-quark propagator 
\begin{equation}\label{eq:shat}
\begin{split}
\hat{L}(x,y)=&\sum_{i=1}^{N_{\rm{ev}}}v_i(x)\otimes [w_i]^*(y)
-\frac{1}{N_{\rm{hits}}}\sum_{i=1}^{N_{\rm{ev}}}\sum_{j=1}^{N_{\rm{hits}}}\sum_{z\in\Lambda} \lla w_i(z), \xi_j(z) \rra v_i(x)\otimes \xi_j^*(y)
\\ &+
\frac{1}{N_{\rm{hits}}}\sum_{j=1}^{N_{\rm{hits}}}\sum_{z\in\Lambda}L(x,z)\xi_j(z)\otimes \xi_j^*(y)\,.
\end{split}
\end{equation}
For our setup, we construct the vectors $\{v_i\}_{1\leq i \leq N_{\rm ev}}$ and $\{w_i\}_{1\leq i \leq N_{\rm ev}}$ from the (approximated) low modes of the preconditioned Dirac operator as explained in Appendix~\ref{sect:a2a}.
Those will be denoted with a superscript `l' [Eqs.~\eqref{eq:vlow} and \eqref{eq:wlow}].
We exploit the following identity valid for two spin-color vectors $V$ and $W$ and a spin-color matrix $B$ at a given site
\begin{equation}\label{eq:trpdt}
\tr\left[(V\otimes W^*) B\right] = 
\lla B^\dagger W, V \rra\,,
\end{equation}
where $\langle\cdot,\cdot\rangle$ is the usual scalar product defined for spin-color vectors.

Define
\begin{equation}
\Phi(u,v)\equiv \gamma_\mu L(u,v) \gamma_\nu K_{\mu\nu}(u-v)\,,
\end{equation} 
and
\begin{equation}
\mathcal{G}^{1}_\rho(x,t_K)\equiv\tr[\gL_\rho G(x,t_K)]\,,
\end{equation}
\begin{equation}
\mathcal{G}^{2}_\rho(x,t_K)\equiv\gL_\rho G(x,t_K)\,.
\end{equation}
With Eq.~\eqref{eq:shat} and Eq.~\eqref{eq:trpdt}, we can express the reduced amplitudes Eqs.~(\ref{eq:a1t3d1a}-\ref{eq:a2t3d1b}) as 
\begin{equation}
\bar{\mathcal{A}}^{\rm{T3D1a}}_1 = \sum_{d\leq\dmax}e^{M_K( \tsep+d)}\;\hat{\mathcal{C}}^{a}_1(d,\tsep,v)\,,
\end{equation}
\begin{equation}
\bar{\mathcal{A}}^{\rm{T3D1a}}_2 = \sum_{d\leq\dmax}e^{M_K( \tsep+d)}\;\hat{\mathcal{C}}^{a}_2(d,\tsep,v)\,,
\end{equation}
\begin{equation}
\bar{\mathcal{A}}^{\rm{T3D1b}}_1 = \sum_{d\leq\dmax}e^{M_K( \tsep+d)}\;\hat{\mathcal{C}}^{b}_1(d,\tsep,v)\,,
\end{equation}
\begin{equation}
\bar{\mathcal{A}}^{\rm{T3D1b}}_2 = \sum_{d\leq\dmax}e^{M_K( \tsep+d)}\;\hat{\mathcal{C}}^{b}_2(d,\tsep,v)\,,
\end{equation}
where
\begin{equation}
\begin{split}
\hat{\mathcal{C}}^a_q(d,\tsep,v) = &\quad
\sum_{i=1}^{N_{\rm{ev}}}\lla M_i(v), N^a_{q,i}(d,\tsep,v) \rra
-\frac{1}{N_{\rm{hits}}}\sum_{i=1}^{N_{\rm{ev}}}\sum_{j=1}^{N_{\rm{hits}}}\lla w^{\rm l}_i(z),\xi_j(z)\rra
\lla P_j(v), N^a_{q,i}(d,\tsep,v)\rra
\\ &
+\frac{1}{N_{\rm{hits}}}\sum_{j=1}^{N_{\rm{hits}}}\lla P_j(v), Q^a_{q,j}(d,\tsep,v)\rra\,,
\end{split}
\end{equation}
\begin{equation}
\begin{split}
\hat{\mathcal{C}}^b_q(d,\tsep,v) = &\quad
\sum_{i=1}^{N_{\rm{ev}}}\lla M_i(v), N^b_{q,i}(d,\tsep,v) \rra^*
-\frac{1}{N_{\rm{hits}}}\sum_{i=1}^{N_{\rm{ev}}}\sum_{j=1}^{N_{\rm{hits}}}\lla w^{\rm l}_i(z),\xi_j(z)\rra^*
\lla P_j(v), N^b_{q,i}(d,\tsep,v)\rra^*
\\ &
+\frac{1}{N_{\rm{hits}}}\sum_{j=1}^{N_{\rm{hits}}}\lla P_j(v), Q_{q,j}^b(d,\tsep,v)\rra^*\,,
\end{split}
\end{equation}
\begin{equation}
M_i(v) \equiv \sum_{u\in\Lambda}\Phi^\dagger(u,v)w_i^{\rm l}(u)\,,
\end{equation}
\begin{equation}
N^a_{q,i}(d,\tsep,v) \equiv \sum_{x\in\Lambda}\delta_{v_0-x_0,d}\;\gamma_5 L^\dagger(x,v) \mathcal{G}^q_\rho(x,t_K)\Gamma_\rho^L v_i^{\rm l}(x)\,,
\end{equation}
\begin{equation}
N^b_{q,i}(d,\tsep,v) \equiv \sum_{x\in\Lambda}\delta_{v_0-x_0,d}\;\gamma_5 L^\dagger(x,v) \left[\mathcal{G}^q_\rho(x,t_K)\Gamma_\rho^L\right]^\dagger v_i^{\rm l}(x)\,,
\end{equation}
\begin{equation}
P_j(v) \equiv \sum_{u\in\Lambda}\Phi^\dagger(u,v)\xi_j(u)\,,
\end{equation}
\begin{equation}
Q^a_{q,j}(d,\tsep,v)\equiv \sum_{x\in\Lambda}\delta_{v_0-x_0,d}\;\gamma_5 L^\dagger(x,v)\mathcal{G}^q_\rho(x,t_K)\Gamma_\rho^L\Omega_j(x)\,,
\end{equation}
\begin{equation}
Q^b_{q,j}(d,\tsep,v)\equiv \sum_{x\in\Lambda}\delta_{v_0-x_0,d}\;\gamma_5 L^\dagger(x,v)\left[\mathcal{G}^q_\rho(x,t_K)\Gamma_\rho^L\right]^\dagger\Omega_j(x)\,.
\end{equation}
\begin{equation}\label{eq:omega}
\Omega_j(x) \equiv \sum_{z\in\Lambda}L(x,z)\xi_j(z)\,.
\end{equation}

Among the above quantities, we have a clear separation between the kernel-dependent quantities and the $\tsep$-dependent ones. 
This factorization helps reduce the overhead for getting results at various $(\tsep,R_{\rm max})$.
With the available precomputed point-source propagators, we can construct the kernel-dependent quantities $M_i(v)$ and $P_j(v)$ for a fixed reference point $v$. 
Although the contraction cost scales linearly with the numbers of vectors used, ie. $N_{\rm ev}$ and $N_{\rm hits}$, with our typical number of low modes of O(1000), the calculation is still dominated by the inversion of the Dirac operator on random vectors to get $\Omega_j(x)$ [Eq.~\eqref{eq:omega}].

\subsection{Type-4}
\begin{figure}[h!]
\centering
\includegraphics[scale=0.8]{./Figures/type4d1a}
\includegraphics[scale=0.8]{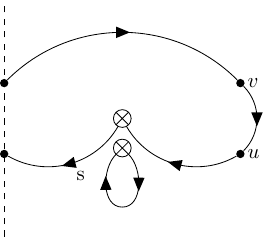}
\includegraphics[scale=0.8]{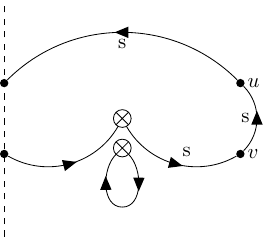}
\includegraphics[scale=0.8]{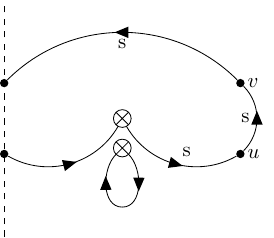}
\caption{Type-4 diagrams. From left to right: T4D1a, T4D1b, T4D2a, T4D2b.}
\label{fig:t4diag}
\end{figure}

The diagrams for this contraction class, T4D1a, T4D1b, T4D2a and T4D2b, are depicted in Figure~\ref{fig:t4diag}. 
The corresponding contraction functions entering Eq.~\eqref{eq:ampT34} are defined as 
\begin{equation}
\begin{split}
\mathcal{C}_{1}^{\mathrm{T4D1a}} = 
\tr[ \gL_\rho  L(x,x)]
\,\tr[\gL_\rho\left\{
L(x,v)\gamma_\nu L(v,u)\gamma_\mu L(u,t_K)\gamma_5 H(t_K,x)
-\mathrm{h.c.}
\right\}]\,,
\end{split}
\end{equation}
\begin{equation}
\begin{split}
\mathcal{C}_{2}^{\mathrm{T4D1a}} = 
\tr\Big[ \gL_\rho  L(x,x)
\gL_\rho\left\{
L(x,v)\gamma_\nu L(v,u)\gamma_\mu L(u,t_K)\gamma_5 H(t_K,x)
-\mathrm{h.c.}
\right\}\Big]\,,
\end{split}
\end{equation}
\begin{equation}
\begin{split}
\mathcal{C}_{1}^{\mathrm{T4D2a}} = 
\tr[ \gL_\rho L(x,x) ]
\,\tr[\gL_\rho\left\{
L(x,t_K)\gamma_5 H(t_K, u)\gamma_\mu H(u,v)\gamma_\nu H(v,x)
-\mathrm{h.c.}
\right\}]\,,
\end{split}
\end{equation}
\begin{equation}
\begin{split}
\mathcal{C}_{2}^{\mathrm{T4D2a}} = 
\tr\Big[ \gL_\rho L(x,x) 
\gL_\rho\left\{
L(x,t_K)\gamma_5 H(t_K, u)\gamma_\mu H(u,v)\gamma_\nu H(v,x)
-\mathrm{h.c.}
\right\}\Big]\,,
\end{split}
\end{equation}
plus the `b' diagrams with $(\mu,u)\leftrightarrow(\nu,v)$.

We use the parametrization Eq.~\eqref{eq:ampT34} with the reference point at $v$ (or with $u$ and $v$ swapped for the `b'-diagrams) so that we can compute these diagrams with several kernel with different $R_{\rm max}$ without involving sequential propagators.
With this parametrization, we need to know the tadpole $\tr[L(x,x)]$ at the weak Hamiltonian for all $x\in\Lambda$. 
This can be achieved using a similar low-mode improved stochastic estimator Eq.~\eqref{eq:shat} discussed in the Type-3 case
\begin{equation}\label{eq:lxx}
\tr[L(x,x)] =\mathbb{E}\left[\tr\left[\hat{L}(x,x)\right]\right]\,.
\end{equation}
For this quantity, we use full spacetime volume sources for the random vector instead, which is supposed to be more efficient due to the exponentially suppressed off-diagonal terms in the propagator.
We can decompose the amplitudes into blocks with explicit $\tsep$- or kernel-dependence to reduce the computational overhead.

Define
\begin{equation}
B_1(v,t_K) = (-1)\times \sum_{u\in\Lambda} K_{\mu\nu}(u-v)\gfive_\mu L^\dagger(u,v)\gfive_\nu L(u,t_K)\,,
\end{equation}
\begin{equation}
B_2(v,t_K) = (-1)\times\sum_{u\in\Lambda}H^\dagger(u,t_K) K_{\mu\nu}(u-v)\gfive_\mu H(u,v)\gfive_\nu \,,
\end{equation}
\begin{equation}
\begin{split}
B_3(u,v_0,t_K) = 
& (-1)\times\sum_{\vec{v}\in\Lambda_0}K_{\mu\nu}(u-v)\gfive_\mu L^\dagger(v,u)\gfive_\nu L(v,t_K)
\,,
\end{split}
\end{equation}
\begin{equation}
\begin{split}
B_4(u,v_0,t_K) = 
& (-1)\times\sum_{\vec{v}\in\Lambda_0}H^\dagger(v,t_K)K_{\mu\nu}(u-v)\gfive_\mu H(v,u)\gfive_\nu
\,,
\end{split}
\end{equation}
\begin{equation}\label{eq:t4g1}
G_1(x) = \tr[\gL_\rho L(x,x)] \gL_\rho\,,
\end{equation}
\begin{equation}\label{eq:t4g2}
G_2(x) = \gL_\rho L(x,x) \gL_\rho\,.
\end{equation}
and for $i=1,2$, 
\begin{equation}
E_{i}^{(1)}(x,t_K)= H^\dagger(x,t_K)\gamma_5 G_i(x)\,,
\end{equation}
\begin{equation}
E_{i}^{(2)}(x,t_K)= G_i(x)\gamma_5H(x,t_K)\,,
\end{equation}
\begin{equation}
F_{i}^{(1)}(x,t_K)= \gamma_5 G_i(x)L(x,t_K)\,,
\end{equation}
\begin{equation}
F_{i}^{(2)}(x,t_K)= L^\dagger(x,t_K)G_i(x)\gamma_5\,.
\end{equation}
We can then write the reduced amplitudes of the Type-4 `a'-diagrams as
\begin{equation}
\bar{\mathcal{A}}_i^{\mathcal{D}} = \sum_{d\leq\dmax}\sum_{x\in\Lambda}\delta_{v_0-x_0,d}\;e^{M_K(\tsep+d)}\;\tilde{\mathcal{C}}_i^{\mathcal{D}}(x,v,\tsep)\,,
\end{equation}
with, for $i=1,2$,
\begin{equation}
\tilde{\mathcal{C}}_{i}^{\mathrm{T4D1a}}(x,v,\tsep) = \tr\left[
E_{i}^{(2)}(x,t_K)B_1^\dagger(v,t_K)L^\dagger(x,v)
-
E_{i}^{(1)}(x,t_K)L(x,v)B_1(v,t_k)
\right]\,,
\end{equation}
\begin{equation}
\tilde{\mathcal{C}}_{i}^{\mathrm{T4D2a}}(x,v,\tsep) = \tr\left[
F^{(1)}_i(x,t_K)B_2(v,t_K)H^\dagger(x,v)
-
F^{(2)}_i(x,t_K)H(x,v)B_2^\dagger(v,t_K)
\right]\,.
\end{equation}
Similarly, for the `b'-diagrams, we write 
\begin{equation}
\bar{\mathcal{A}}_i^{\mathcal{D}} = \sum_{d\leq\dmax}\sum_{v_0=0}^{T-1}\sum_{x\in\Lambda}\delta_{v_0-x_0,d}\;e^{M_K(\tsep+d)}\;\tilde{\mathcal{C}}_i^{\mathcal{D}}(x,u,\tsep)\,,
\end{equation}
where
\begin{equation}
\tilde{\mathcal{C}}_{i}^{\mathrm{T4D1b}}(x,u,\tsep) = \tr\left[ 
E^{(1)}_i(x,t_K)L(x,u)B_3(u,v_0,t_K)
-E^{(2)}_i(x,t_K)B_3^\dagger(u,v_0,t_K)L^\dagger(x,u)
\right]\,,
\end{equation}
\begin{equation}
\tilde{\mathcal{C}}_{i}^{\mathrm{T4D2b}}(x,u,\tsep) = \tr\left[ 
F^{(2)}_i(x,t_K)H(x,u)B_4^\dagger(u,v_0,t_K)
-F^{(1)}_i(x,t_K)B_4(u,v_0,t_K)H^\dagger(x,u)
\right]\,,
\end{equation}
for $i=1,2$.
\subsection{Type-5}\label{sect:t5}
\begin{figure}[h!]
\centering
\includegraphics[scale=1]{./Figures/type5d1a}
\includegraphics[scale=1]{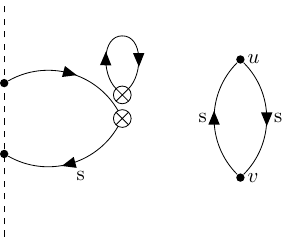}
\caption{T5D1 (left) and T5D2 (right). The point source is put at $v$ if a point-source propagator is used.}
\label{fig:t5diag}
\end{figure}

The Type-5 diagrams, `T5D1' and `T5D2' are depicted in Figure~\ref{fig:t5diag}.
The reduced amplitudes Eq.~\eqref{eq:ampT34} for these diagrams are 
\begin{equation}
\bar{\mathcal{A}}^{\rm T5D1}_{i} = 
\sum_{d\leq\dmax}\sum_{u\in\Lambda} e^{M_K(\tsep+d)}\ F_i(\tsep,d,v_0)K_{\mu\nu}(u-v)\Pi^{\textrm{T5D1}}_{\mu\nu}(u,v)\,,
\end{equation}
\begin{equation}
\bar{\mathcal{A}}^{\rm T5D2}_{i} = 
\sum_{d\leq\dmax}\sum_{u\in\Lambda} e^{M_K(\tsep+d)}\ F_i(\tsep,d,v_0)K_{\mu\nu}(u-v)\Pi^{\textrm{T5D2}}_{\mu\nu}(u,v)\,,
\end{equation}
where 
\begin{equation}
F_i(\tsep, d,v_0) \equiv\sum_{\substack{\vec{x}\in\Lambda_0(x_0), \\ x_0-t_K=\tsep, \\ v_0-x_0=d}} \tr\left[
\mathcal{G}_i(x)\left\{
H(x,t_K)\gamma_5 L(t_K, x)
-\textrm{h.c.}
\right\}
\right]\,,
\end{equation}
\begin{equation}
\mathcal{G}_1 = \tr[\gL_\rho L(x,x)]\gL_\rho\,,
\end{equation}
\begin{equation}
\mathcal{G}_2 = \gL_\rho L(x,x)\gL_\rho\,,
\end{equation}
\begin{equation}
\Pi_{\mu\nu}^{\rm T5D1}(u,v)\equiv \tr\left[
L(u,v)\gamma_\nu L(v,u)\gamma_\mu
\right]\,,
\end{equation}
\begin{equation}
\Pi_{\mu\nu}^{\rm T5D2}(u,v)\equiv \tr\left[
H(u,v)\gamma_\nu H(v,u)\gamma_\mu
\right]\,.
\end{equation}

Due to the quark-disconnected structure, these diagrams are intrinsically noisy. 
For T5D2, we find it sufficient to sample the reduced amplitude at many different reference points $v$ as done for the Type-3 and Type-4 diagrams.
T5D1 is much noisier due to the light-quark propagators.
To improve the statistics, we adapt the Low-Mode Averaging method~\cite{Giusti:2004yp} for our calculation of $\Pi^{\rm T5D1}$.
With the available (Z-M\"obius-approximated) low modes, we decompose a propagator into a low-mode (`l') and a high-mode (`h') part, defined according to
\begin{equation}
L(x,y) = L^{\rm l}(x,y) + L^{\rm h}(x,y)\,,
\quad
L^{\rm l}(x,y)\equiv \sum_{i=1}^{N_{\rm ev}} v_i^{\rm l}(x)\otimes[w_i^{\rm l}]^*(y)\,,
\end{equation}
where the vectors $v^{\rm l }_i$ and $w^{\rm l }_i$ are given in Eq.~\eqref{eq:vlow} and Eq.~\eqref{eq:wlow}.
We then decompose $\Pi^{\rm T5D1}$ into
\begin{equation}\label{eq:pilowhigh}
\Pi^{\rm T5D1}_{\mu\nu}(u,v) = \Pi^{\rm l }_{\mu\nu}(u,v) + \Pi^{\rm h}_{\mu\nu}(u,v)\,,\quad
\Pi^{\rm l}_{\mu\nu}(u,v) = \tr[L^{\rm l}(u,v)\gamma_\nu L^{\rm l}(v,u)\gamma_\mu]\,.
\end{equation}
For the contribution of the low-mode part, we can do an exact volume-average via
\begin{equation}\label{eq:at5d1l}
\begin{split}
\mathcal{A}^{\rm T5D1,l}_{i} \equiv & \,
\frac{1}{V}\sum_{d\leq \dmax}\sum_{u,v\in\Lambda} e^{M_K(\tsep+d)}\ F_i(\tsep,\delta,v_0)K_{\mu\nu}(u-v)\Pi^{\rm l}_{\mu\nu}(u,v)
\\=&\, \frac{1}{V}\sum_{d\leq \dmax}\sum_{u,v\in\Lambda}\sum_{i,j=1}^{N_{\rm ev}} e^{M_K(\tsep+d)}F_i(\tsep,\delta, v_0)K_{\mu\nu}(u-v)G^{(i,j)}_\nu(v) G^{(j,i)}_\mu(u)\,.
\end{split}
\end{equation}
Here $V$ is the spacetime volume and 
\begin{equation}
G^{(i,j)}_{\nu}(u)\equiv
\lla w^{\rm l }_i(u),\gamma_\nu v^{\rm l}_j(u)\rra\,.
\end{equation}
We recognize the sum over $u$ in Eq.~\eqref{eq:at5d1l} is a convolution with the leptonic kernel, which can be done efficiently with a Fast Fourier Transform.
On the other hand, while the contribution of high-mode part $\Pi^{\rm h}$ to the reduced amplitude can be sampled using stochastic methods, we find it more cost-efficient to do it with point-source propagators at many different source positions $v$.

\section{Precondition of the Dirac operator and the Z-M\"obius low modes}\label{sect:a2a}
For part of the calculations reported in Appendix~\ref{sect:comp}, Z-M\"obius approximated low modes are exploited in order to improve sampling efficiency. 
We give some details about the computation of these low modes in this appendix.

Using the notation of Ref.~\cite{RBC:2014ntl}, we solve the preconditioned system with the M\"obius Domain Wall Dirac operator $D_{\rm MDWF}$ 
\begin{equation}
D_{\textrm{prec}}\psi = D_- \eta\,,\quad
D_{\textrm{prec}}\equiv D_-D_{\rm MDWF}\,,
\end{equation}
With the even-odd preconditioning 
\begin{equation}
D_{\rm prec} = \begin{pmatrix}
M_{ee} & M_{eo} \\ 
M_{oe} & M_{oo}
\end{pmatrix}\,,
\end{equation}
we can schur-decompose the inverse of $D_{\textrm{prec}}$ as 
\begin{equation}\label{eq:schur}
D_{\textrm{prec}}^{-1} = 
\begin{pmatrix}
1 & -M_{ee}^{-1}M_{eo} \\
0 & 1 
\end{pmatrix}
\mathcal{M}
\begin{pmatrix}
1 & 0 \\
0 & D_{oo}^\dagger
\end{pmatrix}
\begin{pmatrix}
1 & 0 \\
-M_{oe}M_{ee}^{-1} & 1
\end{pmatrix}\,,
\end{equation}
where 
\begin{equation}
\mathcal{M}\equiv\begin{pmatrix}
M_{ee}^{-1} & 0 \\
0 & \left( D_{oo}^\dagger D_{oo}\right)^{-1}
\end{pmatrix}\,,
\end{equation}
\begin{equation}
D_{oo} \equiv M_{oo} - M_{oe}M_{ee}^{-1}M_{eo}\,.
\end{equation}
We use the \texttt{SchurDiagTwo} scheme defined in the \texttt{Grid} library~\cite{Boyle:2016lf}, which further re-expresses Eq.~\eqref{eq:schur} in terms of 
\begin{equation}
D^\prime_{oo}\equiv D_{oo}M_{oo}^{-1}\,.
\end{equation}
We define low-mode part of the five-dimensional $D_{\rm MDWF}$ by replacing $\mathcal{M}$ in the definition of $D_{\rm prec}$ with the first $N_{\rm ev}$ orthonormal eigenvectors $\{h_i\}$ of $(D_{oo}^\prime)^\dagger D^\prime_{oo}$ with associated eigenvalues $\{\lambda_i\}_{1\leq i\leq N_{\rm ev} }$
\begin{equation}
\mathcal{M}^{\rm l} = 
\begin{pmatrix}
0 & 0 \\ 
0 & M_{oo}^{-1}
\end{pmatrix}
\begin{pmatrix}
0 & 0 \\
0 & \sum_{i=1}^{N_{\rm ev}}\lambda_i^{-1}h_i\otimes h_i^* 
\end{pmatrix}
\begin{pmatrix}
0 & 0 \\ 
0 & [M_{oo}^{-1}]^\dagger
\end{pmatrix}
\,,
\end{equation}
Denote $V^{45}$ and $U^{54}$ the maps projecting to and from the physical four-dimensional space on the fifth-dimension boundary, we define the low-mode part of the physical, four-dimensional propagator as 
\begin{equation}
L^{\rm l}(x,y) = \sum_{i=1}^{N_{\rm ev}}v_i^{\rm l}(x)\otimes [w_i^{\rm l}]^*(y)\,,
\end{equation}
where
\begin{equation}\label{eq:vlow}
v_i^{\rm l} = V^{45}\lambda_i^{-1}
\begin{pmatrix}
-M_{ee}^{-1}M_{eo}M_{oo}^{-1}h_i \\
M_{oo}^{-1}h_i
\end{pmatrix}\,,
\end{equation}
\begin{equation}\label{eq:wlow}
w_i^{\rm l} = [U^{54}]^\dagger D_-^\dagger
\begin{pmatrix}
-[M_{ee}^{-1}]^\dagger M_{oe}^\dagger D^\prime_{oo}h_i\\
D^\prime_{oo}h_i
\end{pmatrix}\,.
\end{equation}
For this work, the eigenvectors involved are not generated for the original $D_{\rm MDWF}$ but for its Z-M\"obius approximated counterpart (see Ref.~\cite{Mcglynn:2015uwh} and references therein) to reduce computational cost as it has a shorter extent in the fifth dimension.

\section{Eliminating the unphysical intermediate state contribution without introducing O($a$) errors}\label{sect:unphys}

In this section we describe how the continuum Eqs.~\eqref{eq:Result_4pt}
and~\eqref{eq:Result_sub} of Section~\ref{sect:form} are implemented in our lattice calculation in a way that avoids introducing O($a$) errors.  Since our gauge and fermion actions introduce finite lattice spacing errors only at O($a^2$) and above, we must take care that any space-time integrals we perform, such as the integral over $v_0$  in Eq.~\eqref{eq:Result_4pt}, do not introduce errors of O($a$).

Consider the case of a single state $|n\rangle$ with energy $E_n < M_K$ and represent Eqs.~\eqref{eq:Result_4pt}
and~\eqref{eq:Result_sub} schematically by the two equations
\begin{eqnarray}
    A^{\rm lat}(\delta_{\max}) &=& \int_{-\delta_{\min}}^{\delta_{\max}} f(v_0) dv_0\,, \label{eq:Result_4pt-sch}\\
    A^{\rm sub}(\delta_{\max}) &=& A_n\frac{e^{-(E_n-M_K)\delta_{\max}}}{M_K-E_n}\,,
    \label{eq:Result_sub-sch}
\end{eqnarray}
where for large $v_0$ the integrand in Eq.~\eqref{eq:Result_4pt-sch} behaves as 
\begin{equation}
f(v_0) \sim A_ne^{-(E_n-M_K)v_0}.
\end{equation}

We can avoid introducing errors of O($a$) if we use the trapezoidal rule when replacing the integral over $v_0$ in Eq.~\eqref{eq:Result_4pt-sch} by a sum over lattice times:
\begin{equation}
    A^{\rm lat}(\delta_{\max}) = +\frac{1}{2}f(\delta_{\max}) + \sum_{v_0=-\delta_{\min}}^{\delta_{\max}-1} f(v_0). \label{eq:Result_4pt-sch-2}
\end{equation}
Here $v_0$, $\delta_{\max}$ and $\delta_{\min}$ are now integers corresponding to discrete values of the time variable $v_0$.  Note, we have not replaced the term $f(-\delta_{\min})$ evaluated at the most negative value of $v_0$ by a term reduced by a factor of 1/2 because we assume that $\delta_{\min}$ is sufficiently large that $f(-\delta_{\min})=0$.  The lattice version of Eq.~\eqref{eq:Result_total} can then be written:
\begin{equation}
    A_{\rm LD2\gamma}(\delta_{\max}) = +\frac{1}{2}f(\delta_{\max}) + \sum_{v_0=-\delta_{\min}}^{\delta_{\max}-1} f(v_0) - A_n\frac{e^{-(E_n-M_K)\delta_{\max}}}{M_K-E_n}. \label{eq:Result_total-sch1}
\end{equation}

Equation~\eqref{eq:Result_total-sch1} provides the physical result for LD2$\gamma$ amplitude by removing the unphysical exponentially-growing contribution from the intermediate state $|n\rangle$ from the O($a^2$)-accurate sum over lattice four-point Green's function results.  For the case that more than a single state $|n\rangle$ has energy below $M_K$, the right-most subtraction term on the right-hand side of Eq.~\eqref{eq:Result_total-sch1} must be replaced by a sum over all such states.   

In Eq.~\eqref{eq:Result_total-sch1} the upper limit $\delta_{\max}$ must be chosen sufficiently large that the unphysical, exponentially falling terms associated with states with energies larger than $M_K$ can be neglected.  In the case of the $\eta$ meson, the noise associated with the disconnected diagrams does not allow such a large choice for $\delta_{\max}$ and the unphysical, exponentially falling contribution of the $\eta$ should also be included in subtraction term on the right-hand side of Eq.~\eqref{eq:Result_total-sch1}.

While Eq.~\eqref{eq:Result_total-sch1} is all that is needed to obtain a physical result, it may also be useful to have an expression for the four-point Green's function as a function of $v_0$ from which these unphysical terms have been subtracted.  However, for $v_0$ near zero, specifying such a subtracted Green's function is somewhat arbitrary since the subtraction term $A^{\rm sub}(\delta)$ is meaningful only when $\delta$ is large compared to the lattice spacing.  For example when $v_0=0$ we encounter the product $J_\nu({\bf v},0)\chw({\bf 0},0)$. If written in this order we can insert a $\pi^0$ intermediate state while if the factors $J$ and $H$ are reversed we cannot.

For the purposes of plotting the subtracted Green's function as a function of $v_0$, we make the choice 
\begin{equation}
f^{\rm sub}(v_0) = f(v_0)- w(v_0) A_n e^{-(E_n-M_K)v_0}
\end{equation}
where the subtraction weight function $w(v_0)$ is given by
\begin{eqnarray}
&\omega(v_0)& = 
\begin{cases} 
e^{-(E_n-M_K)v_0} & \text{if } v_0 \geq 1, \\
\hspace{0.8 cm}\frac{1}{2} & \text{if } v_0 = 0, \\
\hspace{0.8 cm}0 & \text{if } v_0 < 0. \\
\end{cases}  
\end{eqnarray}
While this choice of $w(v_0)$ may be somewhat arbitrary, the subtracted Green's function $f^{\rm sub}(v_0)$ is a reasonable choice and does obey the necessary requirement that when summed over $v_0$ the result agrees with the O($a^2$)-correct result in Eq.~\eqref{eq:Result_total-sch1}:
\begin{equation}
A_{\rm LD2\gamma}(\delta_{\max}) = \sum_{v_0=-\delta_{\min}}^{\delta_{\max}} f^{\rm sub}(v_0)
\label{eq:Result_total-sch2}
\end{equation}
for sufficiently large $\delta_{\max}$, up to terms of O($a^2$).

\section{Systematic error from the three-pion intermediate state}\label{sec:3pi}
In the framework discussed in Section~\ref{sect:form}, we ignored possible unphysical contributions from three-pion state in Eq.~\eqref{eq:Result_sub}.
Depending on the geometry of the lattice, one might need to subtract the unphysical contributions from the excited three-pion states in addition to the ground state as they could be lighter than the kaon.
Treating the pions as non-interacting, unphysical, exponentially growing contributions from $3\pi$-excited states start to appear at a spatial lattice extent $L\gtrsim 11$ fm with physical pion and kaon.
Nonetheless, already as part of our strategy to avoid the complications associated with $\pi\pi(\gamma)$ intermediate state, such a large volume should not be included.
As a consequence, the unphysical contribution from the excited $3\pi$ states should not be an issue in our proposed lattice calculation.
On the other hand, as seen in Section~\ref{sect:ieq1}, subtracting the unphysical $\pi^0$ contribution seems to result in a stable plateau in the cutoff $\dmax$, suggesting that the unphysical contribution from the $3\pi$ ground state is negligible under the achievable precision.
However, just like the case of the $\eta$-intermediate state described in Section~\ref{sect:ieq0}, as the energy of the $3\pi$ state is fairly close to the mass of the kaon, the amplitude might still be far from saturated at the cutoff.
The objective of this section is to estimate the size of the systematic error related to this issue with the $3\pi$ state.
To this end, we write schematically the spectral decomposition of the contribution from the $3\pi$ channel to the $\kl\to\mu^+\mu^-$ amplitude as
\begin{equation}\label{eq:decomp}
\mathcal{A}_{3\pi} \equiv \left[ \int_{3m_\pi}^{E_{\rm cut}} dE_{3\pi}+ \int_{E_{\rm cut}}^\infty dE_{3\pi}\right]\frac{A(E_{3\pi})}{E_{3\pi} - M_K - i\varepsilon}\,,
\end{equation}
In the above, $E_{\rm cut }$ is a cutoff energy which we cannot probe with a restricted spatial lattice extent.
The $i\varepsilon$ prescription is used here to indicate that we are focusing on the $\kl\to 3\pi\to \gamma^*\gamma^*$ process where special care is required:
analogous to the case of the $\pi^0$, the unphysical $3\pi$ states appear under Wick rotation with the time-ordering where the Weak Hamiltonian is at an earlier time than the two EM currents [Eq.~\eqref{eq:Result_sub}].
A typical gauge ensemble with physical pion mass which we would include in our continuum trajectory would have $m_\pi L\approx 4$.
On such an ensemble, the first excited (non-interacting) three-pion state produced by the initial kaon, comprised of two back-to-back pions and a pion at rest, has an energy around $ 660$ MeV.
This provides a natural energy cutoff $E_{\rm cut}$ in Eq.~\eqref{eq:decomp}.
If computed in Euclidean space, the second integral in the square brackets in Eq.~\eqref{eq:decomp} can be saturated to about $60-70$\% if the data quality allows to reach an Euclidean time of $1.2-1.6$ fm.
Assuming that the latter is feasible, the main source of systematic error with the proposed lattice calculation should come from the first integral in the square brackets of Eq.~\eqref{eq:decomp}.\footnote{In the case of our 24ID study, this range corresponds to $\dmax\in[6,8]$.
}
To proceed further, we make the hypothesis that there is no steep enhancement for $A(E_{3 \pi})$ in the interval $E_{3\pi}\in [3m_\pi, E_{\rm cut}]=[420, 660]$ MeV and that it can be approximated by a constant $A(E_{3\pi})\approx A(M_K)$.
Then,
\begin{equation}\label{eq:scal}
\mathcal{S}\equiv\int_{3m_\pi}^{E_{\rm cut}} dE_{3\pi}\ \frac{A(M_K)}{E_{3\pi}-M_K-i\varepsilon} = A(M_K)\left[
\ln\left(\frac{E_{\rm cut} - M_K}{M_K- 3m_\pi}\right)+i\pi
\right]
\,,
\end{equation}
should give a realistic estimate for the systematic error related to the $3\pi$-intermediate state in our lattice calculation. 

Through out this appendix, we will use the convention Eq.~\eqref{eq:A-def} for the amplitudes.
We will use the shorthand notation $|3\pi\rangle\equiv |\pi^{a_1}(\vec{q_1})\pi^{a_2}(\vec{q_2})\pi^{a_3}(\vec{q_3})\rangle$, where the superscripts indicate the charges, which are kept generic for the moment.
In Minkowski space, this contribution reads
\begin{equation}\label{eq:amp}
\begin{split}
\mathcal{A}_{3\pi} = & \int d\Pi_{3\pi}\int d^4 r L_{\mu\nu}(r)e^{iP\cdot r/2}
\frac{\delta^2}{e^2\delta A_\mu(r) \delta A_\nu(0)}\Big[\lla 0 | e^{iS_{\rm int}}| 3\pi\rra\Big]
\\
&\times
\frac{
\lla 3\pi |\chw(0)|\kl\rra}{E_{3\pi}-M_K-i\varepsilon}
(2\pi)^3\delta^3\left(\sum_{i=1}^3\vec{q}_i\right)
\\ =\ &
-i\ \frac{m_\mu e^2}{4096\ \pi^{10} M_K}\int \frac{dE_1 dE_2 d^2\Omega_1 d^2\Omega_2\ |\vec{q}_1||\vec{q}_2|}{E_3(E_1+E_2+E_3-M_K-i\varepsilon)} \lla 3\pi |\chw(0)|\kl\rra
\\ & \times 
\int d^4 p \frac{\epsilon^{\mu\nu\alpha\beta}p_\alpha P_\beta  \widetilde{J_\mu J_\nu}(p)}{
\left[ \left( P -p \right)^2 + i\varepsilon \right] \left( p^2 + i\varepsilon \right)\left[
\left(p - k^+\right)^2 - m_\mu^2 + i\varepsilon
\right]}\,,
\end{split}
\end{equation}
where $k^+$ is the momentum of the outgoing $\mu^+$, $P=(M_K,\vec{0})$ is the four-momentum of the kaon in the rest frame, $E_i=\sqrt{m_{\pi^{a_i}}^2+\vec{q}_i^{\ 2}}$ and $\Omega_i$ are the energy and the solid angle of the spatial momentum of the $i$-th pion,
\begin{equation}
d\Pi_{3\pi} = \frac{1}{(2\pi)^9}\frac{d^3\vec{q}_1}{2E_1}
\frac{d^3\vec{q}_2}{2E_2}\frac{d^3\vec{q}_3}{2E_3} |3\pi\rangle\langle 3\pi|
\end{equation} 
is the Lorentz-invariant phase-space measure of the three pions and
\begin{equation}\label{eq:jj3pi}
\widetilde{J_\mu J_\nu}(p)\equiv\int d^4 r \ e^{ipr}
\frac{\delta^2}{\delta A_\mu(r) \delta A_\nu(0)}\Big[\lla 0 | e^{iS_{\rm int}}| 3\pi\rra\Big]\,,
\end{equation}
with $S_{\rm int}$ the action from the (effective) interaction Lagrangian and $A_\mu$ a photon field.

The experimental decay amplitude $A_{+-0}^{L}\equiv\lla \pi^+\pi^-\pi^0|\chw(0)|\kl\rra$ admits the following parametrization~\cite{Bijnens:2002vr}\footnote{Here the notation is due to Ref.~\cite{Weinberg:1960zza}, where the third pion is the ``odd'' pion, ie. the first two of them satisfy Bose-symmetry constraint.}
\begin{equation}\label{eq:apm0}
A_{+-0}^{L} = (\alpha_1+\alpha_3)
-(\beta_1+\beta_3)\bar{y} + (\zeta_1 - 2\zeta_3)\left(\bar{y}^2+\frac{1}{3}\bar{x}^2\right)
+(\xi_1 - 2\xi_3)\left(\bar{y}^2 - \frac{1}{3}\bar{x}^2\right)\,,
\end{equation}
where
\begin{equation}
\bar{y}\equiv \frac{s_{3}-s_0}{m_{\pi^+}^2}\,,\quad
\bar{x}\equiv \frac{s_2 - s_1}{m_{\pi^+}^2}\,,\quad
s_{i} \equiv (P_K-p_{i})^2\,,\quad s_0 \equiv \frac{1}{3}\left( M_K^2 + m_{\pi^0}^2 + m_{\pi^+}^2 + m_{\pi^-}^2\right)\,.
\end{equation}
Note that the coefficients $\alpha_1,\ldots , \xi_3$ in Eq.~\eqref{eq:apm0} fitted in Ref.~\cite{Bijnens:2002vr} result in a $\kl\to\pi^+\pi^-\pi^0$ decay rate of $1.57\times  10^{-18}$ GeV -- consistent with the experimental value~\cite{ParticleDataGroup:2022pth} -- if the physical pion masses are used.
In the remainder of our discussion, we will assume isospin symmetry for simplicity. 
In that limit, Eq.~\eqref{eq:apm0} with the same fitted coefficients gives a decay rate of $1.30 \times 10^{-18}$ GeV, suggesting a correction for the strong isospin-breaking effects at the ten percent level on the decay amplitude. 

\begin{figure}[h!]
\centering
\includegraphics[scale=0.45]{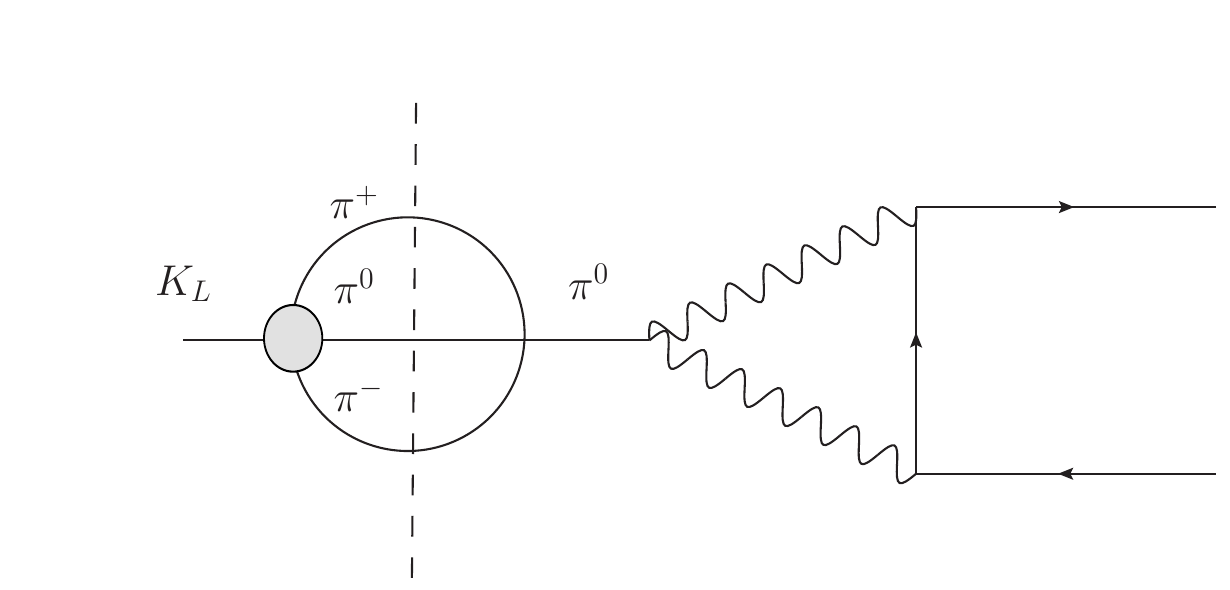}
\caption{Three-pion contribution from the four-pion and WZ $\pi^0\gamma\gamma$ vertices.}\label{fig:feyn4pi}
\end{figure}

\begin{figure}[h!]
\centering
\begin{minipage}{0.45\textwidth}
\centering
\includegraphics[scale=0.35]{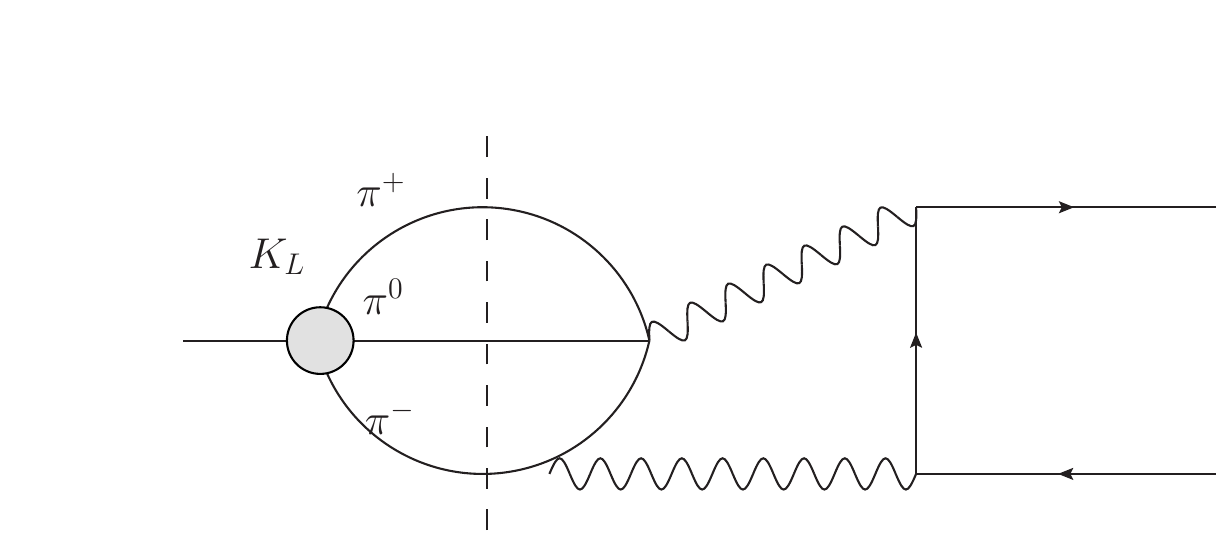}
\end{minipage}
\begin{minipage}{0.45\textwidth}
\centering
\includegraphics[scale=0.35]{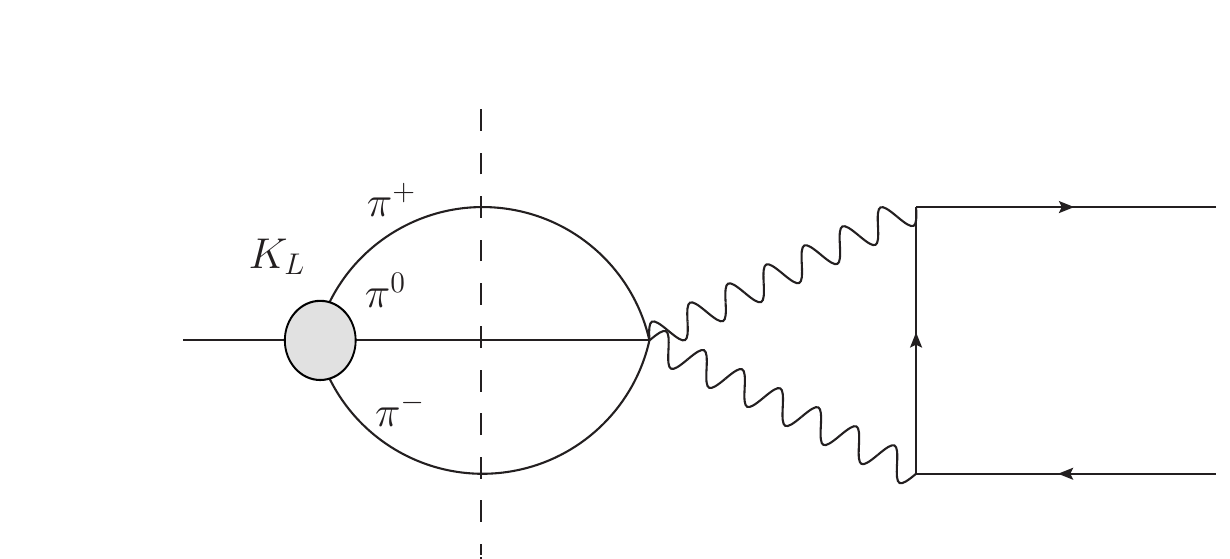}
\end{minipage}
\caption{Three-pion contribution from the WZ $\pi^0\pi^+\pi^-\gamma$ and $\pi^0\pi^+\pi^-\gamma\gamma$ vertices.}\label{fig:feynk3pi}
\end{figure}

As there is no direct experimental data for the quantity Eq.~\eqref{eq:jj3pi}, we estimate it in Chiral Perturbation Theory.
Due to charge and parity symmetries, one needs to include the Wess-Zimino (WZ) term~\cite{Wess:1971yu, Witten:1983tw} for the coupling of an odd number of pions to two photons. 
For simplicity, we will only focus on the pion-pole contribution from the leading O($p^4$) WZ term and the O($p^2$) chiral Lagrangian as displayed in Figure~\ref{fig:feyn4pi}. 
The logarithmic divergence appearing in the $2\gamma\mu$ triangle diagram in Figure~\ref{fig:feyn4pi} can be renormalized by adding a $\pi^0\mu^+\mu^-$ counterterm in the standard way~\cite{Savage:1992ac}.
Following the conventions of Ref.~\cite{Scherer:2012xha}, where the pion decay constant takes the experimental value of $f_\pi\approx 93.3$ MeV, we get
\begin{equation}
\begin{split}
\mathcal{A}_{3\pi}^{\pi^0{\rm-pole}} 
=& i\frac{m_\mu M_K e^4}{6144\ \pi^8 f_\pi^3}\int \frac{dE_1 dE_2 dE_3\ \lla \pi^+\pi^-\pi^0|\chw|\kl\rra}{E_1+E_2+E_3-M_K-i\varepsilon}
\frac{\left(
2M_K^2 - 6M_K E_3 + m_\pi^2
\right)}{M_K^2-m_\pi^2}
A_\mu(M_K^2)
\,,
\end{split}
\end{equation}
where, for a lepton $l$,
\begin{equation}
A_l(q^2) \equiv \frac{2i}{\pi^2 q^2}\int d^4k
\frac{q^2k^2 - (q\cdot k)^2}{k^2 (q-k)^2[(p-k)^2-m_l^2]}\tilde{F}(k^2,(q-k)^2)\,,
\end{equation}
with $\tilde{F}$ the $\pi^0\to\gamma^*\gamma^*$ transition form factor normalized to unity when both photons are on-shell.
Assuming lepton universality, we can fix the scale-dependent low-energy constant $\chi(\mu)$ using the $\pi^0\to e^+ e^-$ process.
Within Chiral Perturbation theory,
\begin{equation}
\textrm{Re}\mathcal{A}_l(q^2)\mid_{\rm ChPT} = 
\frac{\textrm{Li}_2[-y_l(q^2)] + \frac{1}{4}\ln^2[y_l(q^2)]+\frac{\pi^2}{12}}{\sigma_l(q^2)} + 3\ln\left(\frac{m_l}{\mu}\right) - \frac{5}{2} + \chi(\mu)\,,
\end{equation}
where
\begin{equation}
y_l(q^2)\equiv \frac{1-\sigma_l(q^2)}{1+\sigma_l(q^2)}\,,\quad
\sigma_l(q^2)= \sqrt{1-\frac{4m_l^2}{q^2}}
\end{equation}
From Ref.~\cite{Hoferichter:2021lct},
\begin{equation}
\chi(\mu = 0.77\textrm{GeV}) = 2.69(10)\,.
\end{equation}
Altogether, we get
\begin{equation}\label{eq:A4pi}
\mathcal{S} = (1.84 + i\ 2.32)\times 10^{-17} \textrm{ GeV}\,, 
\end{equation}
leading to a ratio to the experimentally-extracted dispersive $\kl\to\mu^+\mu^-$ decay amplitude of $|\mathcal{S}/\textrm{Re}\mathcal{A}^{\rm exp}_{\kl\mu\mu}|  = 1.3\times 10^{-4}$.
As stated earlier, this estimate is sensitive to the strong isospin-breaking effects as they affects the $\kl\to \pi^+\pi^-\pi^0$, but we expect the correction to be minor.
Unless the assumption of $A(E_{3\pi})\approx A(M_K)$ which motivates Eq.~\eqref{eq:scal} suffers from further enhancements of O(100) in the range $E_{3\pi}\in[3m_{\pi}, E_{\rm cut}]$, the missing $3\pi$-contributions from our proposed lattice framework should be less than a percent.
Similar calculations can be performed for the case with a $\pi^0\pi^0\pi^0$ intermediate state, which lead to an estimate at the same order of magnitude.
Finally, we should note that the diagrams in Figure~\ref{fig:feynk3pi}, which are both needed to make a gauge-invariant amplitude, also enter at the same order in Chiral Perturbation Theory. 
We found the total contribution from those diagrams to be about an order of magnitude smaller than the pion-pole diagram in Figure~\ref{fig:feyn4pi} and decided not to report the entire calculation in the present paper to avoid clutter.

\bibliography{references}

\begin{thebibliography}{56}%
\makeatletter
\providecommand \@ifxundefined [1]{%
 \@ifx{#1\undefined}
}%
\providecommand \@ifnum [1]{%
 \ifnum #1\expandafter \@firstoftwo
 \else \expandafter \@secondoftwo
 \fi
}%
\providecommand \@ifx [1]{%
 \ifx #1\expandafter \@firstoftwo
 \else \expandafter \@secondoftwo
 \fi
}%
\providecommand \natexlab [1]{#1}%
\providecommand \enquote  [1]{``#1''}%
\providecommand \bibnamefont  [1]{#1}%
\providecommand \bibfnamefont [1]{#1}%
\providecommand \citenamefont [1]{#1}%
\providecommand \href@noop [0]{\@secondoftwo}%
\providecommand \href [0]{\begingroup \@sanitize@url \@href}%
\providecommand \@href[1]{\@@startlink{#1}\@@href}%
\providecommand \@@href[1]{\endgroup#1\@@endlink}%
\providecommand \@sanitize@url [0]{\catcode `\\12\catcode `\$12\catcode
  `\&12\catcode `\#12\catcode `\^12\catcode `\_12\catcode `\%12\relax}%
\providecommand \@@startlink[1]{}%
\providecommand \@@endlink[0]{}%
\providecommand \url  [0]{\begingroup\@sanitize@url \@url }%
\providecommand \@url [1]{\endgroup\@href {#1}{\urlprefix }}%
\providecommand \urlprefix  [0]{URL }%
\providecommand \Eprint [0]{\href }%
\providecommand \doibase [0]{https://doi.org/}%
\providecommand \selectlanguage [0]{\@gobble}%
\providecommand \bibinfo  [0]{\@secondoftwo}%
\providecommand \bibfield  [0]{\@secondoftwo}%
\providecommand \translation [1]{[#1]}%
\providecommand \BibitemOpen [0]{}%
\providecommand \bibitemStop [0]{}%
\providecommand \bibitemNoStop [0]{.\EOS\space}%
\providecommand \EOS [0]{\spacefactor3000\relax}%
\providecommand \BibitemShut  [1]{\csname bibitem#1\endcsname}%
\let\auto@bib@innerbib\@empty
\bibitem [{\citenamefont {Ambrose}\ \emph {et~al.}(2000)\citenamefont {Ambrose}
  \emph {et~al.}}]{E871:2000wvm}%
  \BibitemOpen
  \bibfield  {author} {\bibinfo {author} {\bibfnamefont {D.}~\bibnamefont
  {Ambrose}} \emph {et~al.} (\bibinfo {collaboration} {E871}),\ }\bibfield
  {title} {\bibinfo {title} {{Improved branching ratio measurement for the
  decay K0(L) --\ensuremath{>} mu+ mu-}},\ }\href
  {https://doi.org/10.1103/PhysRevLett.84.1389} {\bibfield  {journal} {\bibinfo
   {journal} {Phys. Rev. Lett.}\ }\textbf {\bibinfo {volume} {84}},\ \bibinfo
  {pages} {1389} (\bibinfo {year} {2000})}\BibitemShut {NoStop}%
\bibitem [{\citenamefont {Workman}\ \emph {et~al.}(2022)\citenamefont {Workman}
  \emph {et~al.}}]{ParticleDataGroup:2022pth}%
  \BibitemOpen
  \bibfield  {author} {\bibinfo {author} {\bibfnamefont {R.~L.}\ \bibnamefont
  {Workman}} \emph {et~al.} (\bibinfo {collaboration} {Particle Data Group}),\
  }\bibfield  {title} {\bibinfo {title} {{Review of Particle Physics}},\ }\href
  {https://doi.org/10.1093/ptep/ptac097} {\bibfield  {journal} {\bibinfo
  {journal} {PTEP}\ }\textbf {\bibinfo {volume} {2022}},\ \bibinfo {pages}
  {083C01} (\bibinfo {year} {2022})}\BibitemShut {NoStop}%
\bibitem [{\citenamefont {Gorbahn}\ and\ \citenamefont
  {Haisch}(2006)}]{Gorbahn:2006bm}%
  \BibitemOpen
  \bibfield  {author} {\bibinfo {author} {\bibfnamefont {M.}~\bibnamefont
  {Gorbahn}}\ and\ \bibinfo {author} {\bibfnamefont {U.}~\bibnamefont
  {Haisch}},\ }\bibfield  {title} {\bibinfo {title} {{Charm Quark Contribution
  to K(L) ---\ensuremath{>} mu+ mu- at Next-to-Next-to-Leading}},\ }\href
  {https://doi.org/10.1103/PhysRevLett.97.122002} {\bibfield  {journal}
  {\bibinfo  {journal} {Phys. Rev. Lett.}\ }\textbf {\bibinfo {volume} {97}},\
  \bibinfo {pages} {122002} (\bibinfo {year} {2006})},\ \Eprint
  {https://arxiv.org/abs/hep-ph/0605203} {arXiv:hep-ph/0605203} \BibitemShut
  {NoStop}%
\bibitem [{\citenamefont {Martin}\ \emph {et~al.}(1970)\citenamefont {Martin},
  \citenamefont {De~Rafael},\ and\ \citenamefont {Smith}}]{Martin:1970ai}%
  \BibitemOpen
  \bibfield  {author} {\bibinfo {author} {\bibfnamefont {B.~R.}\ \bibnamefont
  {Martin}}, \bibinfo {author} {\bibfnamefont {E.}~\bibnamefont {De~Rafael}},\
  and\ \bibinfo {author} {\bibfnamefont {J.}~\bibnamefont {Smith}},\ }\bibfield
   {title} {\bibinfo {title} {{Neutral kaon decays into lepton pairs}},\ }\href
  {https://doi.org/10.1103/PhysRevD.2.179} {\bibfield  {journal} {\bibinfo
  {journal} {Phys. Rev. D}\ }\textbf {\bibinfo {volume} {2}},\ \bibinfo {pages}
  {179} (\bibinfo {year} {1970})}\BibitemShut {NoStop}%
\bibitem [{\citenamefont {D'Ambrosio}\ \emph {et~al.}(1998)\citenamefont
  {D'Ambrosio}, \citenamefont {Isidori},\ and\ \citenamefont
  {Portoles}}]{DAmbrosio:1997eof}%
  \BibitemOpen
  \bibfield  {author} {\bibinfo {author} {\bibfnamefont {G.}~\bibnamefont
  {D'Ambrosio}}, \bibinfo {author} {\bibfnamefont {G.}~\bibnamefont
  {Isidori}},\ and\ \bibinfo {author} {\bibfnamefont {J.}~\bibnamefont
  {Portoles}},\ }\bibfield  {title} {\bibinfo {title} {{Can we extract short
  distance information from B(K(L) ---\ensuremath{>} mu+ mu-)?}},\ }\href
  {https://doi.org/10.1016/S0370-2693(98)00146-4} {\bibfield  {journal}
  {\bibinfo  {journal} {Phys. Lett. B}\ }\textbf {\bibinfo {volume} {423}},\
  \bibinfo {pages} {385} (\bibinfo {year} {1998})},\ \Eprint
  {https://arxiv.org/abs/hep-ph/9708326} {arXiv:hep-ph/9708326} \BibitemShut
  {NoStop}%
\bibitem [{\citenamefont {Knecht}\ \emph {et~al.}(1999)\citenamefont {Knecht},
  \citenamefont {Peris}, \citenamefont {Perrottet},\ and\ \citenamefont
  {de~Rafael}}]{Knecht:1999gb}%
  \BibitemOpen
  \bibfield  {author} {\bibinfo {author} {\bibfnamefont {M.}~\bibnamefont
  {Knecht}}, \bibinfo {author} {\bibfnamefont {S.}~\bibnamefont {Peris}},
  \bibinfo {author} {\bibfnamefont {M.}~\bibnamefont {Perrottet}},\ and\
  \bibinfo {author} {\bibfnamefont {E.}~\bibnamefont {de~Rafael}},\ }\bibfield
  {title} {\bibinfo {title} {{Decay of pseudoscalars into lepton pairs and
  large N(c) QCD}},\ }\href {https://doi.org/10.1103/PhysRevLett.83.5230}
  {\bibfield  {journal} {\bibinfo  {journal} {Phys. Rev. Lett.}\ }\textbf
  {\bibinfo {volume} {83}},\ \bibinfo {pages} {5230} (\bibinfo {year}
  {1999})},\ \Eprint {https://arxiv.org/abs/hep-ph/9908283}
  {arXiv:hep-ph/9908283} \BibitemShut {NoStop}%
\bibitem [{\citenamefont {Isidori}\ and\ \citenamefont
  {Unterdorfer}(2004)}]{Isidori:2003ts}%
  \BibitemOpen
  \bibfield  {author} {\bibinfo {author} {\bibfnamefont {G.}~\bibnamefont
  {Isidori}}\ and\ \bibinfo {author} {\bibfnamefont {R.}~\bibnamefont
  {Unterdorfer}},\ }\bibfield  {title} {\bibinfo {title} {{On the short
  distance constraints from K(L,S) ---\ensuremath{>} mu+ mu-}},\ }\href
  {https://doi.org/10.1088/1126-6708/2004/01/009} {\bibfield  {journal}
  {\bibinfo  {journal} {JHEP}\ }\textbf {\bibinfo {volume} {01}},\ \bibinfo
  {pages} {009}},\ \Eprint {https://arxiv.org/abs/hep-ph/0311084}
  {arXiv:hep-ph/0311084} \BibitemShut {NoStop}%
\bibitem [{\citenamefont {Hoferichter}\ \emph {et~al.}(2024)\citenamefont
  {Hoferichter}, \citenamefont {Hoid},\ and\ \citenamefont
  {de~Elvira}}]{Hoferichter:2023wiy}%
  \BibitemOpen
  \bibfield  {author} {\bibinfo {author} {\bibfnamefont {M.}~\bibnamefont
  {Hoferichter}}, \bibinfo {author} {\bibfnamefont {B.-L.}\ \bibnamefont
  {Hoid}},\ and\ \bibinfo {author} {\bibfnamefont {J.~R.}\ \bibnamefont
  {de~Elvira}},\ }\bibfield  {title} {\bibinfo {title} {{Improved
  Standard-Model prediction for K$_{L}$ \textrightarrow{}
  \ensuremath{\ell}$^{+}$\ensuremath{\ell}$^{-}$}},\ }\href
  {https://doi.org/10.1007/JHEP04(2024)071} {\bibfield  {journal} {\bibinfo
  {journal} {JHEP}\ }\textbf {\bibinfo {volume} {04}},\ \bibinfo {pages}
  {071}},\ \Eprint {https://arxiv.org/abs/2310.17689} {arXiv:2310.17689
  [hep-ph]} \BibitemShut {NoStop}%
\bibitem [{\citenamefont {Blum}\ \emph
  {et~al.}(2016{\natexlab{a}})\citenamefont {Blum}, \citenamefont {Christ},
  \citenamefont {Hayakawa}, \citenamefont {Izubuchi}, \citenamefont {Jin},\
  and\ \citenamefont {Lehner}}]{Blum:2015gfa}%
  \BibitemOpen
  \bibfield  {author} {\bibinfo {author} {\bibfnamefont {T.}~\bibnamefont
  {Blum}}, \bibinfo {author} {\bibfnamefont {N.}~\bibnamefont {Christ}},
  \bibinfo {author} {\bibfnamefont {M.}~\bibnamefont {Hayakawa}}, \bibinfo
  {author} {\bibfnamefont {T.}~\bibnamefont {Izubuchi}}, \bibinfo {author}
  {\bibfnamefont {L.}~\bibnamefont {Jin}},\ and\ \bibinfo {author}
  {\bibfnamefont {C.}~\bibnamefont {Lehner}},\ }\bibfield  {title} {\bibinfo
  {title} {{Lattice Calculation of Hadronic Light-by-Light Contribution to the
  Muon Anomalous Magnetic Moment}},\ }\href
  {https://doi.org/10.1103/PhysRevD.93.014503} {\bibfield  {journal} {\bibinfo
  {journal} {Phys. Rev. D}\ }\textbf {\bibinfo {volume} {93}},\ \bibinfo
  {pages} {014503} (\bibinfo {year} {2016}{\natexlab{a}})},\ \Eprint
  {https://arxiv.org/abs/1510.07100} {arXiv:1510.07100 [hep-lat]} \BibitemShut
  {NoStop}%
\bibitem [{\citenamefont {Blum}\ \emph {et~al.}(2025)\citenamefont {Blum},
  \citenamefont {Christ}, \citenamefont {Hayakawa}, \citenamefont {Izubuchi},
  \citenamefont {Jin}, \citenamefont {Jung}, \citenamefont {Lehner},\ and\
  \citenamefont {Tu}}]{Blum:2023vlm}%
  \BibitemOpen
  \bibfield  {author} {\bibinfo {author} {\bibfnamefont {T.}~\bibnamefont
  {Blum}}, \bibinfo {author} {\bibfnamefont {N.}~\bibnamefont {Christ}},
  \bibinfo {author} {\bibfnamefont {M.}~\bibnamefont {Hayakawa}}, \bibinfo
  {author} {\bibfnamefont {T.}~\bibnamefont {Izubuchi}}, \bibinfo {author}
  {\bibfnamefont {L.}~\bibnamefont {Jin}}, \bibinfo {author} {\bibfnamefont
  {C.}~\bibnamefont {Jung}}, \bibinfo {author} {\bibfnamefont {C.}~\bibnamefont
  {Lehner}},\ and\ \bibinfo {author} {\bibfnamefont {C.}~\bibnamefont {Tu}}
  (\bibinfo {collaboration} {RBC, UKQCD}),\ }\bibfield  {title} {\bibinfo
  {title} {{Hadronic light-by-light contribution to the muon anomaly from
  lattice QCD with infinite volume QED at physical pion mass}},\ }\href
  {https://doi.org/10.1103/PhysRevD.111.014501} {\bibfield  {journal} {\bibinfo
   {journal} {Phys. Rev. D}\ }\textbf {\bibinfo {volume} {111}},\ \bibinfo
  {pages} {014501} (\bibinfo {year} {2025})},\ \Eprint
  {https://arxiv.org/abs/2304.04423} {arXiv:2304.04423 [hep-lat]} \BibitemShut
  {NoStop}%
\bibitem [{\citenamefont {Asmussen}\ \emph {et~al.}(2023)\citenamefont
  {Asmussen}, \citenamefont {Chao}, \citenamefont {G{\'e}rardin}, \citenamefont
  {Green}, \citenamefont {Hudspith}, \citenamefont {Meyer},\ and\ \citenamefont
  {Nyffeler}}]{Asmussen:2022oql}%
  \BibitemOpen
  \bibfield  {author} {\bibinfo {author} {\bibfnamefont {N.}~\bibnamefont
  {Asmussen}}, \bibinfo {author} {\bibfnamefont {E.-H.}\ \bibnamefont {Chao}},
  \bibinfo {author} {\bibfnamefont {A.}~\bibnamefont {G{\'e}rardin}}, \bibinfo
  {author} {\bibfnamefont {J.~R.}\ \bibnamefont {Green}}, \bibinfo {author}
  {\bibfnamefont {R.~J.}\ \bibnamefont {Hudspith}}, \bibinfo {author}
  {\bibfnamefont {H.~B.}\ \bibnamefont {Meyer}},\ and\ \bibinfo {author}
  {\bibfnamefont {A.}~\bibnamefont {Nyffeler}},\ }\bibfield  {title} {\bibinfo
  {title} {{Hadronic light-by-light scattering contribution to the muon g
  {\ensuremath{-}} 2 from lattice QCD: semi-analytical calculation of the QED
  kernel}},\ }\href {https://doi.org/10.1007/JHEP04(2023)040} {\bibfield
  {journal} {\bibinfo  {journal} {JHEP}\ }\textbf {\bibinfo {volume} {04}},\
  \bibinfo {pages} {040}},\ \Eprint {https://arxiv.org/abs/2210.12263}
  {arXiv:2210.12263 [hep-lat]} \BibitemShut {NoStop}%
\bibitem [{\citenamefont {Chao}\ \emph {et~al.}(2021)\citenamefont {Chao},
  \citenamefont {Hudspith}, \citenamefont {G\'erardin}, \citenamefont {Green},
  \citenamefont {Meyer},\ and\ \citenamefont {Ottnad}}]{Chao:2021tvp}%
  \BibitemOpen
  \bibfield  {author} {\bibinfo {author} {\bibfnamefont {E.-H.}\ \bibnamefont
  {Chao}}, \bibinfo {author} {\bibfnamefont {R.~J.}\ \bibnamefont {Hudspith}},
  \bibinfo {author} {\bibfnamefont {A.}~\bibnamefont {G\'erardin}}, \bibinfo
  {author} {\bibfnamefont {J.~R.}\ \bibnamefont {Green}}, \bibinfo {author}
  {\bibfnamefont {H.~B.}\ \bibnamefont {Meyer}},\ and\ \bibinfo {author}
  {\bibfnamefont {K.}~\bibnamefont {Ottnad}},\ }\bibfield  {title} {\bibinfo
  {title} {{Hadronic light-by-light contribution to $(g-2)_\mu $ from lattice
  QCD: a complete calculation}},\ }\href
  {https://doi.org/10.1140/epjc/s10052-021-09455-4} {\bibfield  {journal}
  {\bibinfo  {journal} {Eur. Phys. J. C}\ }\textbf {\bibinfo {volume} {81}},\
  \bibinfo {pages} {651} (\bibinfo {year} {2021})},\ \Eprint
  {https://arxiv.org/abs/2104.02632} {arXiv:2104.02632 [hep-lat]} \BibitemShut
  {NoStop}%
\bibitem [{\citenamefont {Chao}\ \emph {et~al.}(2022)\citenamefont {Chao},
  \citenamefont {Hudspith}, \citenamefont {G{\'e}rardin}, \citenamefont
  {Green},\ and\ \citenamefont {Meyer}}]{Chao:2022xzg}%
  \BibitemOpen
  \bibfield  {author} {\bibinfo {author} {\bibfnamefont {E.-H.}\ \bibnamefont
  {Chao}}, \bibinfo {author} {\bibfnamefont {R.~J.}\ \bibnamefont {Hudspith}},
  \bibinfo {author} {\bibfnamefont {A.}~\bibnamefont {G{\'e}rardin}}, \bibinfo
  {author} {\bibfnamefont {J.~R.}\ \bibnamefont {Green}},\ and\ \bibinfo
  {author} {\bibfnamefont {H.~B.}\ \bibnamefont {Meyer}},\ }\bibfield  {title}
  {\bibinfo {title} {{The charm-quark contribution to light-by-light scattering
  in the muon $(g-2)$ from lattice QCD}},\ }\href
  {https://doi.org/10.1140/epjc/s10052-022-10589-2} {\bibfield  {journal}
  {\bibinfo  {journal} {Eur. Phys. J. C}\ }\textbf {\bibinfo {volume} {82}},\
  \bibinfo {pages} {664} (\bibinfo {year} {2022})},\ \Eprint
  {https://arxiv.org/abs/2204.08844} {arXiv:2204.08844 [hep-lat]} \BibitemShut
  {NoStop}%
\bibitem [{\citenamefont {Christ}\ \emph
  {et~al.}(2020{\natexlab{a}})\citenamefont {Christ}, \citenamefont {Feng},
  \citenamefont {Jin}, \citenamefont {Tu},\ and\ \citenamefont
  {Zhao}}]{Christ:2020dae}%
  \BibitemOpen
  \bibfield  {author} {\bibinfo {author} {\bibfnamefont {N.~H.}\ \bibnamefont
  {Christ}}, \bibinfo {author} {\bibfnamefont {X.}~\bibnamefont {Feng}},
  \bibinfo {author} {\bibfnamefont {L.}~\bibnamefont {Jin}}, \bibinfo {author}
  {\bibfnamefont {C.}~\bibnamefont {Tu}},\ and\ \bibinfo {author}
  {\bibfnamefont {Y.}~\bibnamefont {Zhao}},\ }\bibfield  {title} {\bibinfo
  {title} {{Calculating the Two-photon Contribution to $\pi^0 \rightarrow e^+
  e^-$ Decay Amplitude}},\ }\href {https://doi.org/10.22323/1.363.0097}
  {\bibfield  {journal} {\bibinfo  {journal} {PoS}\ }\textbf {\bibinfo {volume}
  {LATTICE2019}},\ \bibinfo {pages} {097} (\bibinfo {year}
  {2020}{\natexlab{a}})},\ \Eprint {https://arxiv.org/abs/2001.05642}
  {arXiv:2001.05642 [hep-lat]} \BibitemShut {NoStop}%
\bibitem [{\citenamefont {Christ}\ \emph
  {et~al.}(2020{\natexlab{b}})\citenamefont {Christ}, \citenamefont {Feng},
  \citenamefont {Jin}, \citenamefont {Tu},\ and\ \citenamefont
  {Zhao}}]{Christ:2020bzb}%
  \BibitemOpen
  \bibfield  {author} {\bibinfo {author} {\bibfnamefont {N.~H.}\ \bibnamefont
  {Christ}}, \bibinfo {author} {\bibfnamefont {X.}~\bibnamefont {Feng}},
  \bibinfo {author} {\bibfnamefont {L.}~\bibnamefont {Jin}}, \bibinfo {author}
  {\bibfnamefont {C.}~\bibnamefont {Tu}},\ and\ \bibinfo {author}
  {\bibfnamefont {Y.}~\bibnamefont {Zhao}},\ }\bibfield  {title} {\bibinfo
  {title} {{Lattice QCD calculation of the two-photon contributions to $K_L \to
  \mu^+ \mu^-$ and $\pi^0 \to e^+ e^-$ decays}},\ }\href
  {https://doi.org/10.22323/1.363.0128} {\bibfield  {journal} {\bibinfo
  {journal} {PoS}\ }\textbf {\bibinfo {volume} {LATTICE2019}},\ \bibinfo
  {pages} {128} (\bibinfo {year} {2020}{\natexlab{b}})}\BibitemShut {NoStop}%
\bibitem [{\citenamefont {Christ}\ \emph {et~al.}(2023)\citenamefont {Christ},
  \citenamefont {Feng}, \citenamefont {Jin}, \citenamefont {Tu},\ and\
  \citenamefont {Zhao}}]{Christ:2022rho}%
  \BibitemOpen
  \bibfield  {author} {\bibinfo {author} {\bibfnamefont {N.}~\bibnamefont
  {Christ}}, \bibinfo {author} {\bibfnamefont {X.}~\bibnamefont {Feng}},
  \bibinfo {author} {\bibfnamefont {L.}~\bibnamefont {Jin}}, \bibinfo {author}
  {\bibfnamefont {C.}~\bibnamefont {Tu}},\ and\ \bibinfo {author}
  {\bibfnamefont {Y.}~\bibnamefont {Zhao}},\ }\bibfield  {title} {\bibinfo
  {title} {{Lattice QCD Calculation of \ensuremath{\pi}0\textrightarrow{}e+e-
  Decay}},\ }\href {https://doi.org/10.1103/PhysRevLett.130.191901} {\bibfield
  {journal} {\bibinfo  {journal} {Phys. Rev. Lett.}\ }\textbf {\bibinfo
  {volume} {130}},\ \bibinfo {pages} {191901} (\bibinfo {year} {2023})},\
  \Eprint {https://arxiv.org/abs/2208.03834} {arXiv:2208.03834 [hep-lat]}
  \BibitemShut {NoStop}%
\bibitem [{\citenamefont {Chao}\ and\ \citenamefont
  {Christ}(2024)}]{Chao:2024vvl}%
  \BibitemOpen
  \bibfield  {author} {\bibinfo {author} {\bibfnamefont {E.-H.}\ \bibnamefont
  {Chao}}\ and\ \bibinfo {author} {\bibfnamefont {N.}~\bibnamefont {Christ}},\
  }\bibfield  {title} {\bibinfo {title} {{Calculating the two-photon exchange
  contribution to KL\textrightarrow{}\ensuremath{\mu}+\ensuremath{\mu}-
  decay}},\ }\href {https://doi.org/10.1103/PhysRevD.110.054514} {\bibfield
  {journal} {\bibinfo  {journal} {Phys. Rev. D}\ }\textbf {\bibinfo {volume}
  {110}},\ \bibinfo {pages} {054514} (\bibinfo {year} {2024})},\ \Eprint
  {https://arxiv.org/abs/2406.07447} {arXiv:2406.07447 [hep-ph]} \BibitemShut
  {NoStop}%
\bibitem [{\citenamefont {Bai}\ \emph {et~al.}(2014)\citenamefont {Bai},
  \citenamefont {Christ}, \citenamefont {Izubuchi}, \citenamefont {Sachrajda},
  \citenamefont {Soni},\ and\ \citenamefont {Yu}}]{Bai:2014cva}%
  \BibitemOpen
  \bibfield  {author} {\bibinfo {author} {\bibfnamefont {Z.}~\bibnamefont
  {Bai}}, \bibinfo {author} {\bibfnamefont {N.~H.}\ \bibnamefont {Christ}},
  \bibinfo {author} {\bibfnamefont {T.}~\bibnamefont {Izubuchi}}, \bibinfo
  {author} {\bibfnamefont {C.~T.}\ \bibnamefont {Sachrajda}}, \bibinfo {author}
  {\bibfnamefont {A.}~\bibnamefont {Soni}},\ and\ \bibinfo {author}
  {\bibfnamefont {J.}~\bibnamefont {Yu}},\ }\bibfield  {title} {\bibinfo
  {title} {{$K_L-K_S$ Mass Difference from Lattice QCD}},\ }\href
  {https://doi.org/10.1103/PhysRevLett.113.112003} {\bibfield  {journal}
  {\bibinfo  {journal} {Phys. Rev. Lett.}\ }\textbf {\bibinfo {volume} {113}},\
  \bibinfo {pages} {112003} (\bibinfo {year} {2014})},\ \Eprint
  {https://arxiv.org/abs/1406.0916} {arXiv:1406.0916 [hep-lat]} \BibitemShut
  {NoStop}%
\bibitem [{\citenamefont {Huo}\ \emph {et~al.}(2025)\citenamefont {Huo},
  \citenamefont {Christ},\ and\ \citenamefont {Wang}}]{Huo:2025bhq}%
  \BibitemOpen
  \bibfield  {author} {\bibinfo {author} {\bibfnamefont {Y.}~\bibnamefont
  {Huo}}, \bibinfo {author} {\bibfnamefont {N.~H.}\ \bibnamefont {Christ}},\
  and\ \bibinfo {author} {\bibfnamefont {B.}~\bibnamefont {Wang}},\ }\bibfield
  {title} {\bibinfo {title} {{Enhanced Lattice QCD Studies on $\varepsilon_{K}$
  and $\Delta M_{K}$}},\ }\href {https://doi.org/10.22323/1.466.0239}
  {\bibfield  {journal} {\bibinfo  {journal} {PoS}\ }\textbf {\bibinfo {volume}
  {LATTICE2024}},\ \bibinfo {pages} {239} (\bibinfo {year} {2025})}\BibitemShut
  {NoStop}%
\bibitem [{\citenamefont {Christ}\ \emph
  {et~al.}(2015{\natexlab{a}})\citenamefont {Christ}, \citenamefont {Feng},
  \citenamefont {Martinelli},\ and\ \citenamefont
  {Sachrajda}}]{Christ:2015pwa}%
  \BibitemOpen
  \bibfield  {author} {\bibinfo {author} {\bibfnamefont {N.~H.}\ \bibnamefont
  {Christ}}, \bibinfo {author} {\bibfnamefont {X.}~\bibnamefont {Feng}},
  \bibinfo {author} {\bibfnamefont {G.}~\bibnamefont {Martinelli}},\ and\
  \bibinfo {author} {\bibfnamefont {C.~T.}\ \bibnamefont {Sachrajda}},\
  }\bibfield  {title} {\bibinfo {title} {{Effects of finite volume on the
  $K_L$-$K_S$ mass difference}},\ }\href
  {https://doi.org/10.1103/PhysRevD.91.114510} {\bibfield  {journal} {\bibinfo
  {journal} {Phys. Rev. D}\ }\textbf {\bibinfo {volume} {91}},\ \bibinfo
  {pages} {114510} (\bibinfo {year} {2015}{\natexlab{a}})},\ \Eprint
  {https://arxiv.org/abs/1504.01170} {arXiv:1504.01170 [hep-lat]} \BibitemShut
  {NoStop}%
\bibitem [{\citenamefont {Tuo}\ and\ \citenamefont {Feng}(2025)}]{Tuo:2024bhm}%
  \BibitemOpen
  \bibfield  {author} {\bibinfo {author} {\bibfnamefont {X.-Y.}\ \bibnamefont
  {Tuo}}\ and\ \bibinfo {author} {\bibfnamefont {X.}~\bibnamefont {Feng}},\
  }\bibfield  {title} {\bibinfo {title} {{Finite-volume formalism for physical
  processes with an electroweak loop integral}},\ }\href
  {https://doi.org/10.1103/4mn6-w366} {\bibfield  {journal} {\bibinfo
  {journal} {Phys. Rev. D}\ }\textbf {\bibinfo {volume} {112}},\ \bibinfo
  {pages} {034512} (\bibinfo {year} {2025})},\ \Eprint
  {https://arxiv.org/abs/2407.16930} {arXiv:2407.16930 [hep-lat]} \BibitemShut
  {NoStop}%
\bibitem [{\citenamefont {Isidori}\ \emph {et~al.}(2006)\citenamefont
  {Isidori}, \citenamefont {Martinelli},\ and\ \citenamefont
  {Turchetti}}]{Isidori:2005tv}%
  \BibitemOpen
  \bibfield  {author} {\bibinfo {author} {\bibfnamefont {G.}~\bibnamefont
  {Isidori}}, \bibinfo {author} {\bibfnamefont {G.}~\bibnamefont
  {Martinelli}},\ and\ \bibinfo {author} {\bibfnamefont {P.}~\bibnamefont
  {Turchetti}},\ }\bibfield  {title} {\bibinfo {title} {{Rare kaon decays on
  the lattice}},\ }\href {https://doi.org/10.1016/j.physletb.2005.11.044}
  {\bibfield  {journal} {\bibinfo  {journal} {Phys. Lett. B}\ }\textbf
  {\bibinfo {volume} {633}},\ \bibinfo {pages} {75} (\bibinfo {year} {2006})},\
  \Eprint {https://arxiv.org/abs/hep-lat/0506026} {arXiv:hep-lat/0506026}
  \BibitemShut {NoStop}%
\bibitem [{\citenamefont {Buchalla}\ \emph {et~al.}(1996)\citenamefont
  {Buchalla}, \citenamefont {Buras},\ and\ \citenamefont
  {Lautenbacher}}]{Buchalla:1995vs}%
  \BibitemOpen
  \bibfield  {author} {\bibinfo {author} {\bibfnamefont {G.}~\bibnamefont
  {Buchalla}}, \bibinfo {author} {\bibfnamefont {A.~J.}\ \bibnamefont
  {Buras}},\ and\ \bibinfo {author} {\bibfnamefont {M.~E.}\ \bibnamefont
  {Lautenbacher}},\ }\bibfield  {title} {\bibinfo {title} {{Weak decays beyond
  leading logarithms}},\ }\href {https://doi.org/10.1103/RevModPhys.68.1125}
  {\bibfield  {journal} {\bibinfo  {journal} {Rev. Mod. Phys.}\ }\textbf
  {\bibinfo {volume} {68}},\ \bibinfo {pages} {1125} (\bibinfo {year}
  {1996})},\ \Eprint {https://arxiv.org/abs/hep-ph/9512380}
  {arXiv:hep-ph/9512380} \BibitemShut {NoStop}%
\bibitem [{\citenamefont {Lehner}\ and\ \citenamefont
  {Sturm}(2011)}]{Lehner:2011fz}%
  \BibitemOpen
  \bibfield  {author} {\bibinfo {author} {\bibfnamefont {C.}~\bibnamefont
  {Lehner}}\ and\ \bibinfo {author} {\bibfnamefont {C.}~\bibnamefont {Sturm}},\
  }\bibfield  {title} {\bibinfo {title} {{Matching factors for Delta S=1
  four-quark operators in RI/SMOM schemes}},\ }\href
  {https://doi.org/10.1103/PhysRevD.84.014001} {\bibfield  {journal} {\bibinfo
  {journal} {Phys. Rev. D}\ }\textbf {\bibinfo {volume} {84}},\ \bibinfo
  {pages} {014001} (\bibinfo {year} {2011})},\ \Eprint
  {https://arxiv.org/abs/1104.4948} {arXiv:1104.4948 [hep-ph]} \BibitemShut
  {NoStop}%
\bibitem [{\citenamefont {Blum}\ \emph {et~al.}(2023)\citenamefont {Blum},
  \citenamefont {Boyle}, \citenamefont {Hoying}, \citenamefont {Izubuchi},
  \citenamefont {Jin}, \citenamefont {Jung}, \citenamefont {Kelly},
  \citenamefont {Lehner}, \citenamefont {Soni},\ and\ \citenamefont
  {Tomii}}]{RBC:2023ynh}%
  \BibitemOpen
  \bibfield  {author} {\bibinfo {author} {\bibfnamefont {T.}~\bibnamefont
  {Blum}}, \bibinfo {author} {\bibfnamefont {P.~A.}\ \bibnamefont {Boyle}},
  \bibinfo {author} {\bibfnamefont {D.}~\bibnamefont {Hoying}}, \bibinfo
  {author} {\bibfnamefont {T.}~\bibnamefont {Izubuchi}}, \bibinfo {author}
  {\bibfnamefont {L.}~\bibnamefont {Jin}}, \bibinfo {author} {\bibfnamefont
  {C.}~\bibnamefont {Jung}}, \bibinfo {author} {\bibfnamefont {C.}~\bibnamefont
  {Kelly}}, \bibinfo {author} {\bibfnamefont {C.}~\bibnamefont {Lehner}},
  \bibinfo {author} {\bibfnamefont {A.}~\bibnamefont {Soni}},\ and\ \bibinfo
  {author} {\bibfnamefont {M.}~\bibnamefont {Tomii}} (\bibinfo {collaboration}
  {RBC, UKQCD}),\ }\bibfield  {title} {\bibinfo {title}
  {{\ensuremath{\Delta}I=3/2 and \ensuremath{\Delta}I=1/2 channels of
  K\textrightarrow{}\ensuremath{\pi}\ensuremath{\pi} decay at the physical
  point with periodic boundary conditions}},\ }\href
  {https://doi.org/10.1103/PhysRevD.108.094517} {\bibfield  {journal} {\bibinfo
   {journal} {Phys. Rev. D}\ }\textbf {\bibinfo {volume} {108}},\ \bibinfo
  {pages} {094517} (\bibinfo {year} {2023})},\ \Eprint
  {https://arxiv.org/abs/2306.06781} {arXiv:2306.06781 [hep-lat]} \BibitemShut
  {NoStop}%
\bibitem [{\citenamefont {Christ}\ \emph
  {et~al.}(2015{\natexlab{b}})\citenamefont {Christ}, \citenamefont {Feng},
  \citenamefont {Portelli},\ and\ \citenamefont {Sachrajda}}]{Christ:2015aha}%
  \BibitemOpen
  \bibfield  {author} {\bibinfo {author} {\bibfnamefont {N.~H.}\ \bibnamefont
  {Christ}}, \bibinfo {author} {\bibfnamefont {X.}~\bibnamefont {Feng}},
  \bibinfo {author} {\bibfnamefont {A.}~\bibnamefont {Portelli}},\ and\
  \bibinfo {author} {\bibfnamefont {C.~T.}\ \bibnamefont {Sachrajda}} (\bibinfo
  {collaboration} {RBC, UKQCD}),\ }\bibfield  {title} {\bibinfo {title}
  {{Prospects for a lattice computation of rare kaon decay amplitudes:
  $K\to\pi\ell^+\ell^-$ decays}},\ }\href
  {https://doi.org/10.1103/PhysRevD.92.094512} {\bibfield  {journal} {\bibinfo
  {journal} {Phys. Rev. D}\ }\textbf {\bibinfo {volume} {92}},\ \bibinfo
  {pages} {094512} (\bibinfo {year} {2015}{\natexlab{b}})},\ \Eprint
  {https://arxiv.org/abs/1507.03094} {arXiv:1507.03094 [hep-lat]} \BibitemShut
  {NoStop}%
\bibitem [{\citenamefont {Tu}(2020)}]{Tu:2020vpn}%
  \BibitemOpen
  \bibfield  {author} {\bibinfo {author} {\bibfnamefont {J.}~\bibnamefont
  {Tu}},\ }\emph {\bibinfo {title} {{Lattice QCD Simulations towards Strong and
  Weak Coupling Limits}}},\ \href {https://doi.org/10.7916/d8-74hj-ak60} {Ph.D.
  thesis},\ \bibinfo  {school} {Columbia U.} (\bibinfo {year}
  {2020})\BibitemShut {NoStop}%
\bibitem [{\citenamefont {Blum}\ \emph {et~al.}(2020)\citenamefont {Blum},
  \citenamefont {Christ}, \citenamefont {Hayakawa}, \citenamefont {Izubuchi},
  \citenamefont {Jin}, \citenamefont {Jung},\ and\ \citenamefont
  {Lehner}}]{Blum:2019ugy}%
  \BibitemOpen
  \bibfield  {author} {\bibinfo {author} {\bibfnamefont {T.}~\bibnamefont
  {Blum}}, \bibinfo {author} {\bibfnamefont {N.}~\bibnamefont {Christ}},
  \bibinfo {author} {\bibfnamefont {M.}~\bibnamefont {Hayakawa}}, \bibinfo
  {author} {\bibfnamefont {T.}~\bibnamefont {Izubuchi}}, \bibinfo {author}
  {\bibfnamefont {L.}~\bibnamefont {Jin}}, \bibinfo {author} {\bibfnamefont
  {C.}~\bibnamefont {Jung}},\ and\ \bibinfo {author} {\bibfnamefont
  {C.}~\bibnamefont {Lehner}},\ }\bibfield  {title} {\bibinfo {title}
  {{Hadronic Light-by-Light Scattering Contribution to the Muon Anomalous
  Magnetic Moment from Lattice QCD}},\ }\href
  {https://doi.org/10.1103/PhysRevLett.124.132002} {\bibfield  {journal}
  {\bibinfo  {journal} {Phys. Rev. Lett.}\ }\textbf {\bibinfo {volume} {124}},\
  \bibinfo {pages} {132002} (\bibinfo {year} {2020})},\ \Eprint
  {https://arxiv.org/abs/1911.08123} {arXiv:1911.08123 [hep-lat]} \BibitemShut
  {NoStop}%
\bibitem [{\citenamefont {McGlynn}(2016)}]{Mcglynn:2015uwh}%
  \BibitemOpen
  \bibfield  {author} {\bibinfo {author} {\bibfnamefont {G.}~\bibnamefont
  {McGlynn}},\ }\bibfield  {title} {\bibinfo {title} {{Algorithmic improvements
  for weak coupling simulations of domain wall fermions}},\ }\href
  {https://doi.org/10.22323/1.251.0019} {\bibfield  {journal} {\bibinfo
  {journal} {PoS}\ }\textbf {\bibinfo {volume} {LATTICE2015}},\ \bibinfo
  {pages} {019} (\bibinfo {year} {2016})}\BibitemShut {NoStop}%
\bibitem [{\citenamefont {Yin}\ and\ \citenamefont
  {Mawhinney}(2011)}]{Yin:2011np}%
  \BibitemOpen
  \bibfield  {author} {\bibinfo {author} {\bibfnamefont {H.}~\bibnamefont
  {Yin}}\ and\ \bibinfo {author} {\bibfnamefont {R.~D.}\ \bibnamefont
  {Mawhinney}},\ }\bibfield  {title} {\bibinfo {title} {{Improving DWF
  Simulations: the Force Gradient Integrator and the M\"obius Accelerated DWF
  Solver}},\ }\href {https://doi.org/10.22323/1.139.0051} {\bibfield  {journal}
  {\bibinfo  {journal} {PoS}\ }\textbf {\bibinfo {volume} {LATTICE2011}},\
  \bibinfo {pages} {051} (\bibinfo {year} {2011})},\ \Eprint
  {https://arxiv.org/abs/1111.5059} {arXiv:1111.5059 [hep-lat]} \BibitemShut
  {NoStop}%
\bibitem [{\citenamefont {Chao}(2024)}]{Chao:2024cnu}%
  \BibitemOpen
  \bibfield  {author} {\bibinfo {author} {\bibfnamefont {E.-H.}\ \bibnamefont
  {Chao}},\ }\bibfield  {title} {\bibinfo {title} {{Progress on $K_{\rm
  L}\rightarrow\mu^+\mu^-$ from Lattice QCD}},\ }in\ \href@noop {} {\emph
  {\bibinfo {booktitle} {{12th International Workshop on the CKM Unitarity
  Triangle}}}}\ (\bibinfo {year} {2024})\ \Eprint
  {https://arxiv.org/abs/2403.18885} {arXiv:2403.18885 [hep-lat]} \BibitemShut
  {NoStop}%
\bibitem [{\citenamefont {Bernard}\ \emph {et~al.}(1985)\citenamefont
  {Bernard}, \citenamefont {Draper}, \citenamefont {Soni}, \citenamefont
  {Politzer},\ and\ \citenamefont {Wise}}]{Bernard:1985wf}%
  \BibitemOpen
  \bibfield  {author} {\bibinfo {author} {\bibfnamefont {C.~W.}\ \bibnamefont
  {Bernard}}, \bibinfo {author} {\bibfnamefont {T.}~\bibnamefont {Draper}},
  \bibinfo {author} {\bibfnamefont {A.}~\bibnamefont {Soni}}, \bibinfo {author}
  {\bibfnamefont {H.~D.}\ \bibnamefont {Politzer}},\ and\ \bibinfo {author}
  {\bibfnamefont {M.~B.}\ \bibnamefont {Wise}},\ }\bibfield  {title} {\bibinfo
  {title} {{Application of Chiral Perturbation Theory to K ---\ensuremath{>} 2
  pi Decays}},\ }\href {https://doi.org/10.1103/PhysRevD.32.2343} {\bibfield
  {journal} {\bibinfo  {journal} {Phys. Rev. D}\ }\textbf {\bibinfo {volume}
  {32}},\ \bibinfo {pages} {2343} (\bibinfo {year} {1985})}\BibitemShut
  {NoStop}%
\bibitem [{\citenamefont {Boyle}(2015)}]{Boyle:2015mb}%
  \BibitemOpen
  \bibfield  {author} {\bibinfo {author} {\bibfnamefont {P.~A.}\ \bibnamefont
  {Boyle}},\ }\bibfield  {title} {\bibinfo {title} {{Conserved currents and
  results from 2+1f dynamical Mobius DWF simulations at the physical point}},\
  }\href {https://doi.org/10.22323/1.214.0087} {\bibfield  {journal} {\bibinfo
  {journal} {PoS}\ }\textbf {\bibinfo {volume} {LATTICE2014}},\ \bibinfo
  {pages} {087} (\bibinfo {year} {2015})}\BibitemShut {NoStop}%
\bibitem [{\citenamefont {Blossier}\ \emph {et~al.}(2009)\citenamefont
  {Blossier}, \citenamefont {Della~Morte}, \citenamefont {von Hippel},
  \citenamefont {Mendes},\ and\ \citenamefont {Sommer}}]{Blossier:2009kd}%
  \BibitemOpen
  \bibfield  {author} {\bibinfo {author} {\bibfnamefont {B.}~\bibnamefont
  {Blossier}}, \bibinfo {author} {\bibfnamefont {M.}~\bibnamefont
  {Della~Morte}}, \bibinfo {author} {\bibfnamefont {G.}~\bibnamefont {von
  Hippel}}, \bibinfo {author} {\bibfnamefont {T.}~\bibnamefont {Mendes}},\ and\
  \bibinfo {author} {\bibfnamefont {R.}~\bibnamefont {Sommer}},\ }\bibfield
  {title} {\bibinfo {title} {{On the generalized eigenvalue method for energies
  and matrix elements in lattice field theory}},\ }\href
  {https://doi.org/10.1088/1126-6708/2009/04/094} {\bibfield  {journal}
  {\bibinfo  {journal} {JHEP}\ }\textbf {\bibinfo {volume} {04}},\ \bibinfo
  {pages} {094}},\ \Eprint {https://arxiv.org/abs/0902.1265} {arXiv:0902.1265
  [hep-lat]} \BibitemShut {NoStop}%
\bibitem [{\citenamefont {Gomez~Dumm}\ and\ \citenamefont
  {Pich}(1998)}]{GomezDumm:1998gw}%
  \BibitemOpen
  \bibfield  {author} {\bibinfo {author} {\bibfnamefont {D.}~\bibnamefont
  {Gomez~Dumm}}\ and\ \bibinfo {author} {\bibfnamefont {A.}~\bibnamefont
  {Pich}},\ }\bibfield  {title} {\bibinfo {title} {{Long distance contributions
  to the K(L) ---\ensuremath{>} mu+ mu- decay width}},\ }\href
  {https://doi.org/10.1103/PhysRevLett.80.4633} {\bibfield  {journal} {\bibinfo
   {journal} {Phys. Rev. Lett.}\ }\textbf {\bibinfo {volume} {80}},\ \bibinfo
  {pages} {4633} (\bibinfo {year} {1998})},\ \Eprint
  {https://arxiv.org/abs/hep-ph/9801298} {arXiv:hep-ph/9801298} \BibitemShut
  {NoStop}%
\bibitem [{\citenamefont {Buchalla}\ and\ \citenamefont
  {Buras}(1994)}]{Buchalla:1993wq}%
  \BibitemOpen
  \bibfield  {author} {\bibinfo {author} {\bibfnamefont {G.}~\bibnamefont
  {Buchalla}}\ and\ \bibinfo {author} {\bibfnamefont {A.~J.}\ \bibnamefont
  {Buras}},\ }\bibfield  {title} {\bibinfo {title} {{The rare decays $K^+ \to
  \pi^+ \nu \bar \nu$ and $K_L \to \mu^+ \mu^-$ beyond leading logarithms}},\
  }\href {https://doi.org/10.1016/0550-3213(94)90496-0} {\bibfield  {journal}
  {\bibinfo  {journal} {Nucl. Phys. B}\ }\textbf {\bibinfo {volume} {412}},\
  \bibinfo {pages} {106} (\bibinfo {year} {1994})},\ \Eprint
  {https://arxiv.org/abs/hep-ph/9308272} {arXiv:hep-ph/9308272} \BibitemShut
  {NoStop}%
\bibitem [{\citenamefont {Feng}\ \emph {et~al.}(2022)\citenamefont {Feng},
  \citenamefont {Jin},\ and\ \citenamefont {Riberdy}}]{Feng:2021zek}%
  \BibitemOpen
  \bibfield  {author} {\bibinfo {author} {\bibfnamefont {X.}~\bibnamefont
  {Feng}}, \bibinfo {author} {\bibfnamefont {L.}~\bibnamefont {Jin}},\ and\
  \bibinfo {author} {\bibfnamefont {M.~J.}\ \bibnamefont {Riberdy}},\
  }\bibfield  {title} {\bibinfo {title} {{Lattice QCD Calculation of the Pion
  Mass Splitting}},\ }\href {https://doi.org/10.1103/PhysRevLett.128.052003}
  {\bibfield  {journal} {\bibinfo  {journal} {Phys. Rev. Lett.}\ }\textbf
  {\bibinfo {volume} {128}},\ \bibinfo {pages} {052003} (\bibinfo {year}
  {2022})},\ \Eprint {https://arxiv.org/abs/2108.05311} {arXiv:2108.05311
  [hep-lat]} \BibitemShut {NoStop}%
\bibitem [{\citenamefont {Buchalla}\ and\ \citenamefont
  {Buras}(1998)}]{Buchalla:1997kz}%
  \BibitemOpen
  \bibfield  {author} {\bibinfo {author} {\bibfnamefont {G.}~\bibnamefont
  {Buchalla}}\ and\ \bibinfo {author} {\bibfnamefont {A.~J.}\ \bibnamefont
  {Buras}},\ }\bibfield  {title} {\bibinfo {title} {{Two loop large $m_t$
  electroweak corrections to $K \to \pi \nu \bar \nu$ for arbitrary Higgs boson
  mass}},\ }\href {https://doi.org/10.1103/PhysRevD.57.216} {\bibfield
  {journal} {\bibinfo  {journal} {Phys. Rev. D}\ }\textbf {\bibinfo {volume}
  {57}},\ \bibinfo {pages} {216} (\bibinfo {year} {1998})},\ \Eprint
  {https://arxiv.org/abs/hep-ph/9707243} {arXiv:hep-ph/9707243} \BibitemShut
  {NoStop}%
\bibitem [{\citenamefont {Montvay}\ and\ \citenamefont
  {Munster}(1997)}]{Montvay:1994cy}%
  \BibitemOpen
  \bibfield  {author} {\bibinfo {author} {\bibfnamefont {I.}~\bibnamefont
  {Montvay}}\ and\ \bibinfo {author} {\bibfnamefont {G.}~\bibnamefont
  {Munster}},\ }\href {https://doi.org/10.1017/CBO9780511470783} {\emph
  {\bibinfo {title} {{Quantum fields on a lattice}}}},\ Cambridge Monographs on
  Mathematical Physics\ (\bibinfo  {publisher} {Cambridge University Press},\
  \bibinfo {year} {1997})\BibitemShut {NoStop}%
\bibitem [{\citenamefont {Gattringer}\ and\ \citenamefont
  {Lang}(2010)}]{Gattringer:2010zz}%
  \BibitemOpen
  \bibfield  {author} {\bibinfo {author} {\bibfnamefont {C.}~\bibnamefont
  {Gattringer}}\ and\ \bibinfo {author} {\bibfnamefont {C.~B.}\ \bibnamefont
  {Lang}},\ }\href {https://doi.org/10.1007/978-3-642-01850-3} {\emph {\bibinfo
  {title} {{Quantum chromodynamics on the lattice}}}},\ Vol.\ \bibinfo {volume}
  {788}\ (\bibinfo  {publisher} {Springer},\ \bibinfo {address} {Berlin},\
  \bibinfo {year} {2010})\BibitemShut {NoStop}%
\bibitem [{\citenamefont {Boyle}\ \emph {et~al.}(2016)\citenamefont {Boyle},
  \citenamefont {Cossu}, \citenamefont {Yamaguchi},\ and\ \citenamefont
  {Portelli}}]{Boyle:2016lf}%
  \BibitemOpen
  \bibfield  {author} {\bibinfo {author} {\bibfnamefont {P.~A.}\ \bibnamefont
  {Boyle}}, \bibinfo {author} {\bibfnamefont {G.}~\bibnamefont {Cossu}},
  \bibinfo {author} {\bibfnamefont {A.}~\bibnamefont {Yamaguchi}},\ and\
  \bibinfo {author} {\bibfnamefont {A.}~\bibnamefont {Portelli}},\ }\bibfield
  {title} {\bibinfo {title} {{Grid: A next generation data parallel C++ QCD
  library}},\ }\href {https://doi.org/10.22323/1.251.0023} {\bibfield
  {journal} {\bibinfo  {journal} {PoS}\ }\textbf {\bibinfo {volume} {LATTICE
  2015}},\ \bibinfo {pages} {023} (\bibinfo {year} {2016})}\BibitemShut
  {NoStop}%
\bibitem [{\citenamefont {Yamaguchi}\ \emph {et~al.}(2022)\citenamefont
  {Yamaguchi}, \citenamefont {Boyle}, \citenamefont {Cossu}, \citenamefont
  {Filaci}, \citenamefont {Lehner},\ and\ \citenamefont
  {Portelli}}]{Yamaguchi:2022feu}%
  \BibitemOpen
  \bibfield  {author} {\bibinfo {author} {\bibfnamefont {A.}~\bibnamefont
  {Yamaguchi}}, \bibinfo {author} {\bibfnamefont {P.}~\bibnamefont {Boyle}},
  \bibinfo {author} {\bibfnamefont {G.}~\bibnamefont {Cossu}}, \bibinfo
  {author} {\bibfnamefont {G.}~\bibnamefont {Filaci}}, \bibinfo {author}
  {\bibfnamefont {C.}~\bibnamefont {Lehner}},\ and\ \bibinfo {author}
  {\bibfnamefont {A.}~\bibnamefont {Portelli}},\ }\bibfield  {title} {\bibinfo
  {title} {{Grid: OneCode and FourAPIs}},\ }\href
  {https://doi.org/10.22323/1.396.0035} {\bibfield  {journal} {\bibinfo
  {journal} {PoS}\ }\textbf {\bibinfo {volume} {LATTICE2021}},\ \bibinfo
  {pages} {035} (\bibinfo {year} {2022})},\ \Eprint
  {https://arxiv.org/abs/2203.06777} {arXiv:2203.06777 [hep-lat]} \BibitemShut
  {NoStop}%
\bibitem [{\citenamefont {Jung}(2014)}]{Jung:2014XW}%
  \BibitemOpen
  \bibfield  {author} {\bibinfo {author} {\bibfnamefont {C.}~\bibnamefont
  {Jung}},\ }\bibfield  {title} {\bibinfo {title} {{Overview of Columbia
  Physics System(CPS)}},\ }\href {https://doi.org/10.22323/1.187.0417}
  {\bibfield  {journal} {\bibinfo  {journal} {PoS}\ }\textbf {\bibinfo {volume}
  {LATTICE 2013}},\ \bibinfo {pages} {417} (\bibinfo {year}
  {2014})}\BibitemShut {NoStop}%
\bibitem [{\citenamefont {Edwards}\ and\ \citenamefont
  {Joo}(2005)}]{Edwards:2004sx}%
  \BibitemOpen
  \bibfield  {author} {\bibinfo {author} {\bibfnamefont {R.~G.}\ \bibnamefont
  {Edwards}}\ and\ \bibinfo {author} {\bibfnamefont {B.}~\bibnamefont {Joo}}
  (\bibinfo {collaboration} {SciDAC, LHPC, UKQCD}),\ }\bibfield  {title}
  {\bibinfo {title} {{The Chroma software system for lattice QCD}},\ }\href
  {https://doi.org/10.1016/j.nuclphysbps.2004.11.254} {\bibfield  {journal}
  {\bibinfo  {journal} {Nucl. Phys. B Proc. Suppl.}\ }\textbf {\bibinfo
  {volume} {140}},\ \bibinfo {pages} {832} (\bibinfo {year} {2005})},\ \Eprint
  {https://arxiv.org/abs/hep-lat/0409003} {arXiv:hep-lat/0409003} \BibitemShut
  {NoStop}%
\bibitem [{qdp()}]{qdpxx}%
  \BibitemOpen
  \href@noop {} {\bibinfo {title} {{QDP}++}},\ \bibinfo {howpublished}
  {\url{https://github.com/usqcd-software/qdpxx/tree/master}}\BibitemShut
  {NoStop}%
\bibitem [{\citenamefont {Blum}\ \emph
  {et~al.}(2016{\natexlab{b}})\citenamefont {Blum} \emph
  {et~al.}}]{RBC:2014ntl}%
  \BibitemOpen
  \bibfield  {author} {\bibinfo {author} {\bibfnamefont {T.}~\bibnamefont
  {Blum}} \emph {et~al.} (\bibinfo {collaboration} {RBC, UKQCD}),\ }\bibfield
  {title} {\bibinfo {title} {{Domain wall QCD with physical quark masses}},\
  }\href {https://doi.org/10.1103/PhysRevD.93.074505} {\bibfield  {journal}
  {\bibinfo  {journal} {Phys. Rev. D}\ }\textbf {\bibinfo {volume} {93}},\
  \bibinfo {pages} {074505} (\bibinfo {year} {2016}{\natexlab{b}})},\ \Eprint
  {https://arxiv.org/abs/1411.7017} {arXiv:1411.7017 [hep-lat]} \BibitemShut
  {NoStop}%
\bibitem [{\citenamefont {Zhao}(2022)}]{Zhao:2022njd}%
  \BibitemOpen
  \bibfield  {author} {\bibinfo {author} {\bibfnamefont {Y.}~\bibnamefont
  {Zhao}},\ }\emph {\bibinfo {title} {{Lattice Calculation of the $\pi^0
  \rightarrow e^+ e^-$ and the $K_L \rightarrow \gamma \gamma$ Decays}}},\
  \href {https://doi.org/10.7916/mbap-bh60} {Ph.D. thesis},\ \bibinfo  {school}
  {Columbia U.} (\bibinfo {year} {2022})\BibitemShut {NoStop}%
\bibitem [{\citenamefont {Foley}\ \emph {et~al.}(2005)\citenamefont {Foley},
  \citenamefont {Jimmy~Juge}, \citenamefont {O'Cais}, \citenamefont {Peardon},
  \citenamefont {Ryan},\ and\ \citenamefont {Skullerud}}]{Foley:2005ac}%
  \BibitemOpen
  \bibfield  {author} {\bibinfo {author} {\bibfnamefont {J.}~\bibnamefont
  {Foley}}, \bibinfo {author} {\bibfnamefont {K.}~\bibnamefont {Jimmy~Juge}},
  \bibinfo {author} {\bibfnamefont {A.}~\bibnamefont {O'Cais}}, \bibinfo
  {author} {\bibfnamefont {M.}~\bibnamefont {Peardon}}, \bibinfo {author}
  {\bibfnamefont {S.~M.}\ \bibnamefont {Ryan}},\ and\ \bibinfo {author}
  {\bibfnamefont {J.-I.}\ \bibnamefont {Skullerud}},\ }\bibfield  {title}
  {\bibinfo {title} {{Practical all-to-all propagators for lattice QCD}},\
  }\href {https://doi.org/10.1016/j.cpc.2005.06.008} {\bibfield  {journal}
  {\bibinfo  {journal} {Comput. Phys. Commun.}\ }\textbf {\bibinfo {volume}
  {172}},\ \bibinfo {pages} {145} (\bibinfo {year} {2005})},\ \Eprint
  {https://arxiv.org/abs/hep-lat/0505023} {arXiv:hep-lat/0505023} \BibitemShut
  {NoStop}%
\bibitem [{\citenamefont {Giusti}\ \emph {et~al.}(2004)\citenamefont {Giusti},
  \citenamefont {Hernandez}, \citenamefont {Laine}, \citenamefont {Weisz},\
  and\ \citenamefont {Wittig}}]{Giusti:2004yp}%
  \BibitemOpen
  \bibfield  {author} {\bibinfo {author} {\bibfnamefont {L.}~\bibnamefont
  {Giusti}}, \bibinfo {author} {\bibfnamefont {P.}~\bibnamefont {Hernandez}},
  \bibinfo {author} {\bibfnamefont {M.}~\bibnamefont {Laine}}, \bibinfo
  {author} {\bibfnamefont {P.}~\bibnamefont {Weisz}},\ and\ \bibinfo {author}
  {\bibfnamefont {H.}~\bibnamefont {Wittig}},\ }\bibfield  {title} {\bibinfo
  {title} {{Low-energy couplings of QCD from current correlators near the
  chiral limit}},\ }\href {https://doi.org/10.1088/1126-6708/2004/04/013}
  {\bibfield  {journal} {\bibinfo  {journal} {JHEP}\ }\textbf {\bibinfo
  {volume} {04}},\ \bibinfo {pages} {013}},\ \Eprint
  {https://arxiv.org/abs/hep-lat/0402002} {arXiv:hep-lat/0402002} \BibitemShut
  {NoStop}%
\bibitem [{\citenamefont {Bijnens}\ \emph {et~al.}(2003)\citenamefont
  {Bijnens}, \citenamefont {Dhonte},\ and\ \citenamefont
  {Borg}}]{Bijnens:2002vr}%
  \BibitemOpen
  \bibfield  {author} {\bibinfo {author} {\bibfnamefont {J.}~\bibnamefont
  {Bijnens}}, \bibinfo {author} {\bibfnamefont {P.}~\bibnamefont {Dhonte}},\
  and\ \bibinfo {author} {\bibfnamefont {F.}~\bibnamefont {Borg}},\ }\bibfield
  {title} {\bibinfo {title} {{K ---\ensuremath{>} 3 pi decays in chiral
  perturbation theory}},\ }\href
  {https://doi.org/10.1016/S0550-3213(02)00970-7} {\bibfield  {journal}
  {\bibinfo  {journal} {Nucl. Phys. B}\ }\textbf {\bibinfo {volume} {648}},\
  \bibinfo {pages} {317} (\bibinfo {year} {2003})},\ \Eprint
  {https://arxiv.org/abs/hep-ph/0205341} {arXiv:hep-ph/0205341} \BibitemShut
  {NoStop}%
\bibitem [{\citenamefont {Weinberg}(1960)}]{Weinberg:1960zza}%
  \BibitemOpen
  \bibfield  {author} {\bibinfo {author} {\bibfnamefont {S.}~\bibnamefont
  {Weinberg}},\ }\bibfield  {title} {\bibinfo {title} {{New Test for DeltaI=12
  in K+ Decay}},\ }\href {https://doi.org/10.1103/PhysRevLett.4.87} {\bibfield
  {journal} {\bibinfo  {journal} {Phys. Rev. Lett.}\ }\textbf {\bibinfo
  {volume} {4}},\ \bibinfo {pages} {87} (\bibinfo {year} {1960})},\ \bibinfo
  {note} {[Erratum: Phys.Rev.Lett. 4, 585 (1960)]}\BibitemShut {NoStop}%
\bibitem [{\citenamefont {Wess}\ and\ \citenamefont
  {Zumino}(1971)}]{Wess:1971yu}%
  \BibitemOpen
  \bibfield  {author} {\bibinfo {author} {\bibfnamefont {J.}~\bibnamefont
  {Wess}}\ and\ \bibinfo {author} {\bibfnamefont {B.}~\bibnamefont {Zumino}},\
  }\bibfield  {title} {\bibinfo {title} {{Consequences of anomalous Ward
  identities}},\ }\href {https://doi.org/10.1016/0370-2693(71)90582-X}
  {\bibfield  {journal} {\bibinfo  {journal} {Phys. Lett. B}\ }\textbf
  {\bibinfo {volume} {37}},\ \bibinfo {pages} {95} (\bibinfo {year}
  {1971})}\BibitemShut {NoStop}%
\bibitem [{\citenamefont {Witten}(1983)}]{Witten:1983tw}%
  \BibitemOpen
  \bibfield  {author} {\bibinfo {author} {\bibfnamefont {E.}~\bibnamefont
  {Witten}},\ }\bibfield  {title} {\bibinfo {title} {{Global Aspects of Current
  Algebra}},\ }\href {https://doi.org/10.1016/0550-3213(83)90063-9} {\bibfield
  {journal} {\bibinfo  {journal} {Nucl. Phys. B}\ }\textbf {\bibinfo {volume}
  {223}},\ \bibinfo {pages} {422} (\bibinfo {year} {1983})}\BibitemShut
  {NoStop}%
\bibitem [{\citenamefont {Savage}\ \emph {et~al.}(1992)\citenamefont {Savage},
  \citenamefont {Luke},\ and\ \citenamefont {Wise}}]{Savage:1992ac}%
  \BibitemOpen
  \bibfield  {author} {\bibinfo {author} {\bibfnamefont {M.~J.}\ \bibnamefont
  {Savage}}, \bibinfo {author} {\bibfnamefont {M.~E.}\ \bibnamefont {Luke}},\
  and\ \bibinfo {author} {\bibfnamefont {M.~B.}\ \bibnamefont {Wise}},\
  }\bibfield  {title} {\bibinfo {title} {{The Rare decays pi0 ---\ensuremath{>}
  e+ e-, eta ---\ensuremath{>} e+ e- and eta ---\ensuremath{>} mu+ mu- in
  chiral perturbation theory}},\ }\href
  {https://doi.org/10.1016/0370-2693(92)91407-Z} {\bibfield  {journal}
  {\bibinfo  {journal} {Phys. Lett. B}\ }\textbf {\bibinfo {volume} {291}},\
  \bibinfo {pages} {481} (\bibinfo {year} {1992})},\ \Eprint
  {https://arxiv.org/abs/hep-ph/9207233} {arXiv:hep-ph/9207233} \BibitemShut
  {NoStop}%
\bibitem [{\citenamefont {Scherer}\ and\ \citenamefont
  {Schindler}(2012)}]{Scherer:2012xha}%
  \BibitemOpen
  \bibfield  {author} {\bibinfo {author} {\bibfnamefont {S.}~\bibnamefont
  {Scherer}}\ and\ \bibinfo {author} {\bibfnamefont {M.~R.}\ \bibnamefont
  {Schindler}},\ }\href {https://doi.org/10.1007/978-3-642-19254-8} {\emph
  {\bibinfo {title} {{A Primer for Chiral Perturbation Theory}}}},\ Vol.\
  \bibinfo {volume} {830}\ (\bibinfo {year} {2012})\BibitemShut {NoStop}%
\bibitem [{\citenamefont {Hoferichter}\ \emph {et~al.}(2022)\citenamefont
  {Hoferichter}, \citenamefont {Hoid}, \citenamefont {Kubis},\ and\
  \citenamefont {L\"udtke}}]{Hoferichter:2021lct}%
  \BibitemOpen
  \bibfield  {author} {\bibinfo {author} {\bibfnamefont {M.}~\bibnamefont
  {Hoferichter}}, \bibinfo {author} {\bibfnamefont {B.-L.}\ \bibnamefont
  {Hoid}}, \bibinfo {author} {\bibfnamefont {B.}~\bibnamefont {Kubis}},\ and\
  \bibinfo {author} {\bibfnamefont {J.}~\bibnamefont {L\"udtke}},\ }\bibfield
  {title} {\bibinfo {title} {{Improved Standard-Model prediction for $\pi^0\to
  e^+e^-$}},\ }\href {https://doi.org/10.1103/PhysRevLett.128.172004}
  {\bibfield  {journal} {\bibinfo  {journal} {Phys. Rev. Lett.}\ }\textbf
  {\bibinfo {volume} {128}},\ \bibinfo {pages} {172004} (\bibinfo {year}
  {2022})},\ \Eprint {https://arxiv.org/abs/2105.04563} {arXiv:2105.04563
  [hep-ph]} \BibitemShut {NoStop}%
\end{thebibliography}%

\end{document}